\documentclass{aa}  

\usepackage{graphicx}
\usepackage{amssymb}

\usepackage{color}

\usepackage{hyperref}
\hypersetup{
  colorlinks = true,        % false: boxed links; true: colored links
  linkcolor = blue,         % color of internal links
  citecolor = cyan,         % color of links to bibliography
}

\usepackage{txfonts}
\graphicspath{{./figures/}}
\bibpunct{(}{)}{;}{a}{}{,} % to follow the A&A style

\newcommand{\ie}{i.e.~}
\newcommand{\eg}{e.g.~}
\usepackage{amstext}

\begin{document}

\title{Simulations of recoiling black holes: adaptive mesh refinement and
  radiative transfer}

%   \subtitle{}

\author{Zakaria Meliani\inst{\ref{inst1}}
\and
Yosuke Mizuno\inst{\ref{inst2}}
\and
Hector Olivares\inst{\ref{inst2}}
\and
Oliver Porth\inst{\ref{inst2}}
\and
Luciano Rezzolla\inst{\ref{inst2}, \ref{inst3}}
\and
Ziri Younsi\inst{\ref{inst2}}
}

\institute{LUTH, CNRS UMR 8102, Observatoire de Paris, Universit\'e Paris
 Diderot, 92190 Meudon, France
%% \email{zakaria.meliani@obspm.fr}
\label{inst1}
\and
Institute for Theoretical Physics, Goethe-University, D-60438,
Frankfurt am Main, Germany \label{inst2}
\and
Frankfurt Institute for Advanced Studies, D-60438, Frankfurt am Main,
Germany \label{inst3} 
}

\date{\today}

\abstract
{In many astrophysical phenomena, and especially in those that involve the
  high-energy regimes that always accompany the astronomical
  phenomenology of black holes and neutron stars, physical conditions
  that are achieved are extreme in terms of speeds, temperatures,
  and gravitational fields. In such relativistic regimes, numerical
  calculations are the only tool to accurately model the dynamics of the
  flows and the transport of radiation in the accreting matter.}
{We here continue our effort of modelling the behaviour of matter when
it  orbits or is accreted onto a generic black hole by developing a new
  numerical code that employs advanced techniques geared towards
    solving the equations of general-relativistic hydrodynamics.}
{More specifically, the new code employs a number of high-resolution
  shock-capturing Riemann solvers and reconstruction algorithms,
  exploiting the enhanced accuracy and the reduced computational cost of
  adaptive mesh-refinement (AMR) techniques. In addition, the code makes
  use of sophisticated ray-tracing libraries that, coupled with
  general-relativistic radiation-transfer calculations, allow us to
accurately compute the electromagnetic emissions from such accretion
  flows.}
{We validate the new code by presenting an extensive series of stationary
  accretion flows either in spherical or axial symmetry that
are performed
  either in two or three spatial dimensions. In addition, we consider the
  highly nonlinear scenario of a recoiling black hole produced in the
  merger of a supermassive black-hole binary interacting with the
  surrounding circumbinary disc. In this way, we can present for the
  first time ray-traced images of the shocked fluid and the light
curve
  resulting from consistent general-relativistic radiation-transport
  calculations from this process.}
 {The work presented here lays the ground for the development of a
   generic computational infrastructure employing AMR techniques to   accurately and self-consistently calculate general-relativistic accretion
   flows onto compact objects. In addition to the accurate handling of
   the matter, we provide a self-consistent electromagnetic emission from
   these scenarios by solving the associated radiative-transfer
   problem. While magnetic fields are currently excluded from our
   analysis, the tools presented here can have a number of applications
   to study accretion flows onto black holes or neutron stars.}

   
\titlerunning{GRHD simulations of recoiling black holes: AMR and
  radiation transfer} 
\authorrunning{Z. Meliani et al.}
 
\maketitle

%ssssssssssssssssssssssssssssssssssssssssssssssssssssssssssssssssss
\section{Introduction}\label{sec:intro}
%ssssssssssssssssssssssssssssssssssssssssssssssssssssssssssssssssss

Many astrophysical phenomena are complex and subject to nonlinear
dynamics, making numerical simulations an indispensable tool for their
study. In the high-energy regimes that always accompany the astronomical
phenomenology of compact objects, physical conditions are extreme, with
speeds and temperatures so high and gravitational fields so large that both
relativistic and general-relativistic effects must be taken into account.
In events involving compact objects such as black holes,
Einstein's theory of gravitation plays a crucial role, and it is
imperative to use it to model accretion flows and radiation therein. In
addition to having to model dynamics that are often highly nonlinear,
the simulation of compact objects also requires the ability  
to follow
physical phenomena that occur across multiple scales and so must be
resolved simultaneously on small and large scales, which requires large
amounts of computational resources that cannot be sustained. Adaptive
mesh refinement (AMR) provides an effective solution to the problem of
performing simulations of phenomena where it is necessary to resolve global as well as local scales.
  
Over the past few years, great advances in numerical general relativity
have given rise to the development of numerical schemes employing the
$3+1$ formulation and Godunov schemes based on approximate Riemann
solvers \citet{Rezzolla_book:2013}. These advances in numerical general
relativity are best described in the reviews by \citet{Font03} and by
\citet{Marti2015}, which provide a thorough description of
high-resolution shock-capturing schemes in general-relativistic
hydrodynamics (GRHD). Many general-relativistic hydrodynamic and
magnetohydrodynamic codes have been developed and evolved over the past
three decades \citep{Hawley84a, Koide00, DeVilliers03a,
  Gammie03, Baiotti04, Duez05MHD0, Anninos05c,Anton05,
  Mizuno06, DelZanna2007, Giacomazzo:2007ti, Radice2012a,
  Radice2013b, McKinney2014, Etienne2015, White2015,
  Zanotti2015}. Some of these implementations provide additional
capabilities that incorporate radiation transfer in approximate ways
\citep[\eg][]{Sadowski2013} and/or non-ideal magnetohydrodynamics (MHD) regimes
\citep[\eg][]{Dionysopoulou:2012pp, Foucart2015b}. These codes have been
applied to many astrophysical scenarios involving compact objects and
matter. They have been applied to model accretion-ejection,
magnetospheres, and compact star structure collapse
\citep[\eg][]{Dibiet12,Fragile2014,McKinney2014}.

In some astrophysical scenarios, adequate modelling can become extremely
challenging because of the large disparities in the temporal and spatial
scales that may arise in the problem of interest. Under these conditions,
approaches employing uniform and non-adaptive grids may become less
efficient. These limitations can be overcome by using AMR with adequate refinement or coarsening conditions to
sufficiently capture features of interest. The ideal AMR implementation
is meant to provide high-resolution simulations at much lower
computational cost than uniform-grid methods are capable of. Various AMR strategies exist, such as the patch-based blocks used in
\texttt{ASTROBEAR} \citep{Cunninghametal09}, or the full-octree
implementations employed in \texttt{RAMSES} \citep{Teyssier2002}. The
strategy implemented in the code used in this paper is the block-octree
approach \citep{vanderHolst2008}.

In this paper, we focus on GRHD applications, motivated mainly by our own
continued efforts in augmenting the wealth of community codes available
for astrophysical research. We first discuss the implementation of
general relativistic hydrodynamics with a static background metric. This
is then followed by the test of our shock-capturing scheme for GRHD using
AMR strategies, which constitutes the core component of modern code
development. Code tests with static black hole metrics using two
coordinate systems, namely, Boyer-Lindquist and Kerr-Schild (KS), are
discussed.

Using two-dimensional (2D) and three-dimensional (3D)
general-relativistic numerical simulations that incorporate local AMR, we
study the dynamics of a torus in orbit around a recoiling black hole
\citep[see, \eg][for an introductory review on recoiling black
  holes]{Rezzolla:2008sd}. Such a kick is likely to result from the
merger of supermassive binary black hole systems (SMBBHs). We then
calculate the electromagnetic emission from these simulations (images and light curves).  In addition to being a perfect testbed for its highly
nonlinear and out-of-equilibrium dynamics, the study of the interaction
of a recoiling black hole with the surrounding matter has a precise
astrophysical application. The analysis of the accretion rate
and of the resulting electromagnetic counterparts of recoiling SMBBHs is
of great scientific interest, as it will enable the prediction of
recoiling signatures when signals from these sources will be detected by
the planned space-borne gravitational-wave detector eLISA
\citep{Amaro2012b}.  This is indeed a well-explored area of research,
  and several studies have investigated the 2D dynamics resulting from a
  recoiling black hole \cite[\eg][]{Corrales2009, Zanotti2010,
    Zanotti2012a} and 3D simulations \cite[\eg][]{Lippai:2008,
    Megevand2009, Andersonetal10, Ponce2012}.  We assess the
performance and accuracy of local AMR to perform long-term
recoiling black hole simulations within a reasonable amount of
computational time.

The structure of the paper is as follows: In Sect. 2 we describe the
governing equations, numerical methods for their solution, and numerical
test simulations. In Sect. 3 the results of 2D and 3D GRHD
simulations of recoiling black holes are presented. In Sect. 4 we describe
the general-relativistic radiative transfer formulation and underlying
radiative emission model and apply this to the GRHD simulations of
recoiling black holes described in Sect. 3. In Sect. 5 we present our
conclusions.

Throughout this paper, we use units where the speed of light, $c=1$, the
gravitational constant, $G=1$, and gas mass is normalised to the central
compact object mass. Greek indices run over space and time, that
is, $(0,1,2,3),$ and Roman indices run over space only, that
is, $(1,2,\text{and }3)$. We
assume a signature $(-,+,+,+)$ for the space-time metric. Self-gravity
arising from the gas is neglected, and all simulations presented here
  are made using polar spherical coordinates even though the code
  also allows for other choices of coordinates.

%ssssssssssssssssssssssssssssssssssssssssssssssssssssssssssssssssss
\section{Numerical methods and benchmarks}\label{sec:numerics}
%ssssssssssssssssssssssssssssssssssssssssssssssssssssssssssssssssss

%ssssssssssssssssssssssssssssssssssssssssssssssssssssssssssssssssss
\subsection{GRHD equations and numerical methods}
\label{sec:GRHDeq}
%ssssssssssssssssssssssssssssssssssssssssssssssssssssssssssssssssss

We adopted the 3+1 spacetime decomposition \citep[see,
  \eg][]{Rezzolla_book:2013}, where the metric is given by the line
element with the following form,
\begin{equation}\label{Eq:ds2}
  {\rm d}s^2=-\alpha^2\,{\rm d}t^2+\gamma_{i j}\,\left({\rm d}x^{i} +
  \beta^{i}\,{\rm d} t\right)\,\left({\rm d}x^{j} +\beta^{j}\,{\rm d}
  t\right)\,,
\end{equation}
where $\alpha$ is the lapse function, $\beta^{i}$ is the shift vector, and
$\gamma_{ij}$ is the three-metric on space-like hypersurface of constant
time $t$. In the $3+1$ split of space-time, the metric determinant of
space-time $g={\rm det}(g_{\mu\nu})$ relates to the determinant of the
purely spatial three-metric as $\gamma={\rm det}(\gamma_{i
  j})=-g/\alpha^2$ , and only $\gamma$ is required in what follows.

A perfect non-magnetised fluid is described by four physical variables:
the rest-mass density $\rho$, the thermal pressure $p$, the specific
enthalpy $h,$ and the coordinate-frame four-velocity of the fluid
$u^\mu$. With these variables, we can characterise the fluid through the
energy-momentum tensor \citep{Rezzolla_book:2013}
\begin{equation}
 T_{\mu \nu}\;=\;\rho\,h\,u_{\mu}\,u_{\nu}+p\,g_{\mu \nu}\,,\label{Eq:Tmunu}
\end{equation} 
and an equation of state (EOS), relating the pressure to some of the
other thermodynamical properties of the fluid. We used
  a simple ideal-fluid EOS, $p = (\hat{\gamma}-1) \rho \epsilon$, where
  $\epsilon$ is the specific internal energy, $\hat{\gamma}$ is the
  adiabatic index, and the specific enthalpy is given by $h(\rho,p) = 1 +
  \hat{\gamma}/(\hat{\gamma}-1) p/\rho$ \citep{Rezzolla_book:2013}. We
  here used either $\hat{\gamma}=5/3$ or $\hat{\gamma}=4/3$ when
  modelling an ultrarelativistic fluid.
%%%%%%%%%%%%%%%%%%%%%

The fluid evolution is described by the conservation of mass and
energy-momentum,
 \begin{eqnarray}
  \nabla_{\mu}\left(\rho\,u^{\mu}\right) &=& 0\,, \label{Eq:cons_M} \\
  \nabla_{\mu}  T^{\mu \nu} &=& 0\,, \label{Eq:cons_T}
 \end{eqnarray}
which can be written in a form favourable to conservative numerical
integration as
\begin{equation}\label{Eq:Evol}
\frac{\partial \boldsymbol{U}}{\partial t}+\frac{\partial
   \boldsymbol{F}^{i}}{\partial x^{i}}=\boldsymbol{S}\,.
 \end{equation}
Here, the vector of conserved variables
\begin{equation}
  \boldsymbol{U} \equiv \sqrt{\gamma}\left[
   \begin{tabular}{c}
    $D$\\
    $S_j$\\
    $\tau$\,
   \end{tabular}
   \right]\,,
 \end{equation}
is composed of the mass-density $D \equiv \rho\,W$, of the covariant
spatial momentum density $S_{j} \equiv W\,\rho\,h\,u_{j}$, and of the
total energy density $\tau \equiv W^2\,\rho\,h-p-D,$ where we have
subtracted the mass-density $D$  to improve accuracy in the
nonrelativistic regime. The symbol $W \equiv \alpha\,u^{0} =
1/\sqrt{1-v^{i}v_{i}}$ is the Lorentz factor of the fluid as seen by an
Eulerian observer moving with four-velocity $n_\mu=(-\alpha,0_i)$. The
fluxes $\boldsymbol{F}^i$ are then given by
\begin{equation}
\label{Eq:Flux}
  \boldsymbol{F}^{i} \equiv \sqrt{\gamma}\alpha \left[
   \begin{tabular}{c}
    $\rho u^{i}$\\
    $S_{j}\,u^{i}/W+p\delta^{i}_{j}$\\
    $\tau\,u^{i}/W+p v^i$
  \end{tabular}
   \right]\,,
\end{equation}
and the geometric source terms $\boldsymbol{S}$ are written in terms of
Christoffel symbols $\Gamma^{\delta}_{\ \mu\nu}$
 
\begin{equation}\label{Eq:Source}
    \boldsymbol{S} \equiv \sqrt{\gamma}\alpha\left[
   \begin{tabular}{c}
    $0$\\
    $T^{\mu\nu}\left(\dfrac{\partial g_{\nu j}}{\partial
       x^{\mu}}-\Gamma^{\delta}_{\ \mu\nu} \ \! g_{\delta j}\right)$
     \\
    $\left(T^{\mu 0}\, \dfrac{\partial\alpha}{\partial
       x^{\mu}}-\alpha T^{\mu\nu}\,\Gamma^{0}_{\ \mu\nu}\right)$
   \end{tabular}
   \right]\,,
\end{equation}
as discussed by \cite{Banyuls97}, for example. 
Next to the conserved variables $\boldsymbol{U}$, computation of fluxes
requires a set of ``primitive'' variables
\begin{equation}
\boldsymbol{P} = \left[
\begin{tabular}{c}
$\rho$ \\
$v^i$ \\
$p$
\end{tabular}
\right]\,.
\end{equation}

A well-known problem of any conservative formulation of the GRHD
equations is that while the map $\boldsymbol{P}\to\boldsymbol{U}$
is straightforward, in general the inverse
$\boldsymbol{U}\to\boldsymbol{P}$ follows as solution to a set of
nonlinear equations. We here followed the 1D root-finding algorithm discussed in \cite{vanderHolst2008} as it has
proven a good compromise between speed and robustness.  For a
thorough discussion of various algorithms, we refer to the works of
\citet{Noble2006},\citet{Galeazzi2013},and \citet{HamlinNewman2013} for details.

To evaluate the fluxes $\boldsymbol{F}^i$ in Eq. (\ref{Eq:Flux}),
primitive variables are interpolated to cell interfaces by limited
reconstruction using one of the various flavours of piecewise linear
slope limiters \citep[e.g.][]{Toro99}, the piecewise polynomial method
(PPM) by \cite{Colella84}, or the compact stencil third-order
reconstruction by \cite{cada2009}.  This yields the left- and right-biased states $\boldsymbol{U}^L$ and $\boldsymbol{U}^R$. In the following,
superscripts $L$ and $R$ always refers to quantities derived from these
reconstructed values.  The fluxes are obtained from an approximate
Riemann solver.

The two currently available choices are \textit{1.)}  the Rusanov (LF)
scheme, which is based on the knowledge of the maximum absolute value of
the characteristic waves at the interface in the direction $x$: $c^x={\rm
  max} \left\{ \lambda_{+}^{x,L},{\rm abs}
\left(\lambda_{-}^{x,L}\right),\lambda_{+}^{x,R},{\rm abs}
\left(\lambda_{-}^{x,R}\right) \right\} \,$ , and \textit{2.)}  HLL
\citep{Harten83}, which is based on the knowledge of the two fastest
characteristic waves propagating in both directions, one to the left with
$c_{-}^{x}={\rm min}\left(\lambda_{-}^{x,L}, \lambda_{-}^{x,R}\right)$
, and the other wave to the right with $c_{+}^{x}={\rm
  min}\left(\lambda_{+}^{x,L},\lambda_{+}^{x,R}\right)$. The HLL upwind
fluid flux function for the variable $U$ is calculated as
\begin{equation}
\hat{F}^{x}(U)=\left\{
\begin{array}{lcc}
F^x(U^L) &; & c_{-}^{x} > 0 \\
F^x(U^R) &; & c_{+}^{x} < 0 \\
\bar{F}^x(U^L,U^R) &; & {\rm otherwise} \\
\end{array}
\right.
\end{equation}
where
\begin{equation}
\bar{F}^x(U^L,U^R) \equiv \frac{c_{+}^{x}\,F^{x}
  \left(U^{L}\right)-c_{-}^{x}\,F^{x}\left(U^{R}\right)+c_{+}^{x}\,c_{-}^{x}\,
  \left(U^{R} - U^{L}\right)}{c_{+}^{x}\,-\,c_{-}^{x}} \,,
\end{equation}
and the LF flux is simply
\begin{equation}
\hat{F}^{x}(U) = \frac{1}{2}\left(F^x(U^L) + F^x(U^R)\right) -
\frac{1}{2} c^x \left(U^R-U^L\right)\,.
\end{equation}

According to the chosen stencil, the number of the boundary cells (ghost
cells) changes: two cells for linear reconstruction and three for
parabolic reconstruction. On all interior boundaries, these ghost cells
are filled by copy/prolongation/restriction operations, depending on the
refinement level of bounding grid blocks \citep[see][for
  details]{Keppens2012}.

The characteristic wave speed is also used to determine the explicit time
step obeying the usual Courant-Friedrich-Levy (CFL) conditions, where the
characteristic wave speed $\lambda^{i}_{\pm}$ can be written as
\citep{Banyuls97}
\begin{equation}\label{Eq_lambda}
\lambda_{\pm}^{i} =\alpha\,\lambda_{\pm}^{' i}-\beta^{i}\,,
\end{equation}
where 
\begin{align}
\label{Eq_lambdaprim}
\lambda_{\pm}^{' i}  = &\frac{\left(1-c_{\rm s}^2\right)\,v^{i}
  \pm\sqrt{c_{\rm s}^2\,\left(1-v^2\right)\,\left[\left(1-v^2 c_{\rm
        s}^2\right)\gamma^{ii}-\left(1-c_{\rm
        s}^2\right){\left(v^{i}\right)}^2\right]}}
{(1-v^2\,c_{\rm s}^2),} 
\end{align}
where $c_{\rm s}=\displaystyle\sqrt{\partial\ln{h}/\partial\ln\rho}$
represents the local sound speed and $v=(\gamma_{ij} v^{i}v^{j})^{1/2}$
is the modulus of the local Eulerian velocity of the fluid.

As in any fluid simulation, we cannot handle vacuum, therefore we filled the space
of the vacuum region such as outside the torus with a low-density
``atmosphere''. This atmosphere had a fixed value for the rest-mass
density $\rho_{{\rm atm}}$ chosen to be several orders of magnitude lower
than the highest rest-mass density in the initial disc configuration. A
low fixed value for the pressure $p_{{\rm atm}}$ was chosen, and the
material was set to be static with Eulerian velocity $v^i=0$.  At every
time step, the values
of the primitive variables in the cell were set to the atmospheric
values when the density or gas pressure in a given cell fell below the
threshold value of $f \rho_{{\rm atm}}$ or $f p_{{\rm atm}}$. The factor $f$ was chosen for each problem.

We used a block-tree AMR structure where a refinement ratio by a
factor of $2$ was always ensured between any two successive
levels. The global time step was taken as the shortest time step
computed across all mesh refinements
\citep{Keppens2012}. Prolongation and restriction can be used on
conservative variables or primitive variables, where we typically
chose the latter to ensure that no unphysical state was
encountered in the resolution change. Adaptivity
was decided following the simple second-order error estimator by
L\"ohner \citep{Loehner87} on physical variables to trigger the
AMR. This method is also used in AMR codes such as \texttt{FLASH}
\citep{Calderetal02}, \texttt{RAM} \citep{Zhang2006},
\texttt{MPI-AMRVAC} \citep{Keppens2012,Porth2014}, \texttt{PLUTO}
\citep{Mignone2012}, and \texttt{ECHO} \citep{Zanotti2015b}, which
captures features of interest with sufficient accuracy, as we
demonstrate in more detail in Sect. \ref{sec:recoilbhsetup}.

As described by \cite{Keppens2012}, the code is parallelised through
the message-passing interface (MPI) paradigm.  We used a Morton
Z-order space-filling curve to run through all blocks in the
(oct-) tree data structure.  Parallel load balancing was then
achieved by allocating equal sections of the space-filling curve
to the available processors.  Strong- and weak-scaling tests of
the underlying \texttt{MPI-AMRVAC} toolkit were performed
recently by \cite{Porth2014}.  In particular, excellent weak
scaling to over $30\,\!000$ processors was demonstrated.

%ssssssssssssssssssssssssssssssssssssssssssssssssssssssssssssssssss
\subsection{Spherical accretion: the Michel solution}\label{sec:bondi}
%ssssssssssssssssssssssssssssssssssssssssssssssssssssssssssssssssss

%ffffffffffffffffffffffffffffffffffffffffffffffffffffffffffffffffff
\begin{figure*}[]
\centering \includegraphics[width=0.99\hsize]{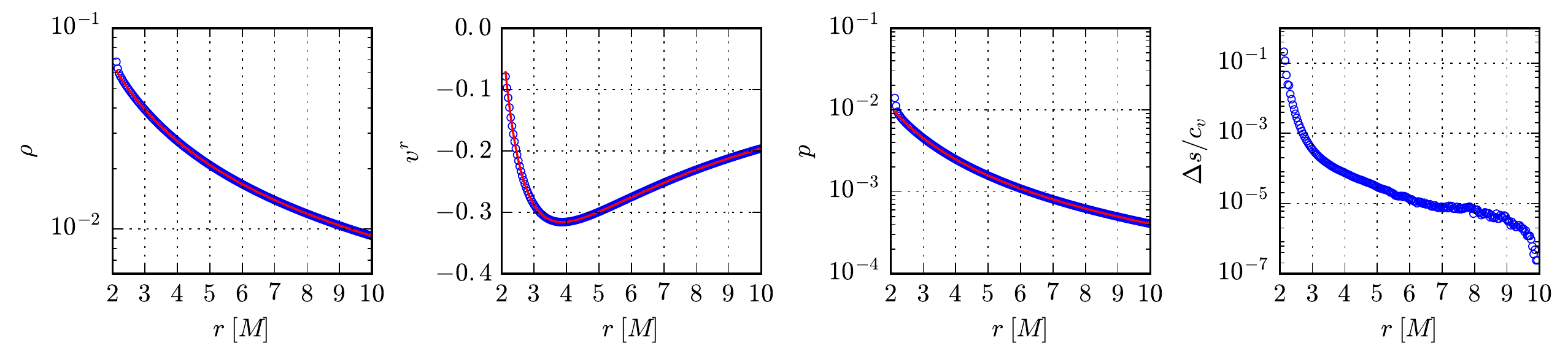}
\includegraphics[width=0.99\hsize]{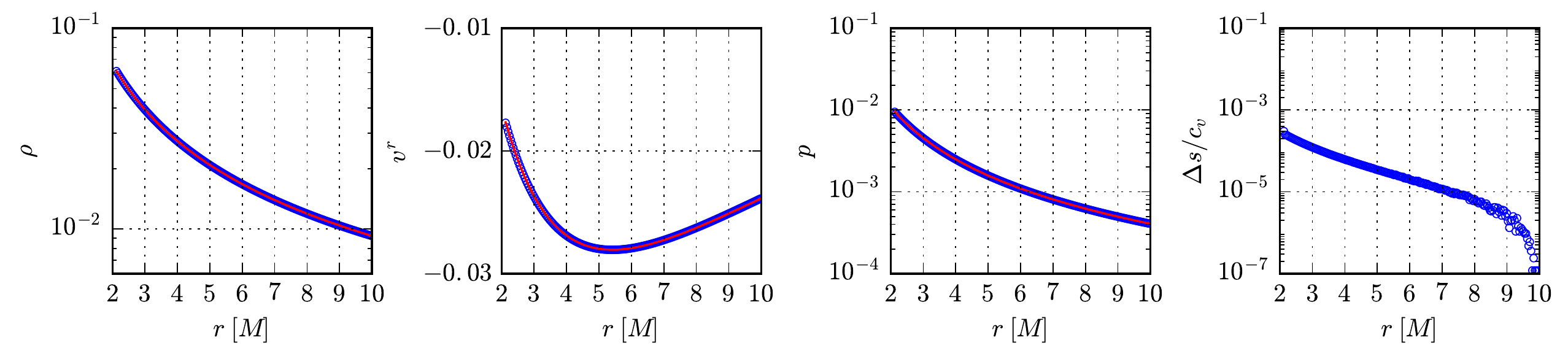}
\caption{First three columns show 1D radial
    profiles of the rest-mass density, of the radial component of the
    four-velocity, and of the gas pressure on the equatorial plane in the
    Michel solution test in Boyer-Lindquist (upper) and Kerr-Schild
    coordinates (lower). The semianalytic solution
is shown in red. This solution
    serves as initial condition, while the numerical results at
    $t=100\,M$ are reported with blue circles. The fourth column shows
    the entropy increase normalised to the specific heat at constant
    volume, $\Delta s/c_v$, from $t=0\,M$ to $t=100\,M$ using each
    coordinate system.  A smaller entropy change is related to a better
    preservation of the stationary solution. }
\label{fig:1D-bondi}
\end{figure*}
%ffffffffffffffffffffffffffffffffffffffffffffffffffffffffffffffffff

%ffffffffffffffffffffffffffffffffffffffffffffffffffffffffffffffffff
\begin{figure}[]
\centering
\includegraphics[width=\hsize]{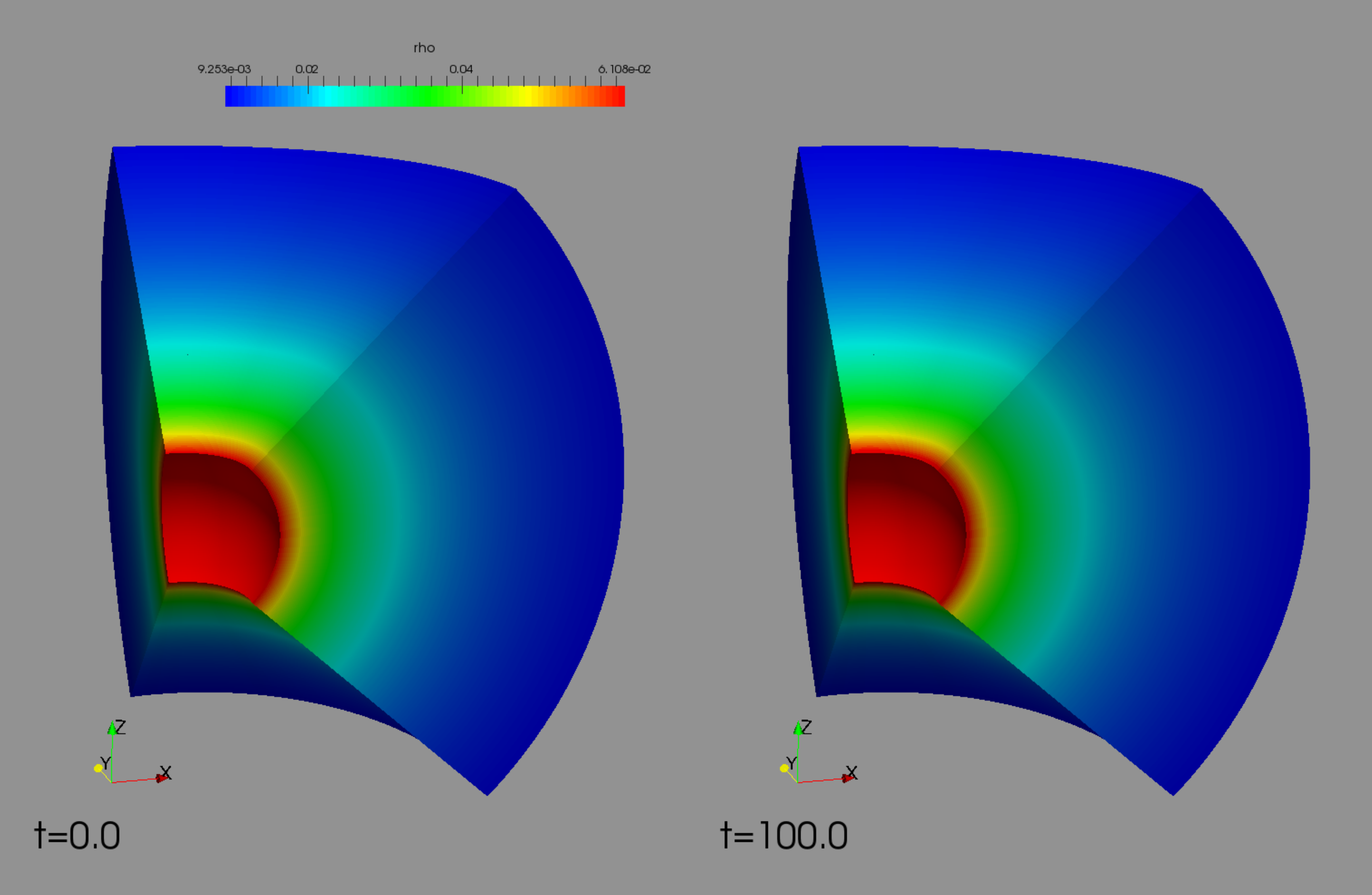}
\caption{Snapshots of the 3D distribution of the rest-mass density in the
  Michel accretion test as computed in Boyer-Lindquist coordinates. The
  solutions are shown at $t=0\,M$ (left) and at $t=100\,M$ (right).}
\label{fig:3D-bondi}
\end{figure}
%ffffffffffffffffffffffffffffffffffffffffffffffffffffffffffffffffff

%ffffffffffffffffffffffffffffffffffffffffffffffffffffffffffffffffff
\begin{figure}[]
\centering
\includegraphics[width=0.8\hsize]{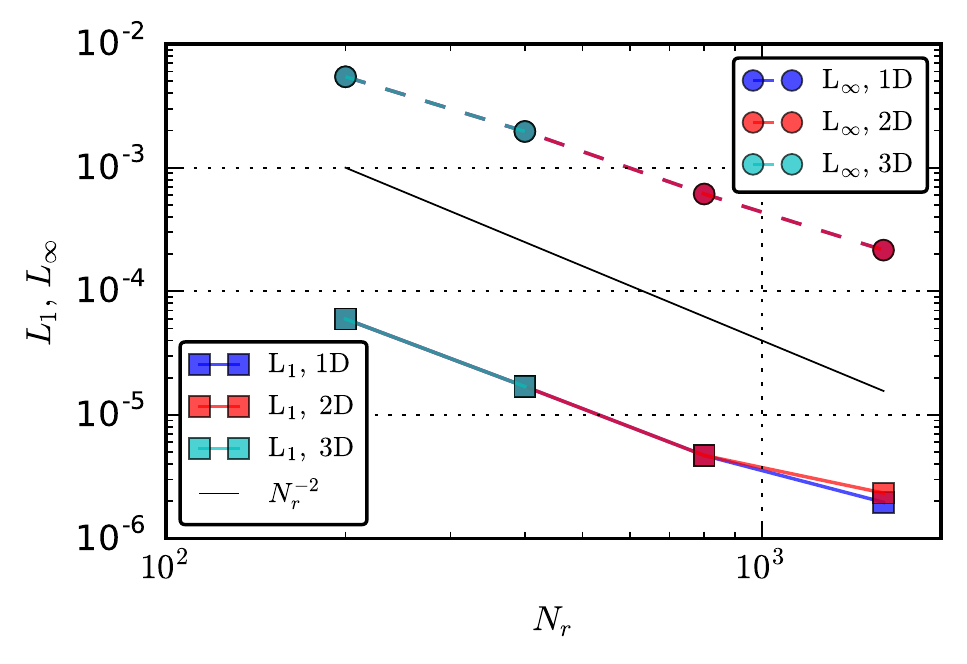}
\caption{Convergence order for the Michel accretion test for 1D,
  2D, and 3D simulations as computed at $t=100\,M$.  Different
  lines refer either to the dimensionality of the simulations or
  to the error indicators used, \ie $L_1$ or $L_{\infty}$ norms;
we  also report as a reference the expected second-order
  convergence slope. }
\label{fig:error-bondi}
\end{figure}
%ffffffffffffffffffffffffffffffffffffffffffffffffffffffffffffffffff

As a first test of the code for the general-relativistic regime, we
considered the stationary solution corresponding to a spherically symmetric
solution onto a Schwarzschild black hole. This is known as the Michel
accretion solution \citep{Michel72} and represents the extension to
general relativity of the corresponding Newtonian solution by
\citet{Bondi52}. The spherical Michel accretion solution is described in
a number of works, \citep[see, \eg][]{Hawley84a, Rezzolla_book:2013}. The
free parameters are the position of the critical radius $r_c$ and the
adiabatic index $\hat{\gamma}$. In this test, we chose $r_c = 8\,M$ and
$\hat{\gamma}=5/3$ using an ideal EOS. The steady Bondi accretion flow was
initialised, and the simulations were run until $t=100M$. We simulated a 
domain spanning from $2.1 \,M \le r \le 10 \,M$, $\pi/4 \le \theta,
\text{and } \phi
\le 3\pi/4$ with 200 cells in all directions.

%ffffffffffffffffffffffffffffffffffffffffffffffffffffffffffffffffff
\begin{figure}[]
\centering
\includegraphics[width=\hsize]{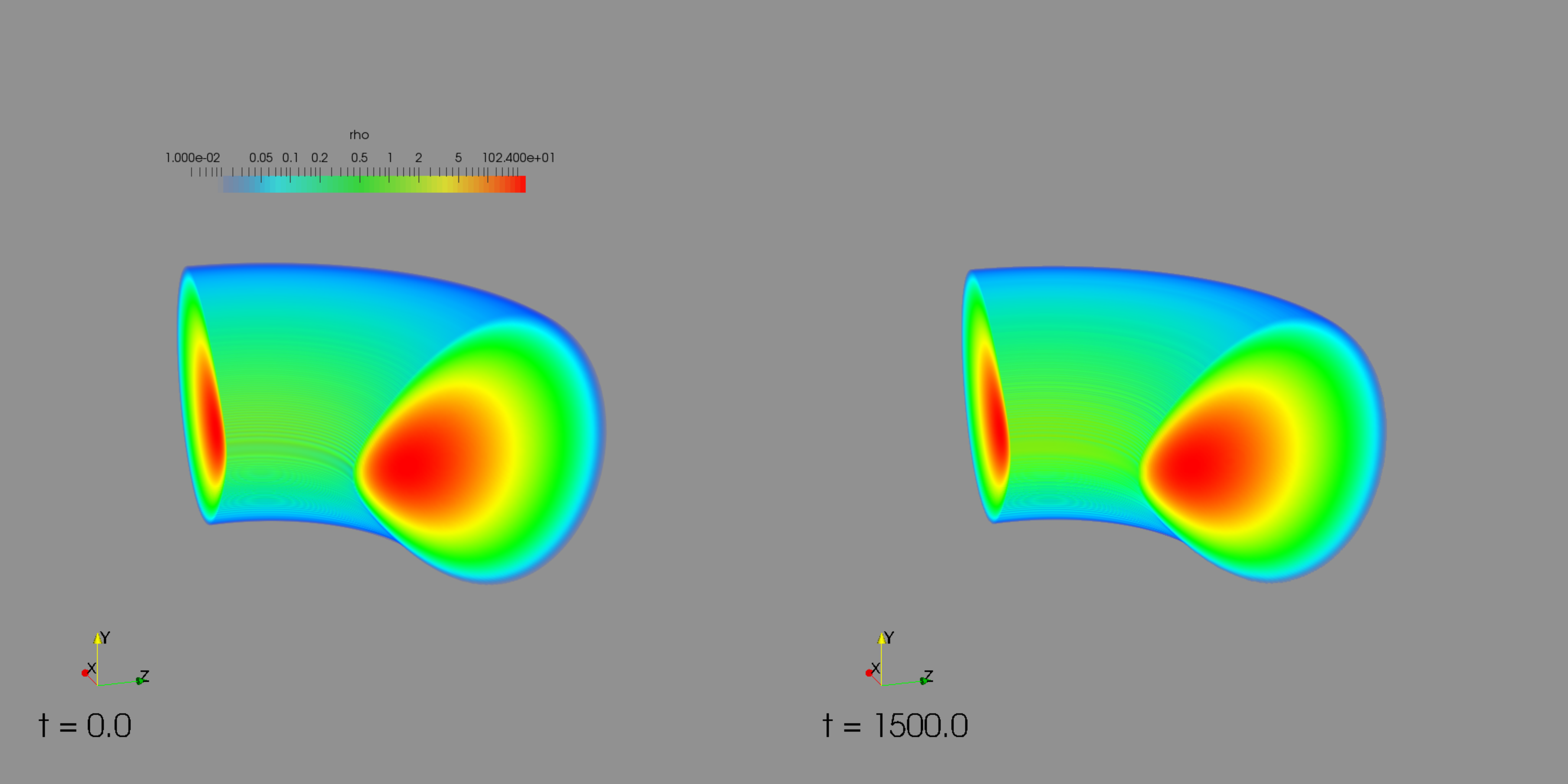}
\caption{3D isovolume density of a stationary torus at $t=0\,M$ (right
  panel) and $t=1500\,M$ (left panel) in Kerr space-time using KS
  coordinates. }
\label{fig:3D-stationary-KerrKS}
\end{figure}
%ffffffffffffffffffffffffffffffffffffffffffffffffffffffffffffffffff

Figure \ref{fig:1D-bondi} shows 1D radial profiles of
the rest-mass density, of the radial component of the four-velocity, of the
pressure, and of the entropy increase normalised to the specific heat
  at constant volume, $\Delta s/c_v$, along the equatorial plane at
$t=0\,M$ and at $t=100\,M$; the top panels in Fig. \ref{fig:1D-bondi}
report the numerical solution in a space-time expressed in Boyer-Lindquist
coordinates, while the bottom panels show it in Kerr-Schild coordinates. The semianalytic solution is shown in red, while the numerical results are
reported with blue circles; we note that although the stationary solution is
isentropic, differences in entropy can result from numerical dissipation
when an ideal EOS is used.

Clearly, the steady accretion flow is well preserved by the numerical
simulations. As is quite common for this test, small differences from
the analytic solution are seen near the inner boundary in the case of
Boyer-Lindquist coordinates. This occurs because in Boyer-Lindquist
coordinates the metric component $g_{rr}$ exhibits severe gradients in
the region near the black hole horizon and is therefore difficult to
resolve numerically. On the other hand, in Kerr-Schild coordinates, the
difference from the analytical solution near the black hole horizon is
far smaller, as the Kerr-Schild coordinates are horizon penetrating and
lead to a superior behaviour in its vicinity. The 3D
structure of the solution is shown in Fig.\ref{fig:3D-bondi} with a
volume-rendering representation of the rest-mass density. The difference
between initial and final simulation times is not recognisable, and
spherical symmetry is well preserved.

To investigate the numerical accuracy, we checked the $L_1$ and $L_\infty$
norms in density between initial ($t=0$) and final simulation time
($t=100M$) with different resolutions in 1D, 2D, and 3D, as seen in
Fig.\ref{fig:error-bondi}. The convergence is second order for all
cases. It is limited by the order of the chosen reconstruction scheme
(here we used a Koren slope limiter function).

%ssssssssssssssssssssssssssssssssssssssssssssssssssssssssssssssssss
\subsection{Stationary tori}\label{sec:stationary}
%ssssssssssssssssssssssssssssssssssssssssssssssssssssssssssssssssss

%ffffffffffffffffffffffffffffffffffffffffffffffffffffffffffffffffff
\begin{figure}[]
\centering
\includegraphics[width=0.9\hsize]{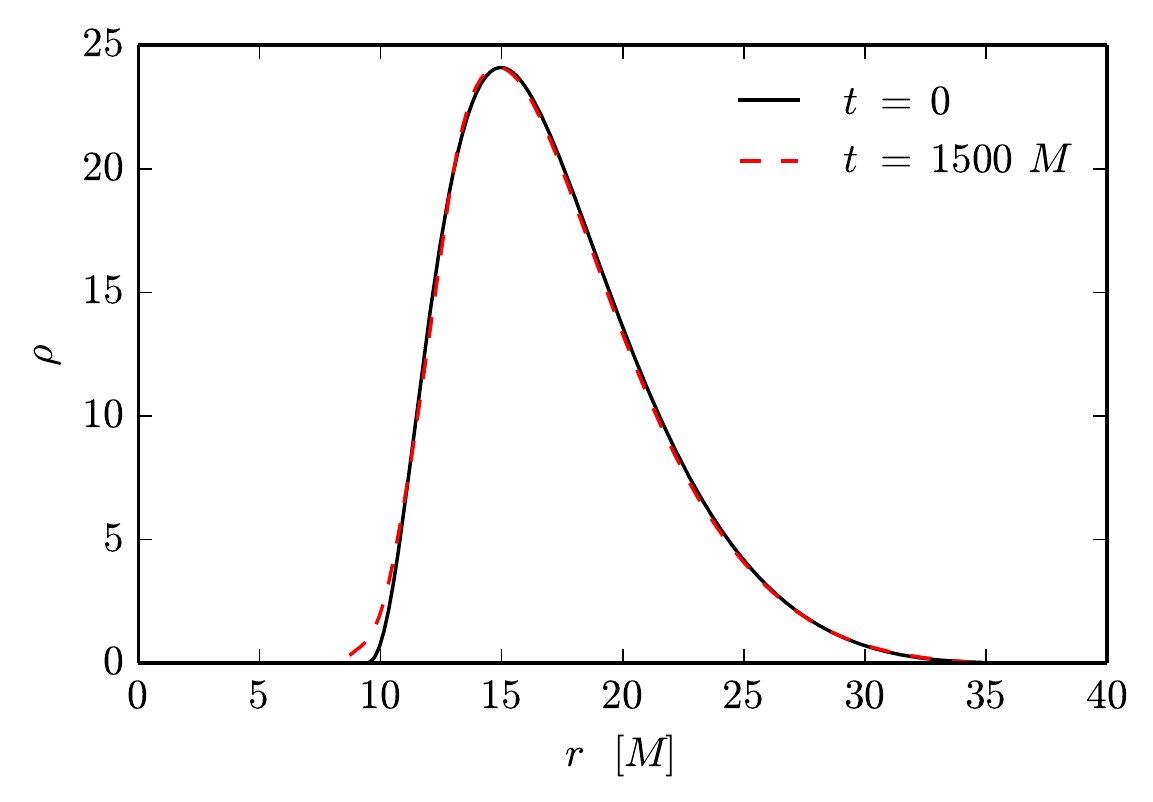}
\caption{1D radial profile of density in the equatorial direction of a
  stationary torus in Kerr space-time using KS coordinates. A resolution
  of $(N_{r},N_{\theta})=(200,100)$ cells was used. }
\label{fig:1Dcut-stationary-KerrKS}
\end{figure}
%ffffffffffffffffffffffffffffffffffffffffffffffffffffffffffffffffff

Before turning to the application of a black hole recoiling in a torus,
we first verify how well the code is able to preserve a
\textit{\textup{stationary}} torus solution. In the regime of small kick
velocities, it is of particular importance to ensure that the evolution
is not governed by numerical artefacts. The hydrodynamic stationary torus
solution was first presented in \cite{Fishbone76} and
\cite{Kozlowski1978}, and is now a standard test, as used for example by
\citet{Font02b}, \citet{Zanotti03} and by \citet{Anton05}. In
particular, following \cite{Font02b}, we adopted a non-accreting
solution that fills its entire Roche lobe, $\Delta W=0$ and constant
specific angular momentum $\ell=4.35$ around a Kerr black hole with
dimensionless spin parameter $a=0.5$. We simulated the evolution for ten
orbital periods, as measured at the inner torus radius $r_{\rm in}=9.34\,
M$. An orbital period at $r_{\rm in}$ corresponds to $P \approx
150$ time units, and the outer radius of the torus is at $r_{\rm out}
\simeq 40\,M$.

In the following we focus on Kerr-Schild coordinates. The domain
covers $r\in [1.85\,M,\,40\,M],$ which extends into the outer black hole horizon. For the 2D cases, the entire meridional plane is simulated
for $\theta\in[0,\pi],$ and in 3D we restrict ourselves to a section with
$\Delta\phi,\Delta\theta=\pi/2$ centred on the equator.

As mentioned in Sect. \ref{sec:GRHDeq}, to avoid the
presence of vacuum regions outside the torus, we applied floor
values for the rest-mass density ($\rho_{{\rm atm}}=10^{-5}
\rho_{{\rm max}}$) and the gas pressure ($p_{\rm{atm}}=p_{{ \rm
    min}}=10^{-3} \rho_{{ \rm atm}}$), where $\rho_{{\rm max}}$
is the maximum rest-mass density inside the torus at $t=0$ (this is
the rest-mass density at the ``centre'' of the torus, which we
took to be $\rho_c=1.38 \times 10^{-10}\,{\rm g\ cm}^{-3}$ .
In
practice, for all numerical cells that satisfy $\rho \le
f\rho_{{ \rm atm}}$ or $p \le fp_{{\rm atm}}$, we set
$\rho=\rho_{{\rm atm}}, p=p_{{\rm atm}}, \text{and } v^i=0$. The factor $f$
is 3.0 for the 2D case and 10.0 for the 3D case.

%ffffffffffffffffffffffffffffffffffffffffffffffffffffffffffffffffff
\begin{figure}[]
\centering
\includegraphics[width=0.90\hsize]{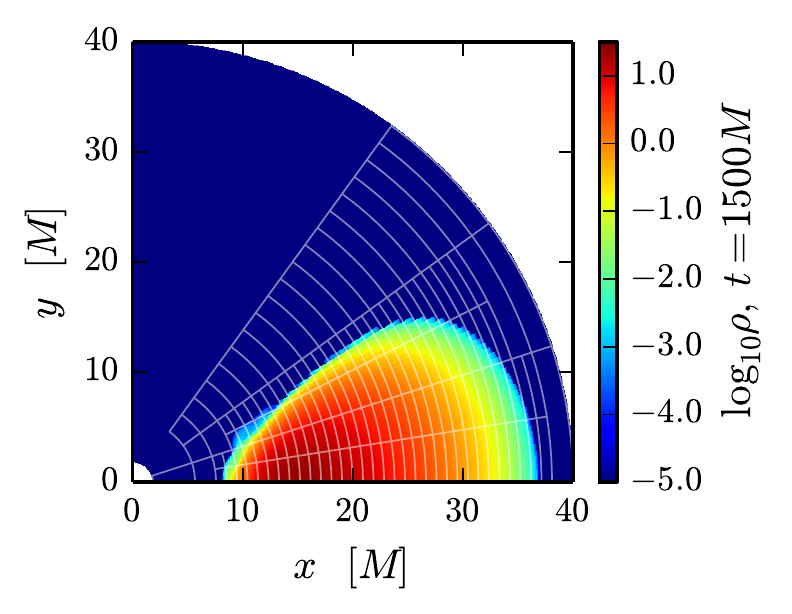}
\caption{2D logarithmic rest-mass density of a stationary torus at
  t=1500M in Kerr space-time using KS coordinates with three different AMR
  levels. }
\label{fig:2D-stationary-KerrKS_AMR}
\end{figure}
%ffffffffffffffffffffffffffffffffffffffffffffffffffffffffffffffffff

%ffffffffffffffffffffffffffffffffffffffffffffffffffffffffffffffffff
\begin{figure}[]
\centering
\includegraphics[width=0.8\hsize]{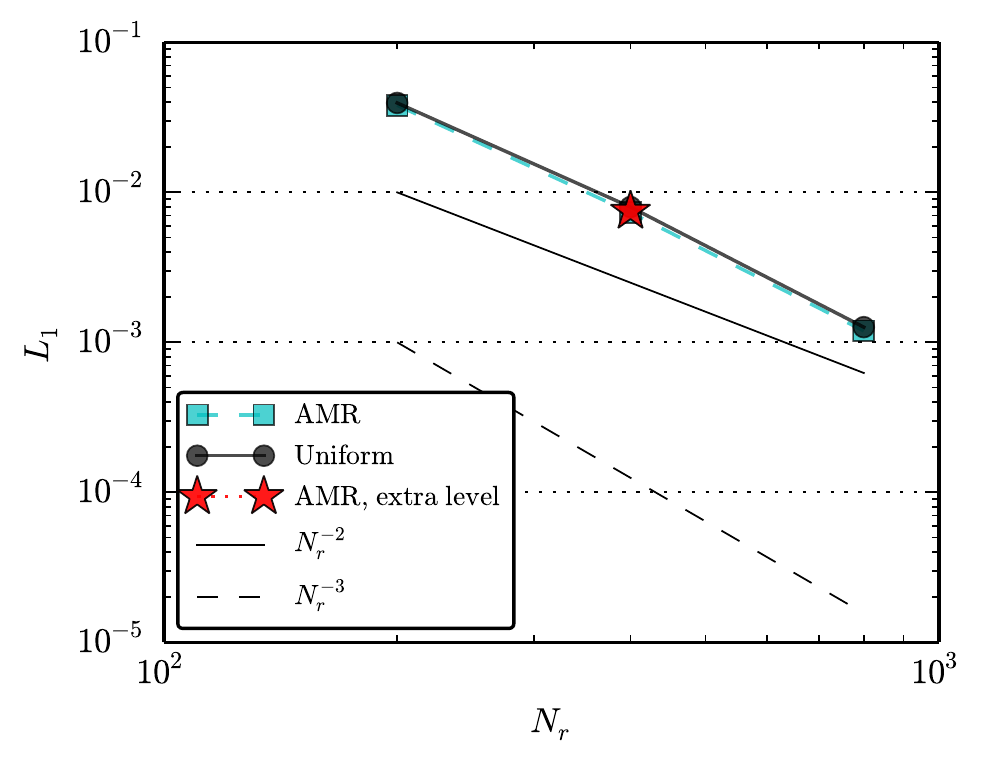}
\caption{$L_1$ norm of the error in the stationary accretion torus in 2D,
  with either a uniform grid or with AMR. The error is measured as the
  difference between the solution at the latest time and the initial
  data.  The red star marks the norm for a test with forced refinement
  jump in the centre of the torus. All curves indicate second-order
  convergence, and the AMR cases compare favourably with their
  corresponding uniform realisations. }
\label{fig:error-stationary}
\end{figure}
%ffffffffffffffffffffffffffffffffffffffffffffffffffffffffffffffffff

Volume renderings of density at the initial state and after $ \text{about ten} $
orbital periods are shown for a resolution of $200^{3}$ cells in Fig.
\ref{fig:3D-stationary-KerrKS}. The torus maintains its integrity, and
only a slight change in density is observable at the inner edge.  An
equatorial cut of the solution is shown for different times in Fig.
\ref{fig:1Dcut-stationary-KerrKS}. As seen already in the 3D rendering,
numerical diffusion leads to a smearing out of the inner edge of the
torus, but overall, the equilibrium is well maintained for the simulated
time. This result gives us confidence about the next set of simulations
of recoiling black holes, in which significant departures from the
initial state occur on the shorter timescale of one orbital period, which
will be captured accurately by the numerical scheme. Similar
  conclusions have also been obtained when considering a kicked
  equilibrium torus in the absence of a black hole (the rest-mass density
  was preserved, but the azimuthal velocity was set to zero and
  the pressure gradient was removed by suitably adjusting the
  specific entropy). When performing this test, we obtained that the
  initial torus shape is well maintained even after being advected
  multiple times across the simulation domain. We hence conclude that the
  numerical effect from a velocity kick alone is negligible compared to
  the physical effects that are due to a recoiling black hole as we study here. 

Another important use of the stationary torus solution is that it
  has allowed us to perform a few controlled experiments with mesh
refinement. In the first experiment we allowed for three mesh refinements
and let the code automatically refine with L\"ohner's error estimator on
the fluid-frame density. The resulting density map including the
grid-structure is illustrated in Fig.  \ref{fig:2D-stationary-KerrKS_AMR}
after ten orbital periods.  Unsurprisingly, the solution in the torus
region is essentially identical to the case where a uniform grid at the
corresponding resolution was employed.

The L\"ohner error estimator in essence measures the smoothness of
the solution for a given variable, as it depends on a weighted sum
of discretised second derivatives. It has the advantage of
being more computationally efficient than other error estimators
that may require, for instance, solutions computed at different times
or different resolutions.

In the case of the recoiling black hole, we relied on automated refinement
based on the L\"ohner error estimator. In a dynamical situation, it
necessarily leads to resolution jumps inside the torus.  To check how
this affects the solution, we performed a second experiment where we
enforced the refinement of a single refinement level at one point only,
namely, at $(r,\theta)=(15\,M,\pi/2)$, which essentially corresponds to
the centre of the torus. For all intents and purposes, the result is
identical to the case without a resolution jump. The global $L_{1}$ error
of this setup is illustrated in Fig. \ref{fig:error-stationary} for the
uniform case, the AMR, and the case with a refinement jump (red star
symbol). Overall, we obtain a very good qualitative and quantitative
agreement with the uniform-grid case and recover the second-order
accuracy of the algorithm.

%ssssssssssssssssssssssssssssssssssssssssssssssssssssssssssssssssss
\section{Recoiling black hole}\label{sec:recoil}
%ssssssssssssssssssssssssssssssssssssssssssssssssssssssssssssssssss

%ssssssssssssssssssssssssssssssssssssssssssssssssssssssssssssssssss
\subsection{Motivation for recoiling black hole research}
\label{sec:recoilbhbackground}
%ssssssssssssssssssssssssssssssssssssssssssssssssssssssssssssssssss

As a full demonstration of the code in a scientific application, we
performed 2D and 3D GRHD simulations of recoiling black holes colliding
with a circumbinary accretion disc.

Most galaxies are expected to contain a central supermassive black hole
that experiences a form of ``co-evolution''  , which is reflected in a rich
phenomenology of black-hole-host galaxy correlations \citep[for a recent
  review, see][]{Kormendy2013}. Cosmological models predict that
galaxies experience several mergers during their evolution
\citep[\eg][]{Haehnelt94,Sesana2004,Volonteri2007}. Following a galactic
merger, the two supermassive black holes will be transported to the
barycentre through dynamical friction and form a binary with a separation of
$\lesssim 1\rm pc$ \citep[\eg][]{Milosavljevic01}. Gas funnelled into the
galactic centre fuels a (truncated) circumbinary accretion disc with
dynamics that are increasingly disconnected from the binary black
hole
pair \citep{Milosavljevic05}. Within the truncated disc, the two black
holes lose energy through stellar three-body encounters and by the
emission of gravitational waves to finally coalesce and form a single
black hole. In the final merger, a significant fraction of the mass, up
to $10\%$ \citep[\eg][]{Reisswig:2009vc,Barausse2012b}, is radiated away,
and the produced single black hole can experience a sudden kick with a
recoil velocity ranging from several hundred $\rm km\, s^{-1}$ up to
$4,\!000\,\rm km\, s^{-1}$ \citep[\eg][]{Bakeretal07, Gonzalez:2006md,
  Campanellietal07a, Koppitz:2007ev}.

Although no direct evidence of the existence of an SMBBHs system has been
found so far, there are several circumstantial possibilities in a number of
candidates, such as the radio galaxy 0402+379 \citep{Rodriguez2006}, the
ultraluminous infrared galaxy NGC6240 \citep{Komossa2003}, and the BL
Lac Object OJ287 \citep{Valtonen2008}. More recently,
\cite{Graham2015} reported strong periodic optical variability of the
quasar PG 1302-102 with an observed period of 5.2 years. When this optical
variability period is matched to the orbital period of the SMBBHs, the
system would be separated by less than 0.01 parsecs. This means that the
system has evolved well into the final parsec scale.

With the recent first detection of gravitational waves from merging
stellar-mass black holes \citep{Abbott2016aa}, the study of SMBBHs is
strongly motivated by the expected detection of their gravitational
signal by the space-based gravitational wave detectors, such as the
planned eLISA detector \citep{Amaro2012b}. Considerable attention has recently been attracted by the possibility of detecting the
electromagnetic signatures of these events \citep[\eg][]{Komossa12,
  Schnittman13}. A number of studies have been carried out to
investigate the properties of these electromagnetic signatures either
during the stages that precede the merger \citep{Palenzuela:2009hx,
  Moesta:2009, Palenzuela:2010, Moesta2011, Alic:2012}, or in
post-merger phase. Several authors have considered the interaction
between the binary and the surrounding stars and gas
\citep[\eg][]{Armitage:2002, Milosavljevic05, vanMeteretal10,
  Farris2010, Farris:2011vx, Farris2012, Bode2012, Giacomazzo2012,
  Noble2012, Gold2014a}. Other scenarios that have not involved
matter have also been considered. In these cases, the supermassive
black hole binary is considered to be in-spiralling in vacuum, but in the
presence of an external magnetic field that is anchored to the
circumbinary disc. \citep[e.g.][]{Palenzuela2009b, Mostaetal10}. In the
post-merger phase, the electromagnetic counterpart is assumed to be
mainly due to the radiation from the circumbinary accretion disc, which
will contain an imprint of any strong dynamical change produced on the
disc by the merger event. There are two main dynamical effects. One is
the abrupt reduction of the rest-mass of the binary that is emitted away
in gravitational waves amounting to up to 10 \% for equal-mass spinning
systems \citep[\eg][]{Reisswig:2009vc}. The second is the recoil velocity
of the merged system, resulting in a kick velocity of the resulting black
hole with respect to the host galaxy \citep[\eg][]{Rezzolla:2008sd}. It is
clear that these two dynamical effects can significantly affect the dynamics
of circumbinary disc, mainly in their contribution to the formation and
propagation of shocks, thereby enhancing the possibility of a strong
electromagnetic signal. Several authors have discussed the dynamics and
related emission from a circumbinary disc with the recoiling central
black hole in the post-merger phase \citep[\eg][]{Lippai:2008,
  Megevand2009, Andersonetal10, Corrales2009, Rossi2010,
  Zanotti2010, Ponce2012, Zanotti2012a, Gold2014b}.

%ssssssssssssssssssssssssssssssssssssssssssssssssssssssssssssssssss
\subsection{Initial setup in 2D}
\label{sec:recoilbhsetup}
%ssssssssssssssssssssssssssssssssssssssssssssssssssssssssssssssssss

For the initial setup of the recoiling black hole in 2D, we followed the
work by \cite{Zanotti2010}. As the initial model of the circumbinary
disc, we adopted a stationary disc with a density and pressure profile
similar to that of the equatorial plane of the torus described in Sect.
\ref{sec:stationary}. Similarly to \citet{Zanotti2010}, we assumed that
the vertical structure of the disc can be neglected and the vertical
thickness can be approximated by a quantity $2H$ that is constant in the
radial direction, as in the standard thin-disc approximation. The spin
parameter of the black hole was $a=0.5$, and the distribution of specific
angular momentum $\ell$ on the equatorial plane was constant, with
$\ell=8$. The inner and outer edges of the torus were located at $r_{\rm
  in}=40\,M$ and $r_{\rm out} \simeq 116\,M$, respectively, and the EOS
of the fluid was that of an ideal gas with adiabatic index
$\hat{\gamma}=4/3$. The setup therefore corresponds to the model referred
to as \texttt{S.50} in the work of \cite{Zanotti2010}. Instead of adding
a recoil velocity to the black hole as described in the previous section,
we evolved the system in a frame where the black hole was fixed and performed
a Lorentz boost on the fluid velocity to account for the recoil velocity
of the black hole. This has the advantage that the black hole remains at
the centre of the coordinate system and the metric functions need not
be updated. In all cases considered, the magnitude of the recoil
  velocity was $v_\mathrm{R} = 10^{-3}$ and was directed along the positive
  $x$ axis.

The simulation domain covers $r \in [1.85, 400]$, which extends into the
outer black hole horizon since we used Kerr-Schild coordinates.  We
  performed the simulations at the equatorial plane ($\theta = \pi/2$) with
  $\phi \in [0,2\pi]$.  To test convergence and the efficiency
of AMR with respect to uniform runs, we performed three simulations with
uniform resolutions, namely $N_r=256$, $512,$ and $1024$, and $N_\phi=
\tfrac{1}{2} N_r$, which we term {\it \textup{low}}, {\it \textup{medium},} and {\it \textup{high}}
resolutions, respectively.

We performed four AMR simulations, two using two refinement levels and two
using three levels. The base level has the same resolution as the lowest
resolution uniform run, and the cell dimensions are halved when a region
moves up one level. In this way, the highest level in a 2 (3) AMR level
run has a resolution equivalent to that of the medium- (high-) resolution
uniform run. As mentioned above, the refinement of the mesh was automated
and the decision of refining or coarsening a given block was taken based
on the L\"ohner estimator. The difference between each of the two pairs
of simulations with the same number of AMR levels lies in the tolerance
prescription. In each pair, one of the runs has a tolerance
$\varepsilon_t=0.1$ and the other has a tolerance
$\varepsilon_t=0.005$. Clearly, having a lower tolerance implies that the
AMR is switched on more frequently, so that in the limit of
$\varepsilon_t \to 0$, the highest refinement level is always present.

The evolution was carried out up to $t=20,\!000\,M$, corresponding to $\sim
15$ orbital periods. We here also made use of an atmosphere with
a fixed value for the rest-mass density $\rho_{{\rm atm}}$ chosen five
orders of magnitude lower than the highest rest-mass density in the
initial disc configuration. A low fixed value for the pressure
$p_{{\rm atm}}$ was chosen, and the material was set to be static with
Eulerian velocity $v^i=0$. 
In every time step, the
values of the primitive variables in the cell were set to the
atmospheric
values when the density in a given
cell fell below the threshold value of $f \rho_{{\rm atm}}$ with $f=1.5$.

%ssssssssssssssssssssssssssssssssssssssssssssssssssssssssssssssssss
\subsection{Results in 2D}\label{sec:recoilbhresult}
%ssssssssssssssssssssssssssssssssssssssssssssssssssssssssssssssssss

%ffffffffffffffffffffffffffffffffffffffffffffffffffffffffffffffffff
\begin{figure*}
\centering
\includegraphics[width=0.8\hsize]{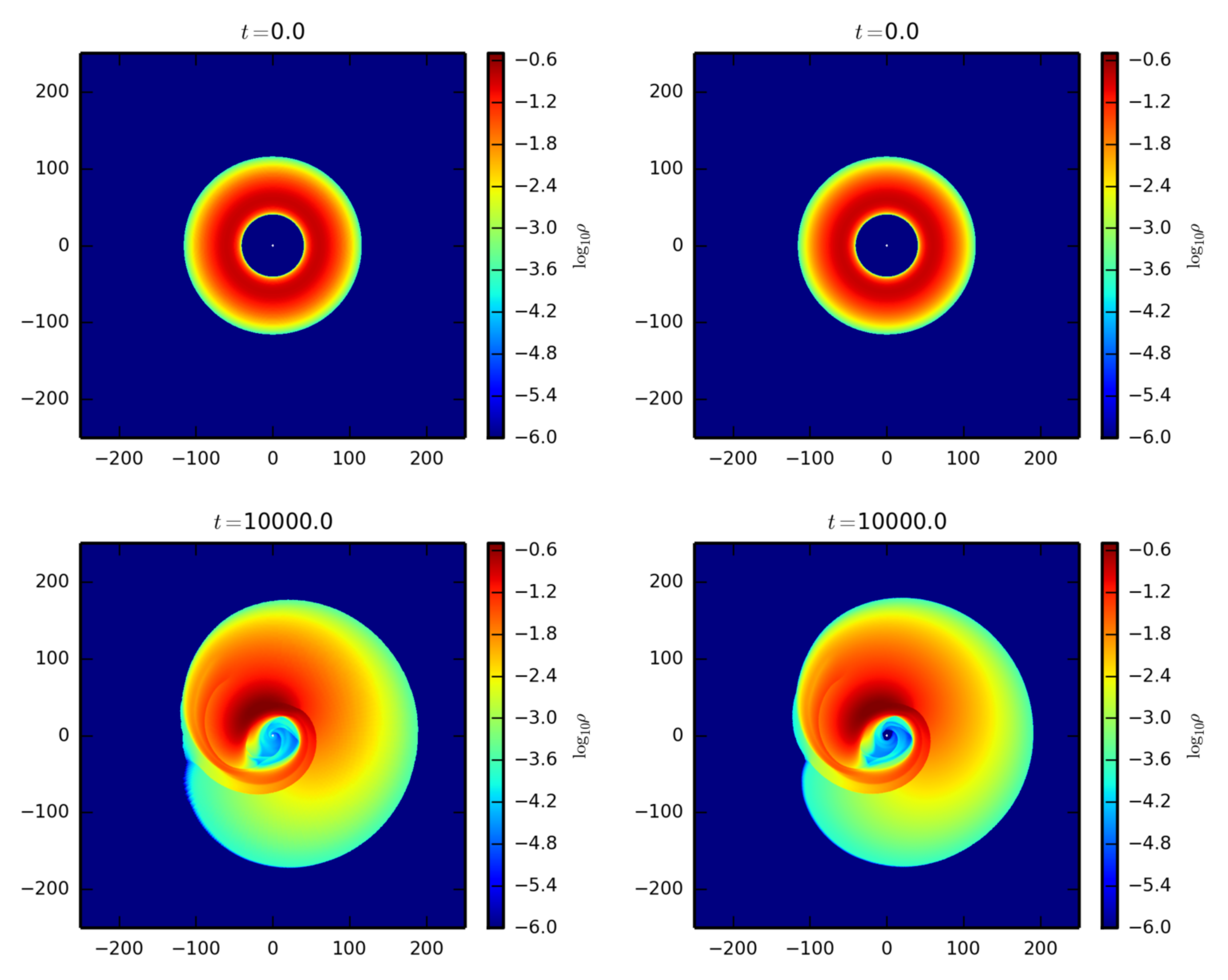}
\includegraphics[width=0.8\hsize]{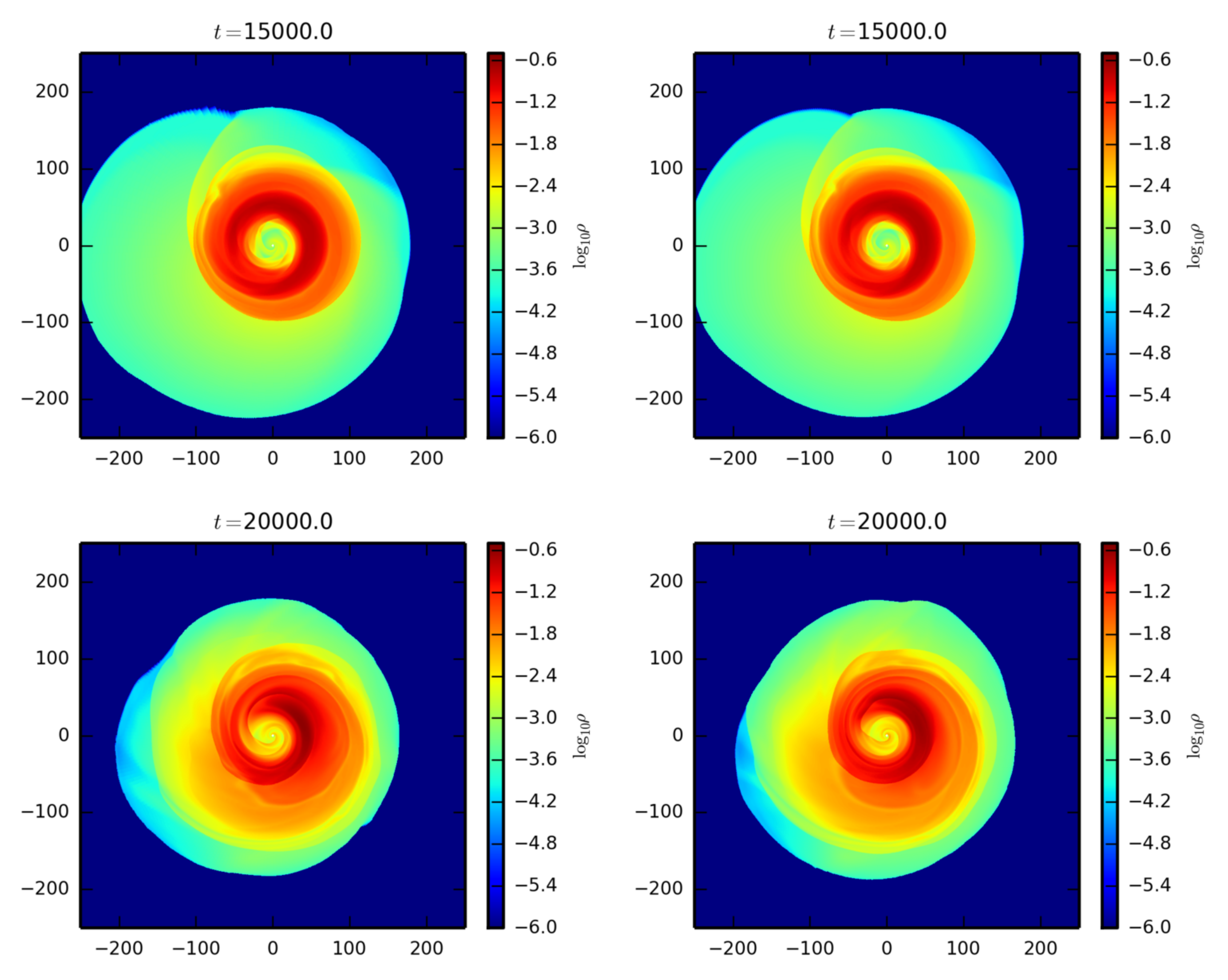}
\caption{Evolution of the logarithmic density in the 2D recoiling
  black hole simulations at times $0$, $10,\!000$, $15,\!000,$ and
  $20,\!000\,M$, with three AMR levels and tolerance $\varepsilon_t=0.1$
  (left columns) and a high-resolution uniform grid ($1024\times512$,
  right columns). At $t=0\,M$, the black hole is moving along the
    positive $x$ direction.}
\label{fig:2Drecoil-rho2D}
\end{figure*}
%ffffffffffffffffffffffffffffffffffffffffffffffffffffffffffffffffff

%ffffffffffffffffffffffffffffffffffffffffffffffffffffffffffffffffff
\begin{figure*}
\centering
\includegraphics[width=0.8\hsize]{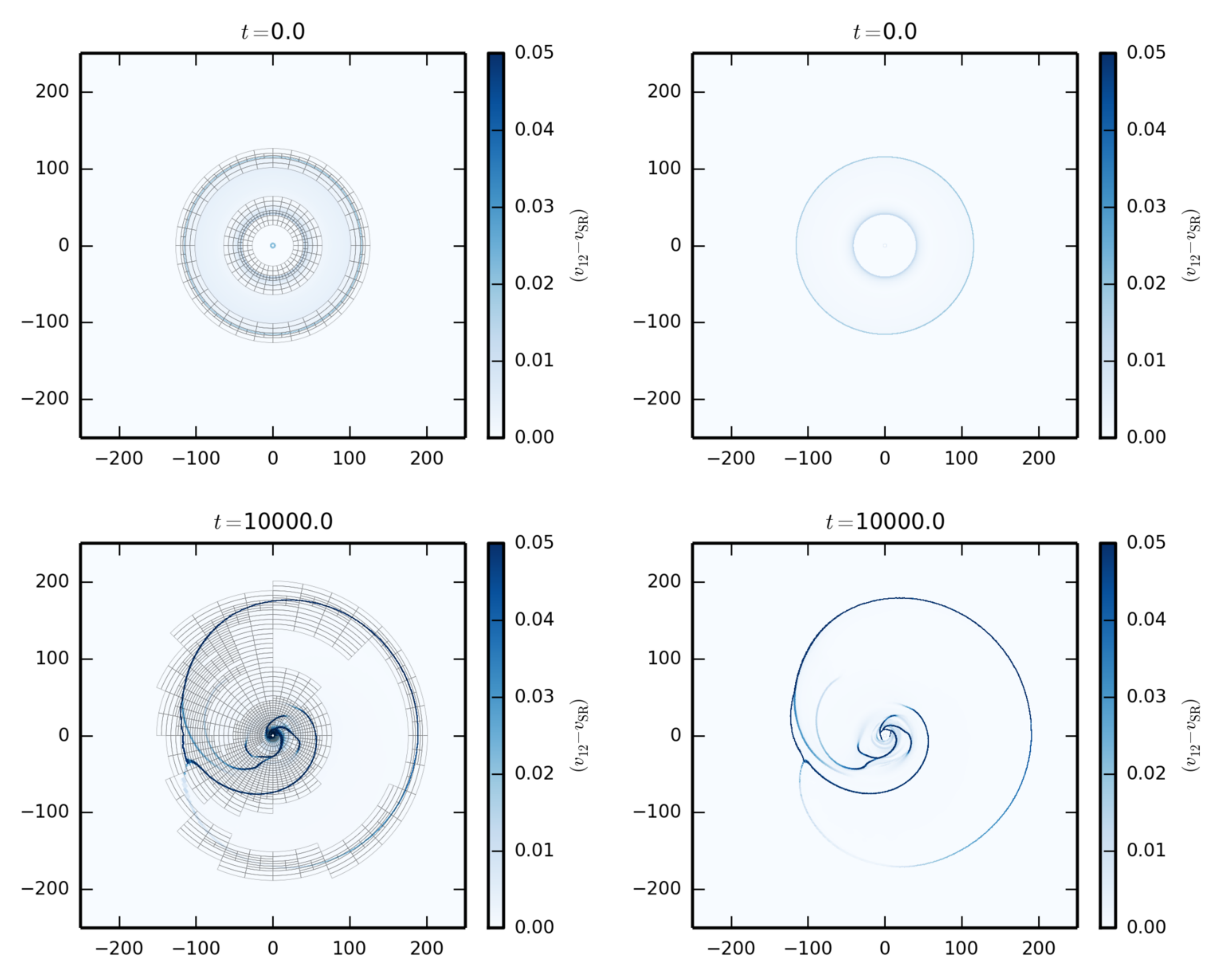}
\includegraphics[width=0.8\hsize]{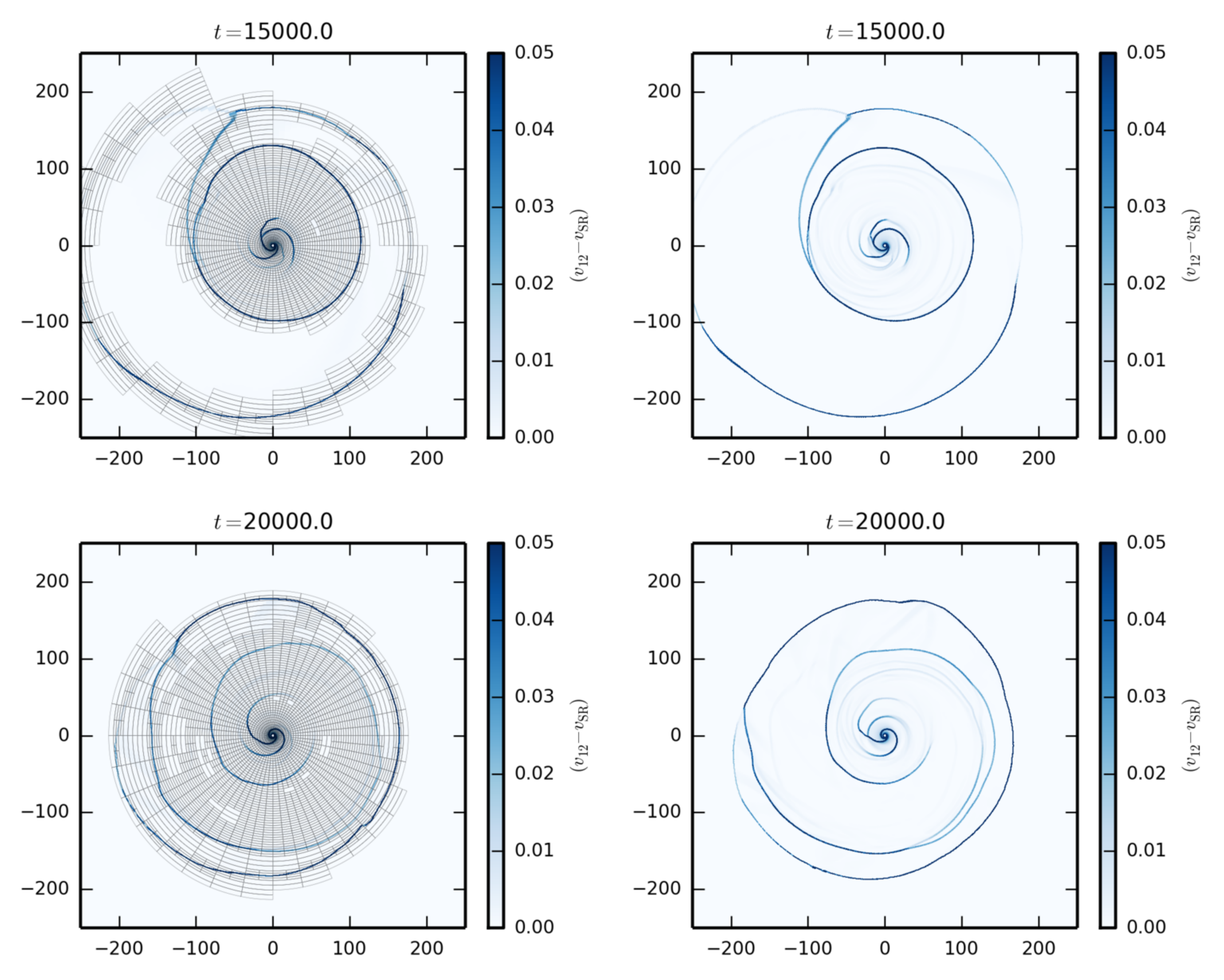}
\caption{Evolution of the shock structure in the 2D recoiling black hole
  simulations at times $0$, $10,\!000$, $15,\!000$ and $20,\!000\,M$, with
  three AMR levels and tolerance $\varepsilon_t = 0.1$ (left columns) and a
  high resolution uniform grid ($1024\times512$, right columns). The
  colour indicates the difference between the relative velocity of the
  sides of the Riemann problem and the threshold velocity for producing a
  shock. For the AMR simulation, the blocks at levels higher than 1 are
  shown.}
\label{fig:2Drecoil-shock2D}
\end{figure*}
%ffffffffffffffffffffffffffffffffffffffffffffffffffffffffffffffffff
%
Figure \ref{fig:2Drecoil-rho2D} shows the logarithmic density of the
fluid at four different simulation times for a three-level AMR
simulation and a high-resolution uniform simulation. AMR and uniform
grid cases exhibit very similar features. The asymmetry introduced by the
kick direction induces an accumulation of gas on one side of the disc,
with a corresponding significant decrease in density on the opposite
side of the disc. As time progresses, the variation in density and size
of the disc increases. Around $t=10,\!000\,M$, a part of the
disc matter
accretes onto the central black hole. By that time, the motion of the
recoiling black hole in the plane of the accretion disc has induced
spiral shocks that move outwards on a timescale that is comparable with
the orbital timescale. These shocks expand from the inner parts of the
disc and help to transport angular momentum outwards in the later
evolutionary stage. It is worth mentioning that even for an AMR
simulation with relatively high tolerance, a good qualitative agreement
with the equivalent high-resolution run can be seen, despite the presence
of strong shocks and complex dynamics.

The accurate determination of the position of the shock is important for
studying the dynamics of the recoiling black hole and for a correct
calculation of the emitted radiation. In previous studies
\citep{Lippai:2008, oneill2009, Megevand2009}, the propagation of a
spiral caustic and a possible shock was inferred only by checking the
density and/or pressure gradients. \cite{Corrales2009} introduced a
more accurate shock detector presented in the \texttt{FLASH}
code. However, these methods are rather empirical criteria and cannot be
used to detect weak shocks.

To improve the sensitivity in the determination of the shock position,
here we used a relativistic shock detector that exploits an idea proposed
in \citet{Zanotti2010} \citep[see also][for more
  details]{Rezzolla02,Rezzolla03}. It consists of the possibility of
predicting the outcome of the wave pattern in a Riemann problem. In
brief, given the left and right states of a Riemann problem, it is
possible to compute the threshold relative velocity between them
that are required to
produce a shock. The actual relative velocity between the two states is
compared to this value, and when it exceeds it, the region is marked as
shocked. The shock location obtained in this way is shown in
Fig. \ref{fig:2Drecoil-shock2D}. The development of a spiral shock in the
accretion disc is clearly seen. In the left panels of
Fig. \ref{fig:2Drecoil-shock2D}, we also plot the AMR blocks at higher
refinement levels (levels two and three). It is also quite clear from
Fig. \ref{fig:2Drecoil-shock2D}  that the L\"ohner scheme
\citep{Loehner87} used for estimating the error and triggering refinement
is very effective and triggers a refinement level even when
the shock is rather weak (cf. the trailing edge of the spiral
shock). Because of its intrinsic simplicity, it may be preferable to the
Rezzolla-Zanotti shock detector when the location of the shock is not of
paramount importance.

%ffffffffffffffffffffffffffffffffffffffffffffffffffffffffffffffffff
\begin{figure}[]
\centering
\includegraphics[width=\hsize]{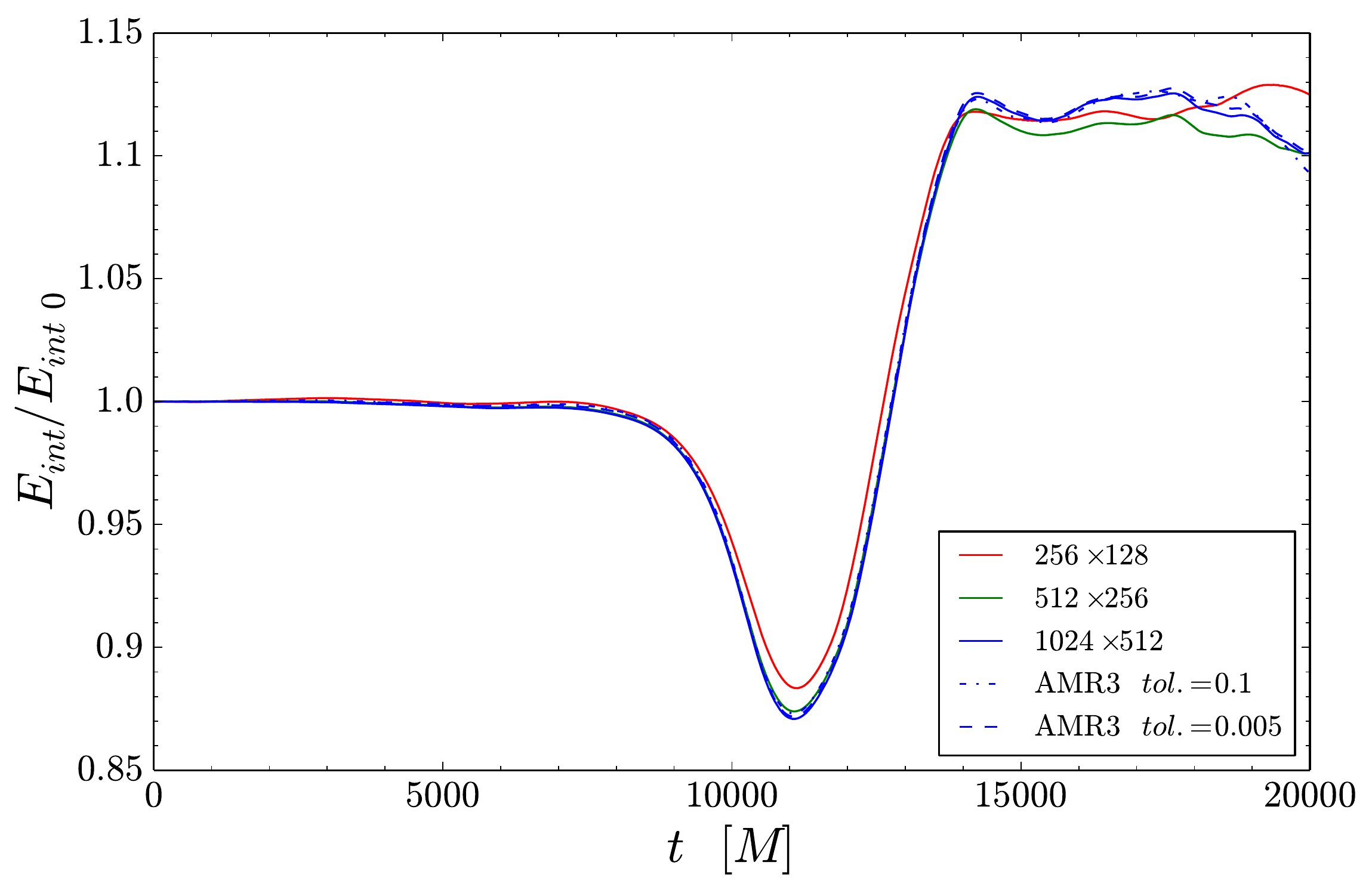}
\caption{Evolution of the internal energy of the disc normalised with
  respect to the
  initial internal energy of the disc and relative to the 2D simulations
  of a recoiling black
  hole. Different lines refer to simulations with uniform grid using
  three different resolutions: $256 \times 128$ (red solid line), $512
  \times 256$ (green solid line), and $1024 \times 512$ (blue solid line).
  We also show the internal energy for AMR simulations using three levels with
different tolerances of $\varepsilon_t=0.1$ (blue dashed) and
  $\varepsilon_t=0.005$ (blue dash-dotted).}
\label{fig:2Drecoil-Eint}
\end{figure}
%ffffffffffffffffffffffffffffffffffffffffffffffffffffffffffffffffff

To analyse the effect of AMR on the dynamics of the disc, we calculated
the internal energy (\ie the volume integral of the internal energy
density $\rho\epsilon$), which can later be compared to the light
curves of
the calculated thermal radiation. In Fig. \ref{fig:2Drecoil-Eint} we show
the evolution of this quantity using uniform grids with different
resolution and using AMR with different tolerances. The overall behaviour
is similar for all cases. Initially, very small oscillations are
seen. This is related to the epicyclic picture of the density maximum in
the disc. The sharp drop occurs when the disc matter is accreted onto the
central black hole. Then a sharp rebound occurs as a result of the development of
the shock in the disc. In the later evolutionary stage, the internal
energy maintains a higher value with small oscillations. A similar
behaviour for the volume-integrated internal energy has been reported in
\cite{Megevand2009}. We also show in Fig. \ref{fig:2Drecoil-Eint}  the
late-time behaviour of the volume-integrated internal energy, which
changes slightly with grid resolution.
However, the AMR cases are in very good agreement with the corresponding high-resolution run, demonstrating the successful capturing of the shock.

We also checked the convergence of the simulations with different
resolutions for the uniform grid and AMR runs and obtained the expected
convergence order in both cases (see Appendix \ref{sec:convergence} for
more details).

%ffffffffffffffffffffffffffffffffffffffffffffffffffffffffffffffffff
\begin{figure}
\centering
\includegraphics[width=\hsize]{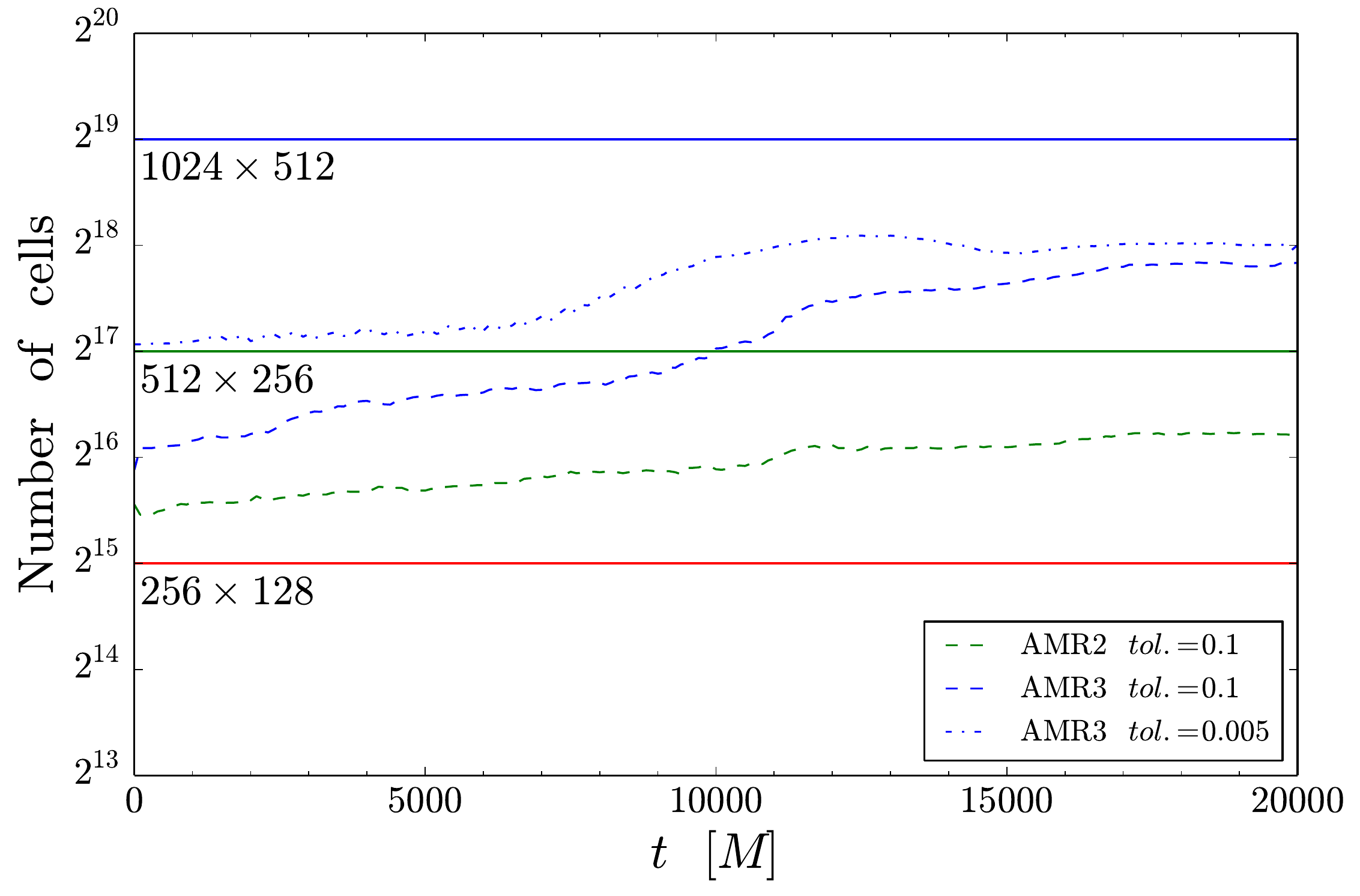}
\caption{Total number of cells during the simulation of the 2D recoiling
  black hole with uniform grid using three different resolutions ($256 \times
  128$: red, $512 \times 256$: green, and $1024 \times 512$: blue) and
  AMR with different tolerances of $\varepsilon_t = 0.1$ (two levels:
  green dashed, three levels: blue dashed) and $\varepsilon_t = 0.005$
  (three levels: blue dash-dotted).}
\label{fig:2Drecoil-grid-occupation}
\end{figure}
%ffffffffffffffffffffffffffffffffffffffffffffffffffffffffffffffffff

As mentioned above, the value of the tolerance for switching on a
refinement level has directly affects whether new cells are
introduced where the equations are to be solved, thus translating into
additional computational cost. Figure \ref{fig:2Drecoil-grid-occupation}
shows the evolution of the total number of cells during each simulation
for each of the different cases. The solid lines correspond to
simulations with uniform grids, while the dashed lines indicates AMR
cases. Initially, $2^{17} \sim 131,\!000$ cells were used even when we used
three AMR levels, which is similar to the number of cells for the
simulation having uniform and medium resolution. When the simulations
enter the accretion phase, however, the total number of cells rapidly
increases because the spiral shock forms, triggering higher refinement
and expanding within the disc. We note that the total number of cells is
still smaller than or nearly half of the total number of cells in the
corresponding high-resolution simulation (blue solid line), thus
resulting in a direct reduction of the computational cost.

%tttttttttttttttttttttttttttttttttttttttttttttttttttttttttttttt
\begin{table}
\caption{CPU hours (CPUH) spent by the simulations of the 2D recoiling
  black hole at uniform resolutions, and fraction of that time spent by
  the equivalent AMR runs.}
\label{tab:2Dcomp-normalized-time}
\centering
\begin{tabular}{cccc}
\hline \hline
Grid size &  CPU time & Equiv. AMR & Equiv. AMR \\
($N_r\times N_\phi$) & uniform  & time fraction & time fraction  \\
 & [CPUH] & [$\varepsilon_t=0.1$] & [$\varepsilon_t=0.005$] \\
\hline
$256\times128$  & $55.0$    & ---    & ---    \\
$512\times256$  & $443.1$   & $0.65$ & $0.70$ \\
$1024\times512$ & $3,377.4$ & $0.47$ & $0.57$ \\
\hline
\end{tabular}
\end{table}
%tttttttttttttttttttttttttttttttttttttttttttttttttttttttttttttt

A comparison of the computational time for each of the 2D recoiling black
hole simulations is shown in Table \ref{tab:2Dcomp-normalized-time}. It
is remarkable that even the three-level AMR simulation could obtain
results of an accuracy comparable to the high-resolution uniform run, but
spent only slightly more than half the computational time used in the
uniform run. In Sect. \ref{sec:ray-recoil2D} we show that the
agreement between this three-level AMR and its corresponding uniform run was
also excellent for the general-relativistic radiative transfer
calculation.

In summary, the AMR employed in our code has proven to be essential for
physical scenarios such as the recoiling black hole, where the dynamics of
the kicked accretion disc are very sensitive to the underlying numerical
resolution. The AMR refinement strategy, triggered by the
L\"ohner scheme, effectively
captures the spiral shock structure developed in the accretion
disc. Moreover, the simulations using AMR require only roughly half of
the computational time of the corresponding uniform grid cases with
highest resolution, making AMR a very useful tool for 3D simulations
involving large-scale shocks.

%ssssssssssssssssssssssssssssssssssssssssssssssssssssssssssssssssss
\subsection{Initial setup in 3D}\label{sec:recoilbhsetup3d}
%ssssssssssssssssssssssssssssssssssssssssssssssssssssssssssssssssss

In contrast to the previous section and to \cite{Zanotti2010}, here we
dropped the assumption that the disc is geometrically thin and evolved the
dynamics in full 3D. The initial setup was now a geometrically thick torus
with a constant angular momentum distribution as described in Sect.
\ref{sec:stationary}. The parameters of this torus and the black hole are
the same as for the 2D case, so that the densities, pressures, and fluid
velocities on the equatorial plane match those of the 2D simulation at
$t=0M$. Specifically, they are $a=0.5$, $\ell=8$, $r_{\rm in}=40\,M$,
$r_{\rm out} = 116\,M,$ and $\hat{\gamma}=4/3$.

The numerical domain
extends over $r \in
[1.85\,M, 400\,M]$ and $\phi \in [0, 2\pi]$. Taking advantage of the
symmetry of the problem with respect to the equatorial plane, we
considered only the upper half of the torus. The domain spanned the
region $\theta \in [\pi/8, \pi/2]$. At the equatorial plane, symmetric
boundary conditions were applied to all variables except for the vertical
component of the velocity, for which an antisymmetric boundary condition
was applied, all to account for a perfectly symmetric lower half
of the torus. We note that during the simulation the fluid never reaches
the outer boundaries of the domain. In this case, the region outside
of the torus was also filled with a tenuous atmosphere following the same
prescription as employed in the 2D case.

We again quantified convergence and compared the performance of AMR to that
of a high-resolution uniform grid simulation. To this end, we performed
three simulations at uniform resolutions of $N_r=128$, $256$, and $512$,
$N_\phi= \tfrac{1}{2} N_r$ , and $N_\theta = \tfrac{1}{8} N_r$, to which we
refer as {\it \textup{low}}, {\it \textup{medium}}, and {\it \textup{high}}.

We performed three simulations using AMR with two and three levels,
for which the base level had the same resolution as the low-resolution
uniform run, and the highest level had the same resolution as the medium
and high resolutions of the uniform cases, respectively.
Owing to the higher computational cost of 3D simulations, this time we
used AMR tolerances higher than in the 2D cases. Instead,
for this setup we tested another feature of the implementation of AMR in
the code,
namely the possibility of specifying a different tolerance for triggering
refinement at the various levels. Setting a higher tolerance for the highest
levels results in a lower propensity of the code to refine towards those levels,
which might decrease the computational cost.

More specifically, for the first two runs (2 and 3 levels), we set a tolerance
of $\varepsilon_t = 0.1$ between every level. For the third run,
we used three levels and set $\varepsilon_t = 0.1$ for refining to the second
level and a less demanding $0.3$ for refining to the third level.

In each case the system was
evolved up to a time of $20,\!000\,M$, corresponding to $\sim 15$ orbital
periods.

%ssssssssssssssssssssssssssssssssssssssssssssssssssssssssssssssssss
\subsection{Results in 3D}\label{sec:3dresult}
%ssssssssssssssssssssssssssssssssssssssssssssssssssssssssssssssssss

%ffffffffffffffffffffffffffffffffffffffffffffffffffffffffffffffffff
\begin{figure*}[]
\centering
\includegraphics[width=0.495\hsize]{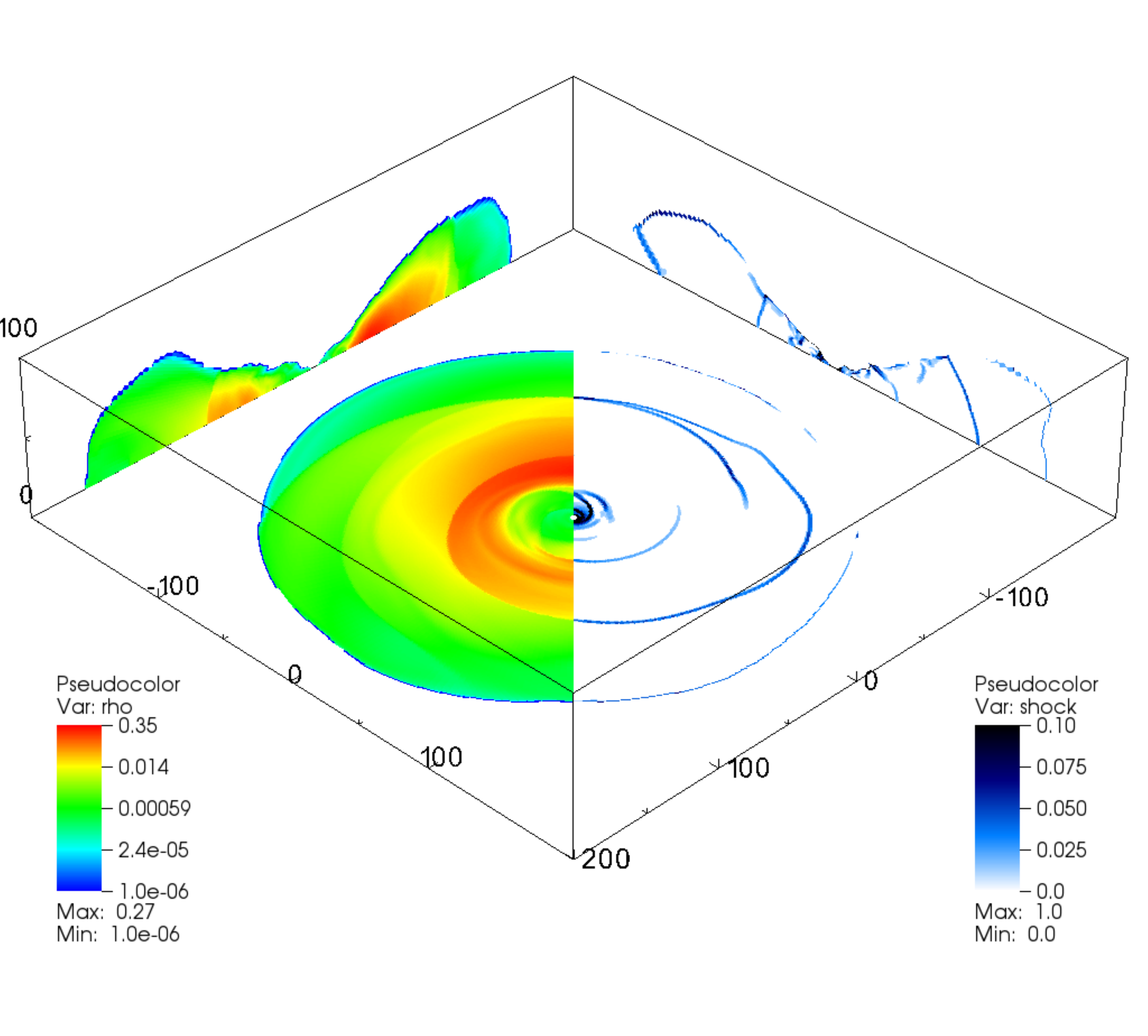}
\includegraphics[width=0.495\hsize]{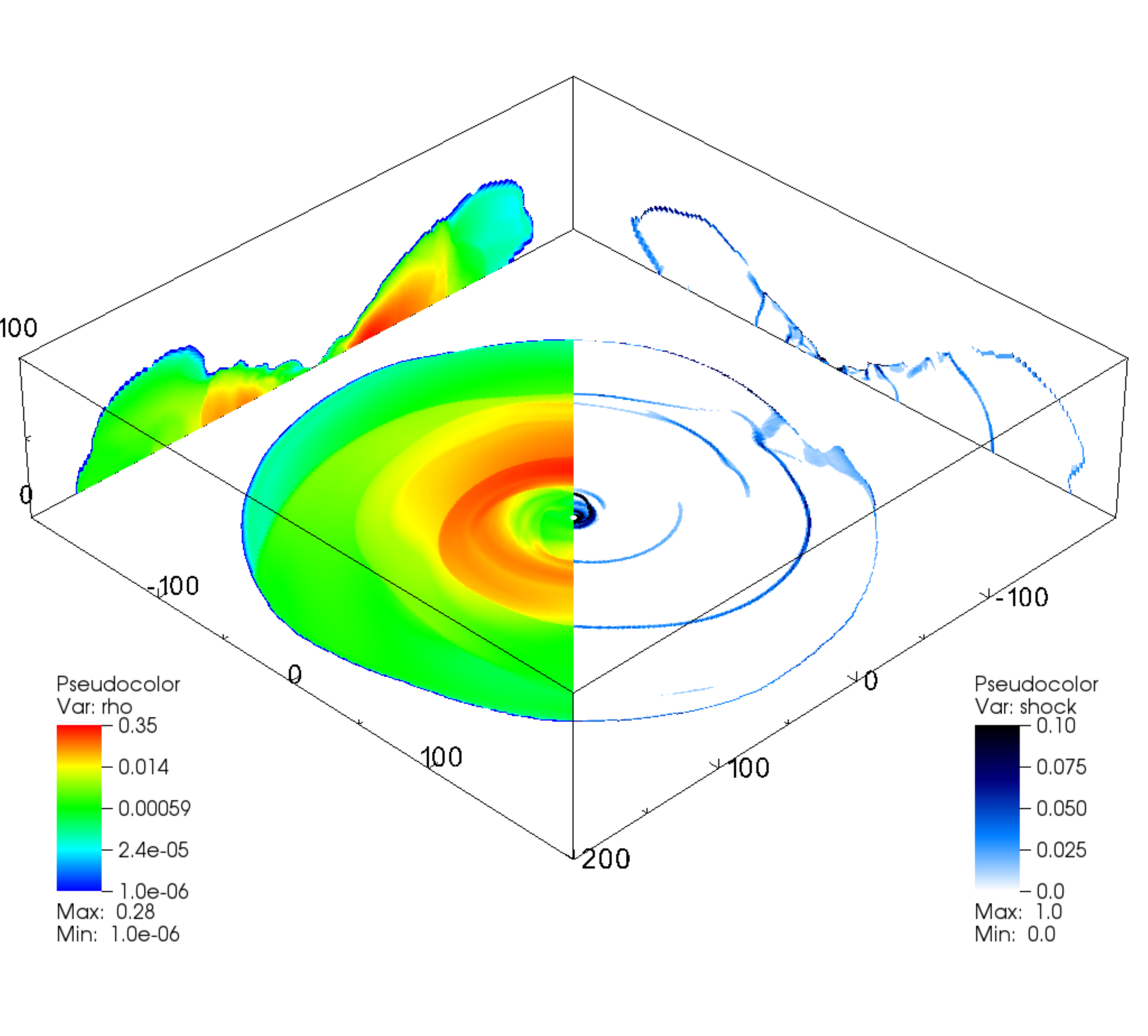} \\
\caption{Snapshots of the logarithmic density and the shock structure of
  the 3D recoiling black hole at the final time, $t=20,\!000\,M$, for the
  uniform run with resolution $512\times256\times64$ ({\it left}) and an
  equivalent three-level AMR run with high tolerance
  ($\varepsilon_t=0.1,0.3$, {\it right}). The floor of the box shows a cut
  through the equatorial plane and the walls show perpendicular slices
  crossing the origin.}
\label{fig:3D-rho-shock}
\end{figure*}
%ffffffffffffffffffffffffffffffffffffffffffffffffffffffffffffffffff

Figure \ref{fig:3D-rho-shock} shows vertical (\ie on the $(x,z)$ plane)
and horizontal (\ie on the $(x,y)$ plane) cuts of the density field
(left half) and the shock structure (right half) of the torus at the
final time of $t=20,\!000\,M$. We also show on the $(x,y)$ and $(y,z)$ planes
the cuts that show the 3D location of the shocks produced by the
recoiling black hole as it interacts with the accreting torus.

Here the kick velocity also breaks the symmetry of the
initial density profile and leads to an accumulation of gas in a small
region of the disc. Similarly to the 2D case, the compression of gas due
to the kick velocity eventually evolves into a spiral shock that is
completely visible around $t=10,\!000\,M$.  As can be appreciated in the
vertical slices of Fig. \ref{fig:3D-rho-shock}, now the spiral shock
also extends in the vertical direction. However, in contrast to the 2D
case, accretion onto the central black hole does not start until $t
\approx 19,\!500\,M$. The reason is that the fluid is no longer confined to
the equatorial plane when the interaction with the black hole produces a
compression wave that, together with the centrifugal force, allows the
fluid to expand in the vertical direction as well.

\begin{figure}[]
\centering
\includegraphics[width=\hsize]{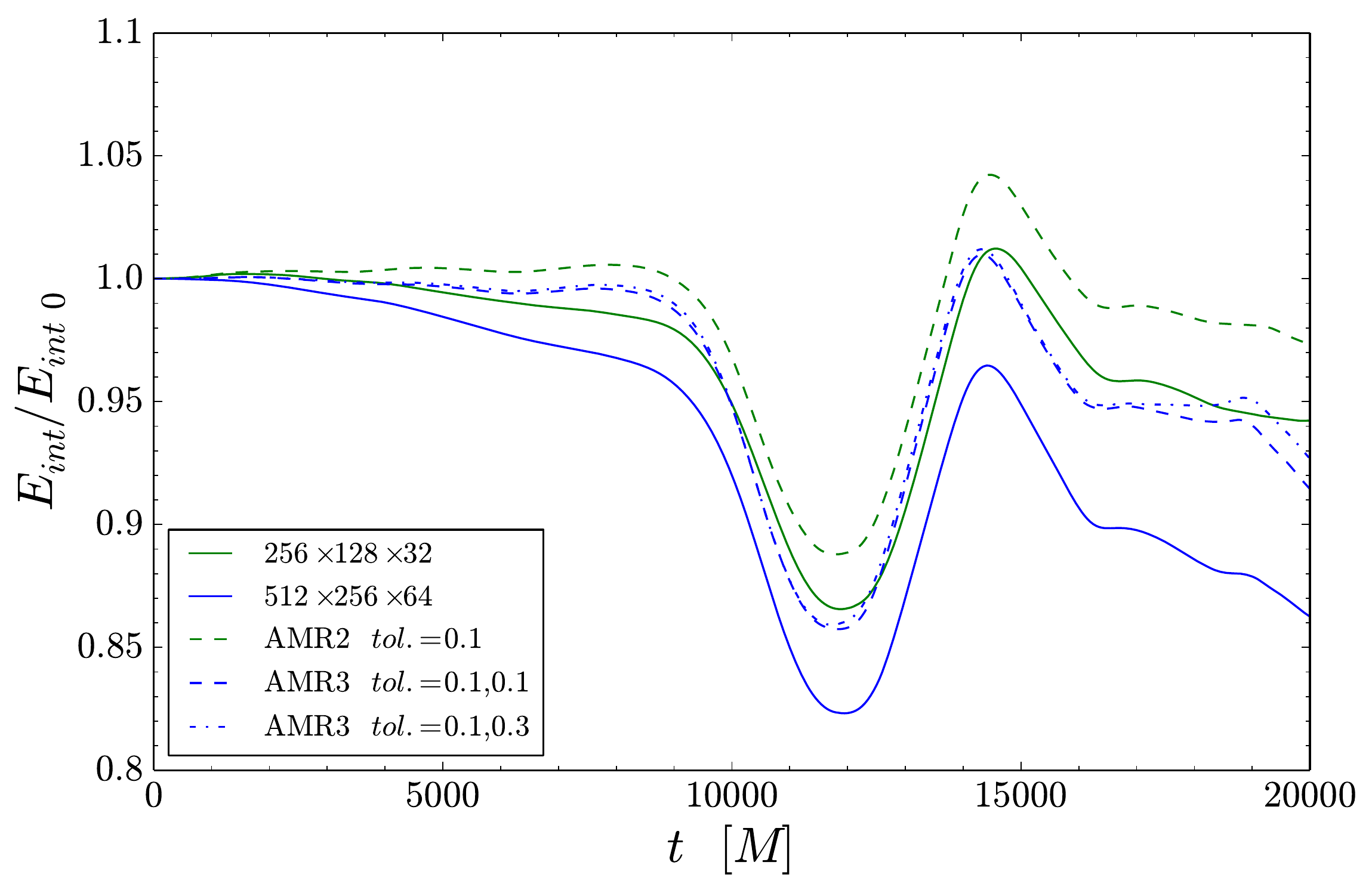}
\caption{Evolution of volume-integrated (normalised) internal energy of
  the torus for the 3D recoiling black hole cases, with uniform grid
  (solid) using the two highest resolutions ($256\times 128\times 32$:
   green, and $512\times 256 \times 64$:
  blue) and AMR (dashed) with two (green) and three (blue) levels.}
\label{fig:3Drecoil-Eint}
\end{figure}

In analogy with the 2D case, Fig. \ref{fig:3Drecoil-Eint} shows the
volume-integrated internal energy of the torus, normalised to its initial
value. By comparing it with Fig. \ref{fig:2Drecoil-Eint}, we can
appreciate some differences between the dynamics in 2D and 3D. While in
2D the internal energy falls at around $t=10,\!000\,M$ as a result
of the expansion
of the disc and rises again to a nearly constant value when the gas is
heated by the shock, in 3D this is no longer possible
because the fluid cools down when it expands in the vertical direction as
a result of the interaction with the shock.

Even though the evolution of the internal energy of the simulation at the
lowest resolution differs significantly from the results of the other
simulations, it is still qualitatively similar. In Appendix
\ref{sec:convergence} we show that in uniform grids as well as
in AMR, the solution of the equations converges at the expected order.

Table \ref{tab:3Dcomp-normalized-time} is the equivalent of Table
\ref{tab:2Dcomp-normalized-time} for the 3D simulations and shows the
CPU time spent by simulations performed at uniform resolutions and the
fraction of that time spent by simulations using AMR.
Two three-level AMR runs were performed that differ only in the
  refinement threshold on the highest level: $\varepsilon_t=0.1$ with
  respect to $\varepsilon_t=0.3$. While the results were practically
  indistinguishable from one another, only a very small saving in
  computational time was achieved, namely 13,550.0 versus 13,416.0 CPU-hours,
  which corresponds to a difference of 0.5 \%.
The AMR simulations obtained an accuracy comparable to
the equivalent uniform-resolution simulations, but with a saving in CPU
time of $\sim 85\%$. In other words, we obtained numerically equivalent
results using slightly more than $1/10$ of the resources. Even though the
three-level AMR run is closer to the medium-resolution run than
to the high-resolution run, both AMR simulations remain close to their
equivalent uniform runs. As a final remark, we emphasise again that as
was shown for the 2D case, an AMR simulation can become as close as
desired to its equivalent high-resolution uniform run by reducing the
tolerance $\varepsilon_t$.

%ffffffffffffffffffffffffffffffffffffffffffffffffffffffffffffffffff
%ffffffffffffffffffffffffffffffffffffffffffffffffffffffffffffffffff

%tttttttttttttttttttttttttttttttttttttttttttttttttttttttttttttt
\begin{table}
\caption{CPU hours (CPUH) spent by the simulations of the 3D recoiling
  black hole at uniform resolutions, and fraction of that time spent by
  the equivalent AMR runs.} 
\label{tab:3Dcomp-normalized-time}
\centering
\begin{tabular}{cccc}
\hline \hline
Grid size &  CPU time & Equiv. AMR & Equiv. AMR \\
$[N_r\times N_\phi\times N_\theta]$ & uniform  & time fraction & time fraction\\
 & [CPUH] & [$\varepsilon_t=0.1,0.1$] & [$\varepsilon_t=0.1,0.3$]\\
\hline
$128\times64\times16$  & 667.1    & ---  & ---  \\
$256\times128\times32$ & 8,557.2  & 0.15 & ---  \\
$512\times256\times64$ & 93,144.8 & 0.14 & 0.14 \\
\hline
\end{tabular}
\end{table}
%tttttttttttttttttttttttttttttttttttttttttttttttttttttttttttttt

%ssssssssssssssssssssssssssssssssssssssssssssssssssssssssssssssssss
\section{Ray-tracing and radiation transfer of solutions}
\label{sec:raytracing}
%ssssssssssssssssssssssssssssssssssssssssssssssssssssssssssssssssss

To accurately compute the electromagnetic emissions from our
simulations, it is necessary to perform ray-tracing calculations coupled
with general-relativistic radiation transfer calculations
\citep[\eg][]{Fuerst2004, Vincent2011, Younsi2012, Younsi2015,
  Dexter2016, Pu2016}. These calculations were performed in
post-processing and therefore the effect of radiation forces coupled with
the hydrodynamic evolution of the material were not included. We also
employed the so-called fast-light approximation, where the dynamical
timescale of the simulation is taken to be much longer than the light-crossing time, and so the finite travel time of photons and their
relative arrival time delays may be neglected. Such an approximation is
acceptable for the large-scale recoiling black hole simulations
considered in this paper.

Electromagnetic radiation follows null geodesics of the space-time, thus
we calculated the geodesics through direct numerical integration of the
geodesic equations of motion. The geodesics were solved using an adaptive
fourth-order Runge-Kutta scheme, integrating backwards in time from an
observer at $10^{3}\,r_{\mathrm{g}}$ from the black hole (where space-time
is practically Euclidean) and assuming that all rays arrive perpendicular
to the observer's image plane.

After calculating the geodesic for each ray, we then solved the
radiation transport equation. We employed the ray-tracing
and radiation transport scheme described in \cite{Younsi2012}. In
covariant form, the general-relativistic radiation transport equation (in
the absence of scattering) may be written as
\begin{equation}
\frac{\mathrm{d}\mathcal{I}}{\mathrm{d}\lambda} = -k_{\mu}u^{\mu}|_{\lambda}\left( -\alpha_{\nu,0}\mathcal{I}+\frac{j_{\nu,0}}{\nu^{3}} \right) \,,
\end{equation}
where the Lorentz-invariant intensity
$\mathcal{I}\equiv I_{\nu}/\nu^{3}$, $\nu$ is the frequency of radiation,
$I_{\nu}$ is the specific intensity, and $\alpha_{\nu,0}$ and $j_{\nu,0}$
are the specific absorption and emission coefficient,
evaluated at frequency $\nu$ and in the local fluid rest frame (hence
denoted by the subscript 0) . Here $k_{\mu}$ is the photon
four-momentum, $u^{\mu}$ is the four-velocity of the emitting medium,
and $\lambda$ is the affine parameter. This equation may be rewritten in
terms of the optical depth of the medium as

\begin{equation}
\frac{\mathrm{d}\mathcal{I}}{\mathrm{d}\tau_{\nu}} = -\mathcal{I} +
\frac{\eta}{\chi} \,, \label{GRRT}
\end{equation}
where $\tau_{\nu}$, the optical depth evaluated at frequency $\nu$, is calculated as
\begin{equation}
\tau_{\nu}\left(\lambda \right) = -\int_{\lambda_{0}}^{\lambda}\mathrm{d}\lambda' \ \! \alpha_{\nu,0} \left(\lambda' \right) k_{\mu}u^{\mu}|_{\lambda'} \,,
\end{equation}
and the invariant emission coefficient $\eta$ and invariant absorption
coefficient $\chi$ are defined as $\eta\equiv j_{\nu}/\nu^{2}$ and
$\chi\equiv\nu\alpha_{\nu}$.

The radiative-transfer equation (\ref{GRRT}) may itself be reduced to two
differential equations \citep[see][for details]{Younsi2012}, yielding
\begin{eqnarray}
\frac{\mathrm{d}\tau_{\nu}}{\mathrm{d}\lambda} &=&
\gamma^{-1}\alpha_{\nu,0} \,, \label{GRRT_1} \\
\frac{\mathrm{d}\mathcal{I}}{\mathrm{d}\lambda} &=&
\gamma^{-1}\frac{j_{\nu,0}}{\nu^{3}} \ \!
\mathrm{exp}\left(-\tau_{\nu}\right) \label{GRRT_2} \,, 
\end{eqnarray}
where the relative energy shift, $\gamma$ (not to be confused with the
determinant of the three-metric), between the radiation emitted from
material orbiting the black hole with four-velocity $u^{\alpha}$ and the
radiation received by a distant observer is given by
\begin{equation}
\gamma^{-1} \equiv \frac{\nu_{0}}{\nu} =
\frac{k_{\alpha}u^{\alpha}|_{0}}{k_{\beta}u^{\beta}|_{\mathrm{obs}}} \,.
\end{equation}
The subscript obs denotes the reference frame of a distant
  observer. Given that the background metric is stationary, the geodesic
  equations of motion are time-symmetric.  The fast-light approximation
  was also adopted, therefore the fluid at each observer time slice is
  stationary, and Eqs. (\ref{GRRT_1})-(\ref{GRRT_2}) are also
  time-symmetric.  We consequently set both the initial intensity
  $\mathcal{I}$ and initial opacity $\tau_{nu}$ to zero, directly
  integrating Eqs. (\ref{GRRT_1})-(\ref{GRRT_2}) together with the
  geodesic equations of motion, backwards in time.

We illustrate the features of this approach. Firstly, it avoids the
process of having to integrate the geodesics backwards in time, store the
geodesics in memory, and then integrate the radiative transfer equations
forward in time towards the observer. Secondly, it offers the option of
specifying a threshold optical depth (typically on the order
of unity) when
encountering optically thick media, enabling the geodesic integration to
be terminated when this optical depth threshold is
exceeded. Consequently, this approach saves significant computational
expense and time.

%ssssssssssssssssssssssssssssssssssssssssssssssssssssssssssssssssss
\subsection{Thermodynamic quantities}\label{sec:ray-thermo}
%ssssssssssssssssssssssssssssssssssssssssssssssssssssssssssssssssss

When we calculate the radiation transport of simulation data, we must
specify the emission and absorption coefficients for all relevant
radiative processes. These coefficients must be calculated in physical
units, whereas the simulation data are output in geometrised units. Length
and times are easily converted into cgs units through re-introducing the
mass, $M$, of the black hole, which hereafter we take to be
$M=10^8\,M_{\odot}$. However, the emission
and absorption coefficients also depend on the density and temperature in
cgs units.  Following \citet{Schnittmanetal2013}, and using the fact that
our EOS is ideal, the conversion between geometrised ($\rm{geo}$) and
cgs units for the fluid temperature is given by
\begin{equation}
T_{\mathrm{cgs}} =
\left(\frac{P_{\mathrm{geo}}}{\rho_{\mathrm{geo}}}\right) \frac{\mu
  m_{\mathrm{p}}}{k_{\mathrm{B}}}c^{2} \,,
\end{equation}
where $\mu$ is the mean molecular
weight of electrons and ions, $m_{\mathrm{p}}$ is the proton rest mass, and
$k_{\mathrm{B}}$ is the Boltzmann constant.
In analogy to \cite{Andersonetal10} and \citet{Zanotti2010}, we
scaled the initial rest mass density at the centre of the torus to
be $\rho_{\mathrm{c}} =1.38\times
10^{-10}~\mathrm{g}~\mathrm{cm}^{-3}$.  The mean molecular weight
was determined from
\begin{equation}
\frac{1}{\mu} \equiv \frac{1}{\mu_{\mathrm{e}}} + \frac{1}{\mu_{\mathrm{i}}} \,,
\end{equation}
where the effective molecular weights of electrons and ions are given by
\begin{align}
\mu_{\mathrm{e}} &\equiv \frac{2}{1+X} \,, &
\mu_{\mathrm{i}} &\equiv \frac{4}{1+3X} \,.
\end{align}
Consequently, the mean molecular weight is given by
\begin{equation}
\mu = \frac{4}{3+5X} \,,
\end{equation}
where $X$, the relative abundance of hydrogen, is set to $3/4$ in all our
calculations, giving $\mu_{\mathrm{e}}=8/7$, $\mu_{\mathrm{i}}=16/13$,
and $\mu=16/27$. With these definitions the electron and ion number
densities are given by
\begin{align}
n_{\mathrm{e}} &= \frac{\rho_{\mathrm{cgs}}}{\mu_{\mathrm{e}} m_{\mathrm{p}}} \,, &
n_{\mathrm{i}} &= \frac{\rho_{\mathrm{cgs}}}{\mu_{\mathrm{i}} m_{\mathrm{p}}} \,.
\end{align}

%ffffffffffffffffffffffffffffffffffffffffffffffffffffffffffffffffff
\begin{figure*}
\centering
\includegraphics[width=0.275\hsize]{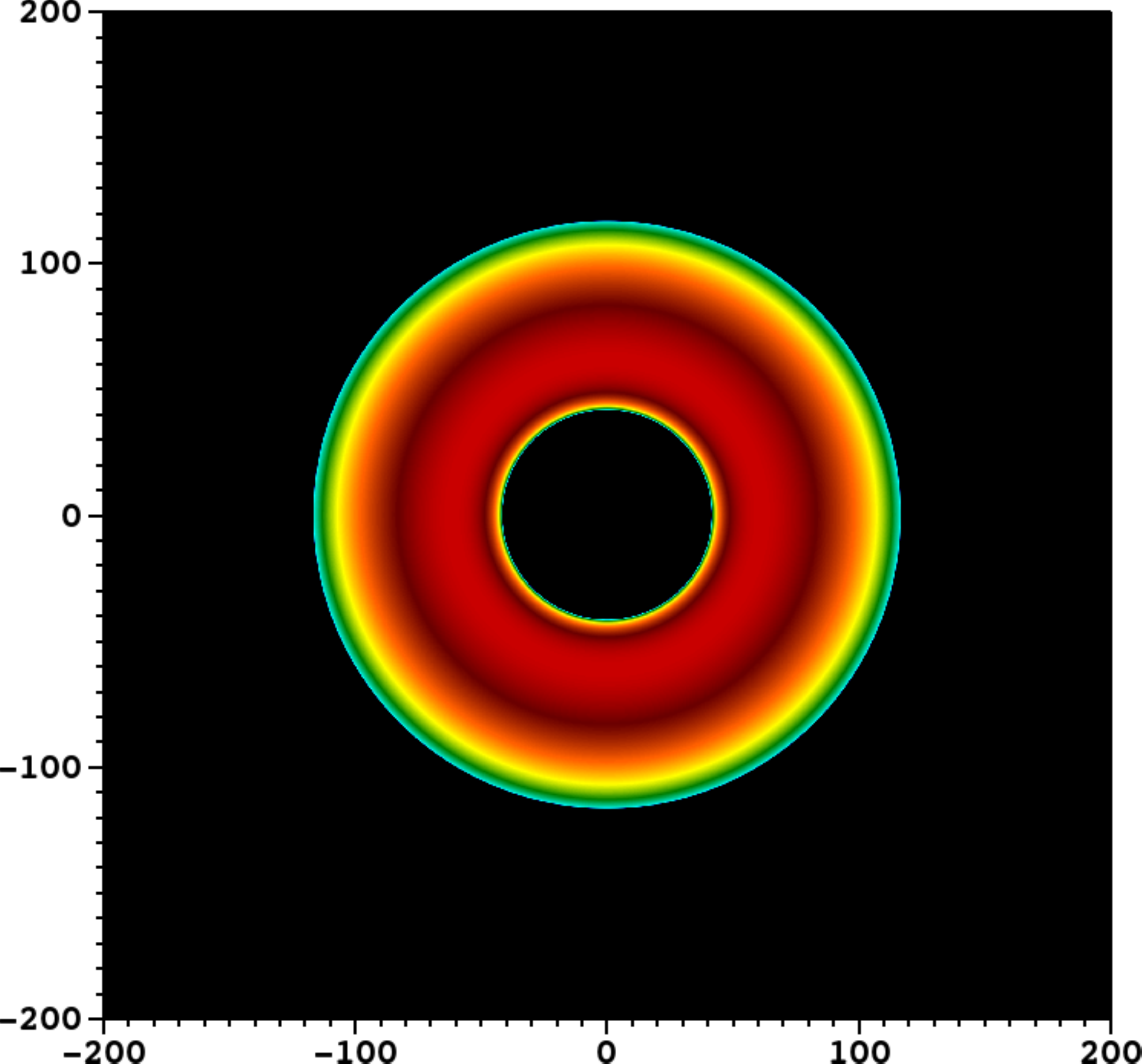}
\includegraphics[width=0.275\hsize]{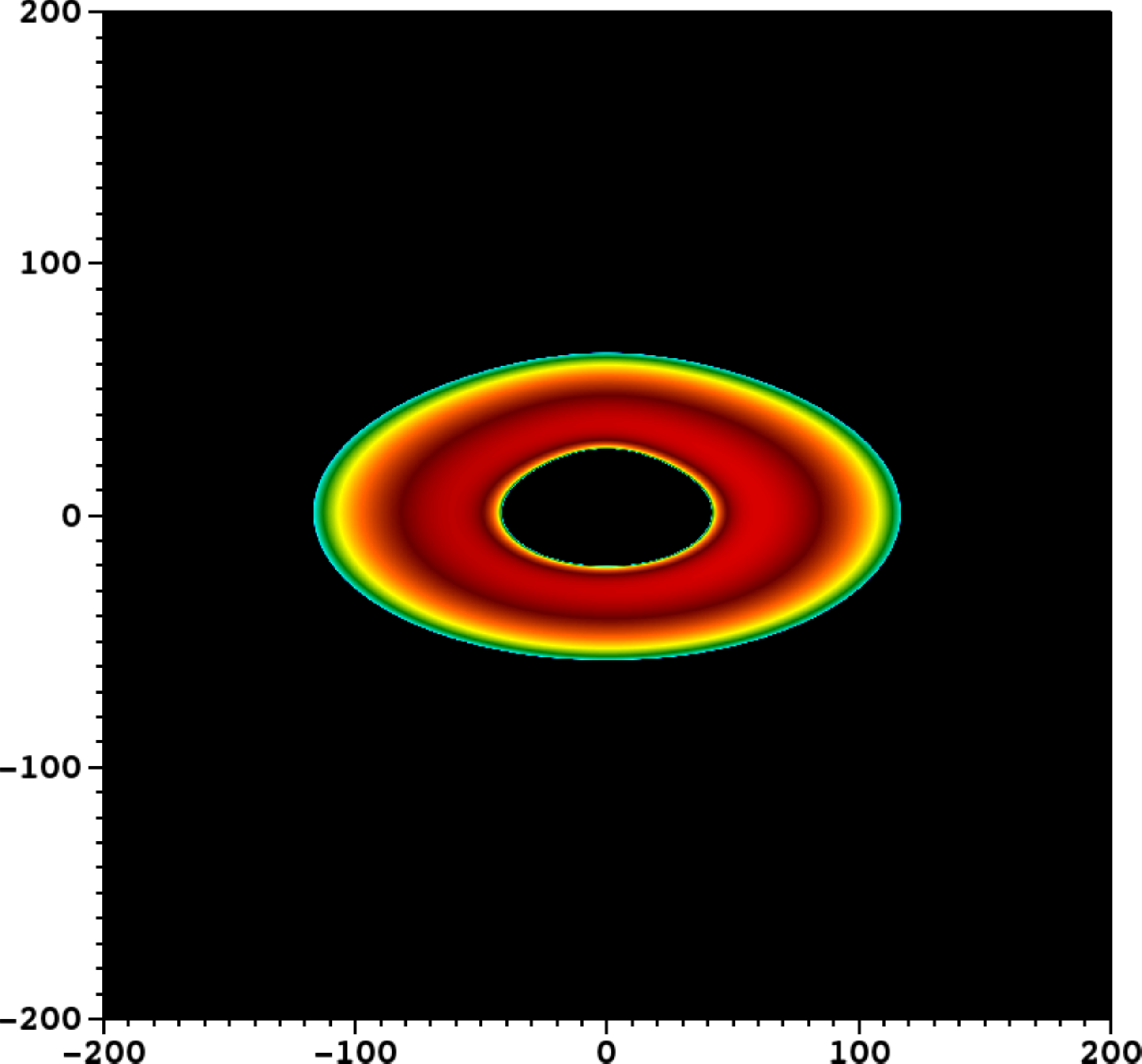} \vspace{4mm} \\
\includegraphics[width=0.275\hsize]{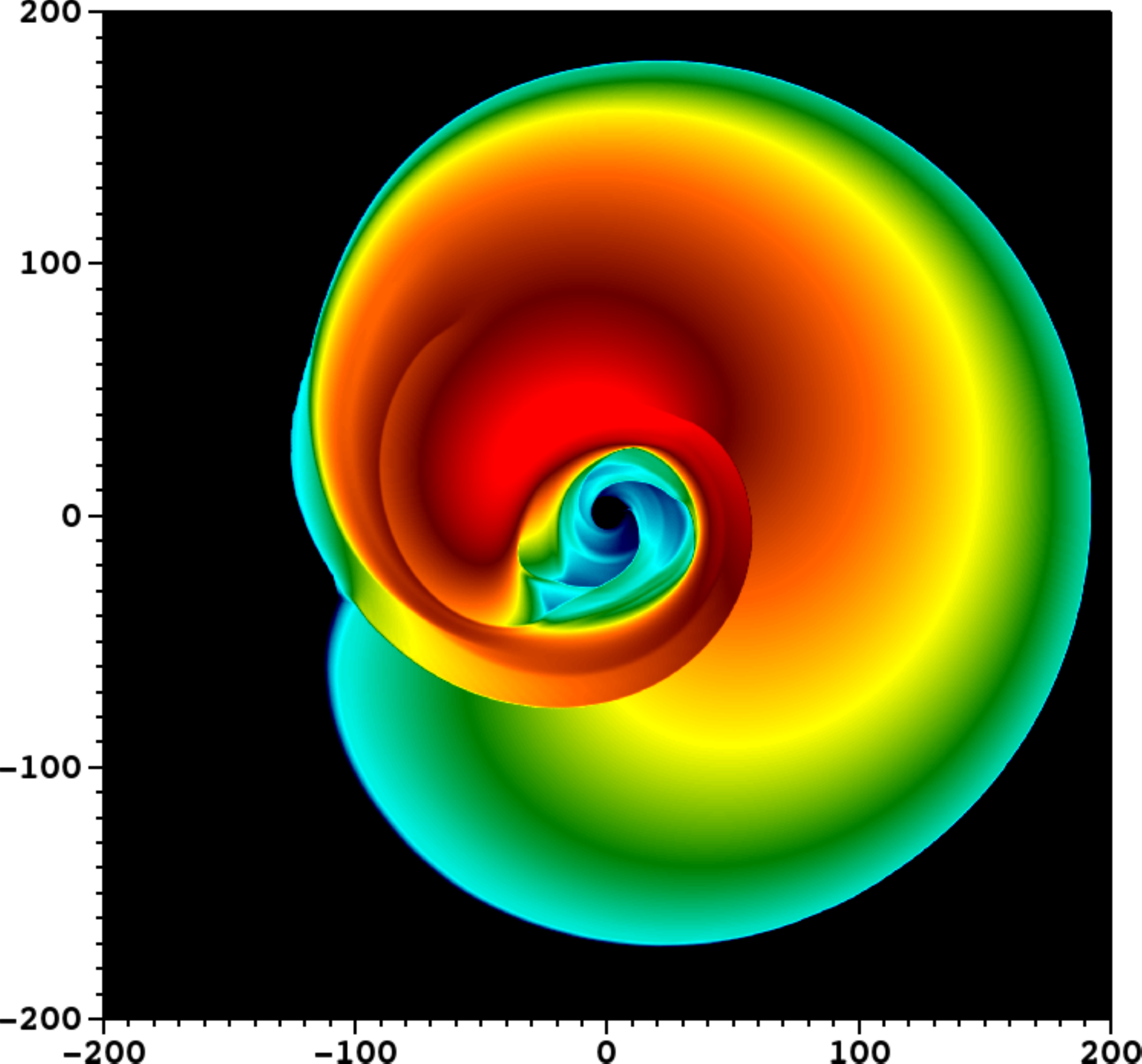} 
\includegraphics[width=0.275\hsize]{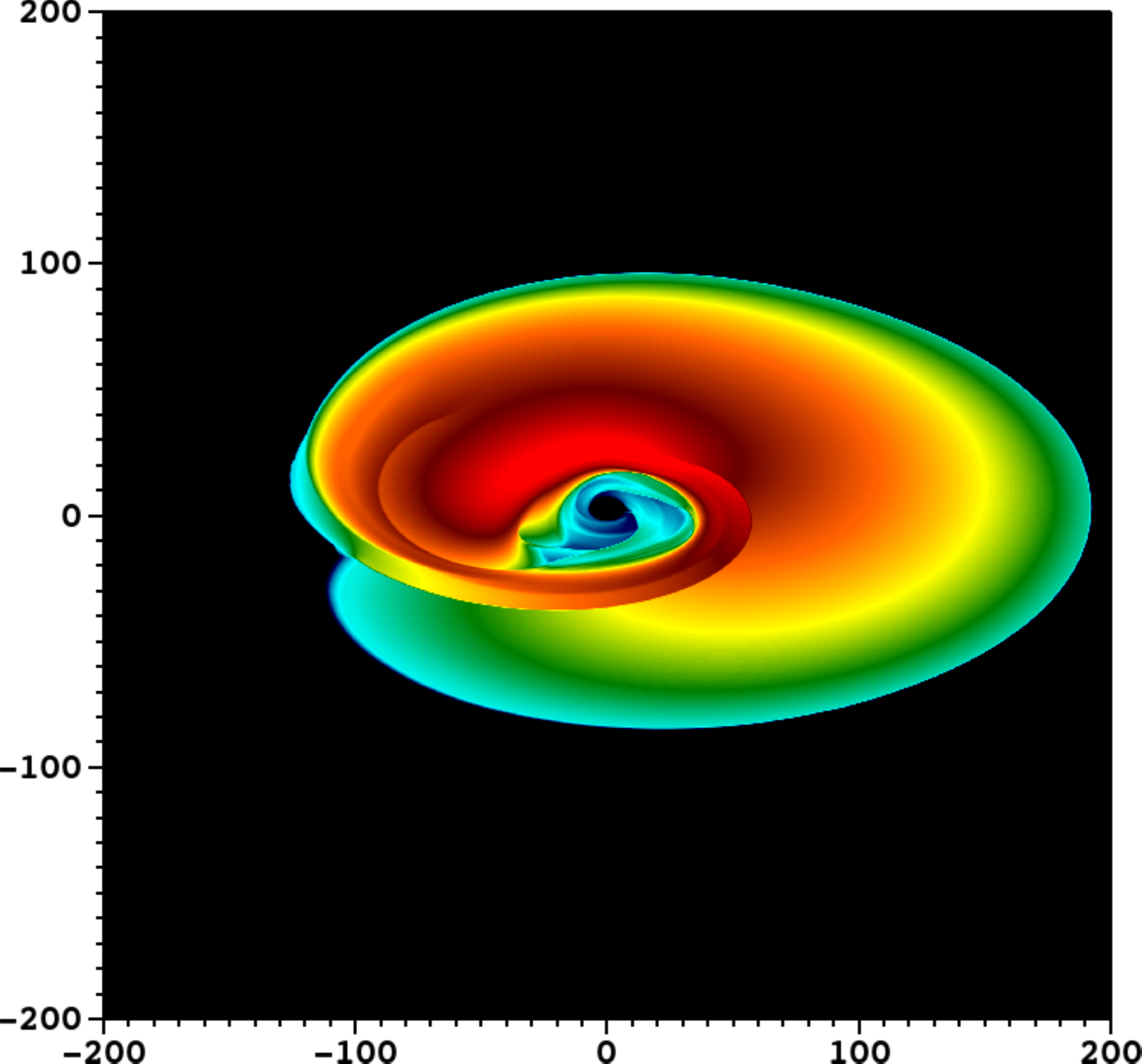} \vspace{4mm} \\
\includegraphics[width=0.275\hsize]{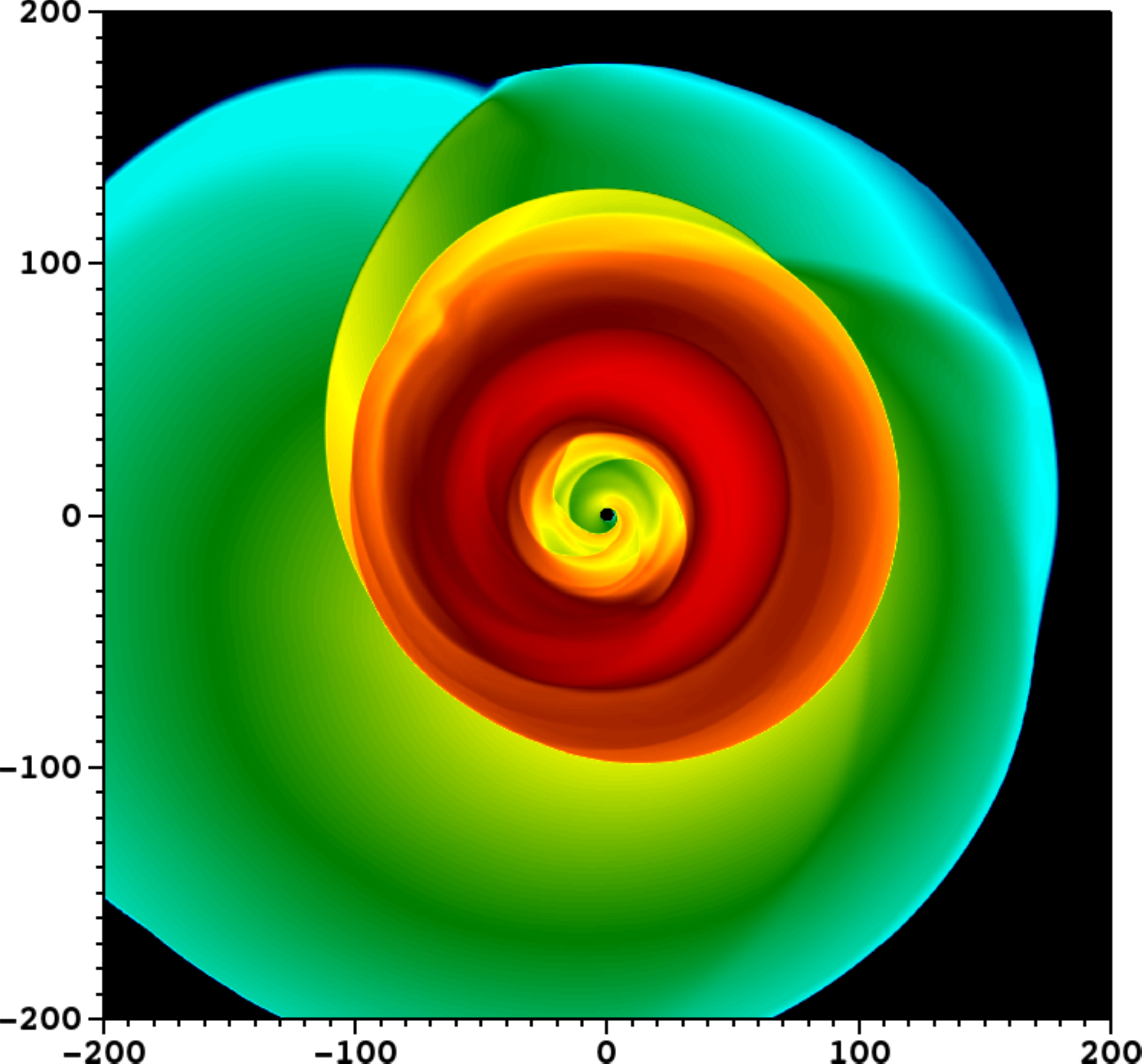}
\includegraphics[width=0.275\hsize]{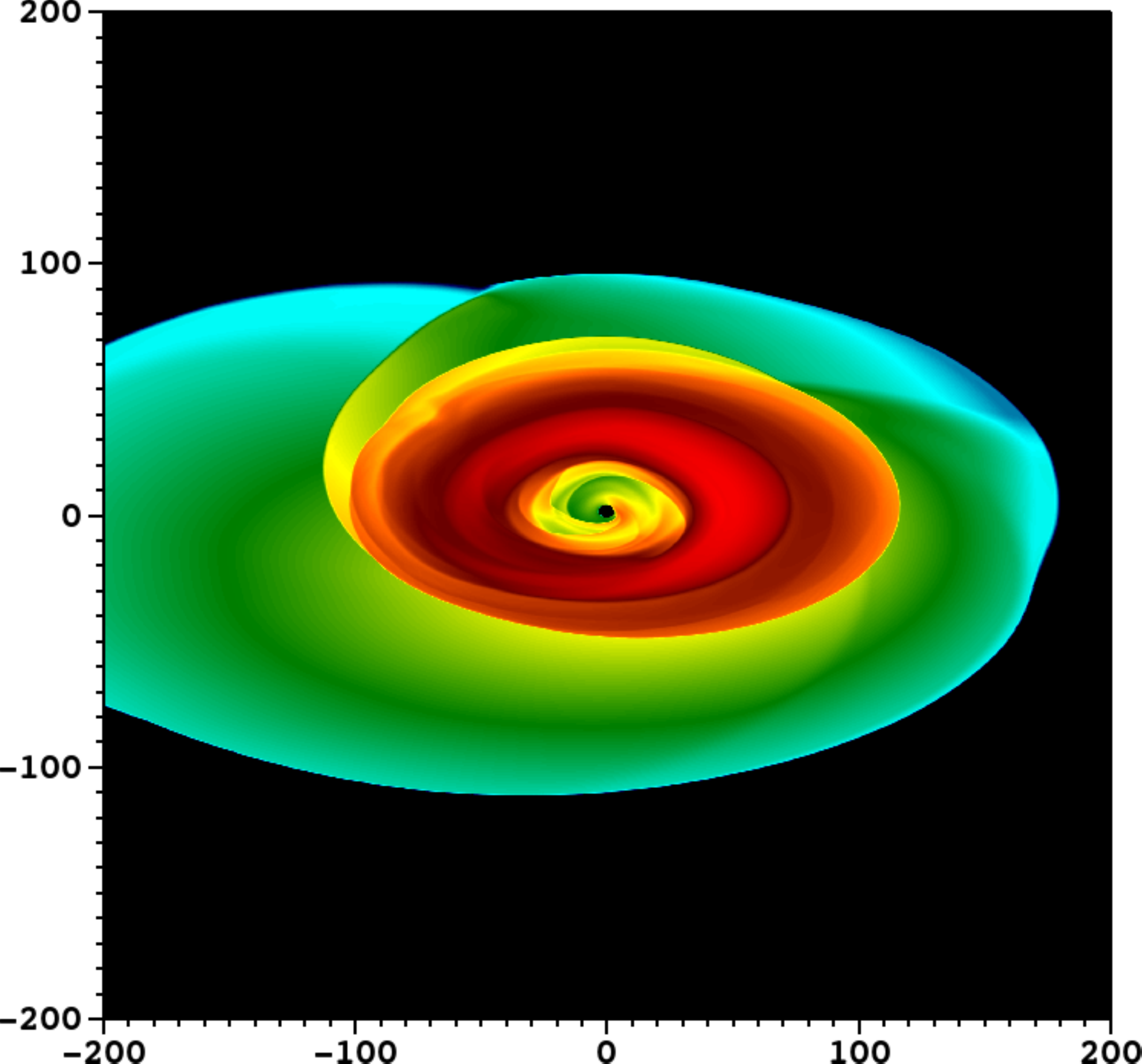} \vspace{4mm} \\
\includegraphics[width=0.275\hsize]{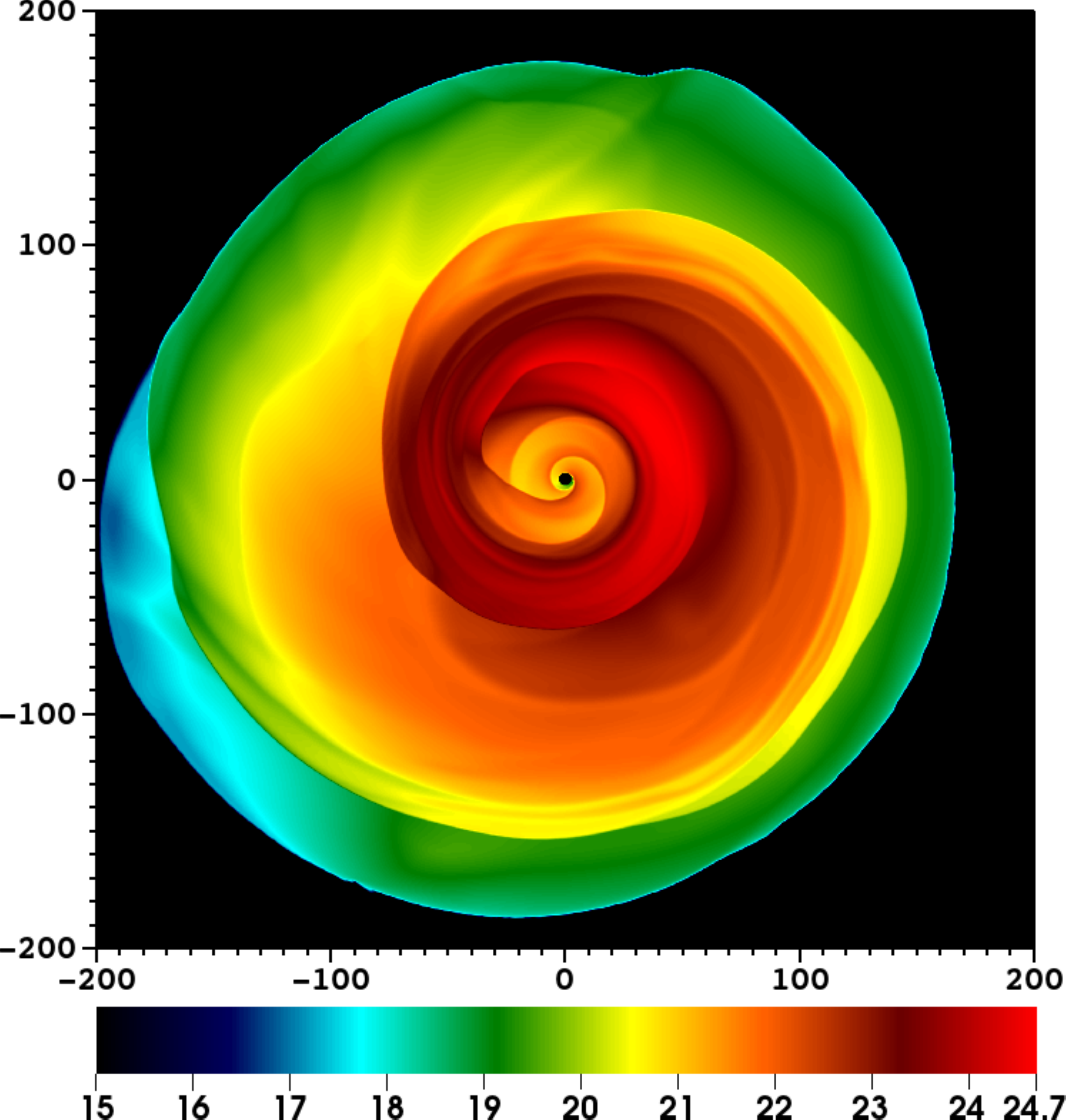}
\includegraphics[width=0.275\hsize]{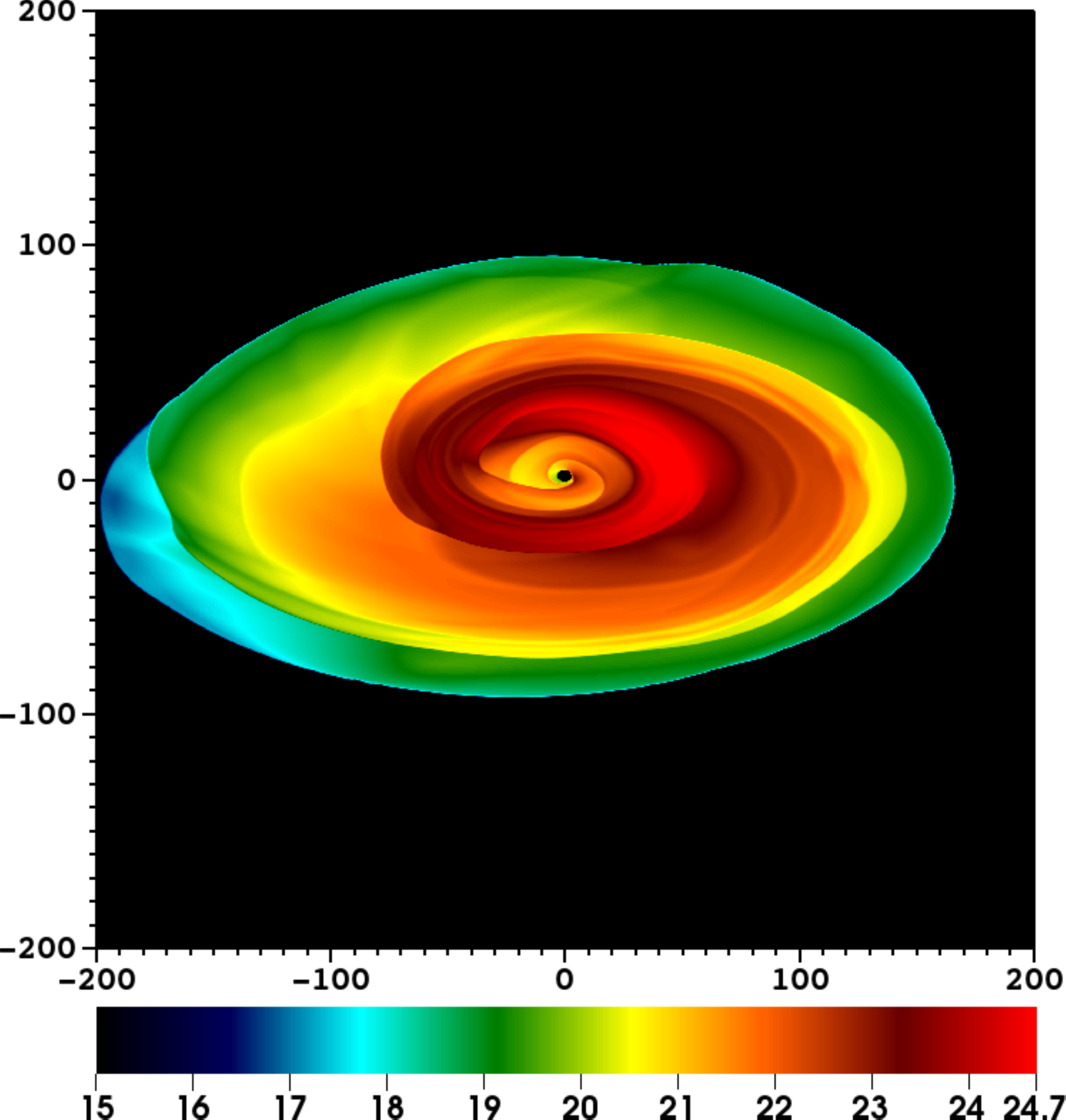}
\caption{Ray-tracing and radiative transfer calculation of 2D recoiling
  black-hole simulation. Left column: as viewed at an observer
  inclination angle of $\theta_{\mathrm{obs}}=0.1^{\circ}$. Right column:
  viewed at an observer inclination angle of $\theta_{\mathrm{obs}} =
  60^{\circ}$. From top row to bottom row, as viewed at $t=0$, $10,\!000$,
  $15,\!000,$ and $20,\!000\,M$ . The colour scale is logarithmic
  in the total intensity, $I$, from each pixel (arbitrary units).}
\label{fig:2D_ray}
\end{figure*}
%ffffffffffffffffffffffffffffffffffffffffffffffffffffffffffffffffff

%ssssssssssssssssssssssssssssssssssssssssssssssssssssssssssssssssss
\subsection{Radiative parameters}\label{sec:ray-radiative}
%ssssssssssssssssssssssssssssssssssssssssssssssssssssssssssssssssss

To calculate the electromagnetic emission from the recoiling black-hole
simulations, we assumed emission primarily in the form of thermal
bremsstrahlung from electron-ion and electron-electron interactions. Owing
to the relativistic equation of state used in these simulations and because the temperature can range between
$\sim10^{7}-10^{11}\,\mathrm{K}$, it is necessary to employ a
relativistic Maxwellian distribution for the population of thermal
electrons. Following \citet{Stepney1983} and \citet{Narayan1995}, the
total thermal bremsstrahlung cooling rate \citep[see
  also][]{Straub2012} may be written as
\begin{equation}
q_{\mathrm{br}}^{-} = q_{\mathrm{ei}}^{-} + q_{\mathrm{ee}}^{-} \,,
\end{equation}
where $q_{\mathrm{ei}}^{-}$ and $q_{\mathrm{ee}}^{-}$ are
the electron-ion and electron-electron cooling terms, respectively. The electron-ion
cooling rate is given by
\begin{equation}
q_{\mathrm{ei}}^{-} = 3.013\times
10^{25}\rho_{\mathrm{cgs}}^{2}F_{\mathrm{ei}}(\Theta_{\mathrm{e}})
\ \mathrm{erg} \ \mathrm{cm^{-3}} \ \mathrm{s^{-1}} \,,
\end{equation}
where the dimensionless electron temperature ($\Theta_{\mathrm{e}}$) is
defined as
\begin{eqnarray}
\Theta_{\mathrm{e}} \equiv \frac{k_{\mathrm{B}}T_{\mathrm{e}}}{m_{\mathrm{e}}c^{2}} \,,
\end{eqnarray}
and where $T_{\mathrm{e}}$ is the electron temperature and
$m_{\mathrm{e}}$ the electron rest mass. The function
$F_{\mathrm{ei}}(\Theta_{\mathrm{e}})$ is given by
\begin{equation}
F_{\mathrm{ei}}(\Theta_{\mathrm{e}}) = 
\begin{cases}
  4\left(\dfrac{2\Theta_{\mathrm{e}}}{\pi^{3}}\right)^{1/2} \left(1+
  1.781\Theta_{\mathrm{e}}^{1.34} \right) \,, &
  \text{$\Theta_{\mathrm{e}}<1$ \,,} \vspace*{2mm} \\
  \dfrac{9\Theta_{\mathrm{e}}}{2\pi} \left[1.5 + \mathrm{ln}\left(
    1.123\Theta_{\mathrm{e}}+0.48 \right)\right] \,, &
  \text{$\Theta_{\mathrm{e}}>1$ \,,}
\end{cases}
\end{equation}
where $F_{\mathrm{ei}}$ is continuous across $\Theta_{\mathrm{e}}=1$. The
electron-electron bremsstrahlung cooling term is given by
\begin{equation}
q_{\mathrm{ee}}^{-} = 
\begin{cases}
  \mathcal{C}_{1}\rho_{\mathrm{cgs}}^{2}\Theta_{\mathrm{e}}^{3/2}\left(1+1.1\Theta_{\mathrm{e}}+\Theta_{\mathrm{e}}^{2}-1.25\Theta_{\mathrm{e}}^{5/2}
  \right) \,, & \text{$\Theta_{\mathrm{e}}<1$ \,,} \vspace*{2mm} \\
  \mathcal{C}_{2}\rho_{\mathrm{cgs}}^{2}\Theta_{\mathrm{e}}\left(\mathrm{ln}1.123\Theta_{\mathrm{e}}+1.28
  \right) \,, & \text{$\Theta_{\mathrm{e}}>1$ \,,}
\end{cases}
\end{equation}
where $\mathcal{C}_{1}=7.028\times 10^{25}$ and
$\mathcal{C}_{2}=9.334\times 10^{25}$, and $q_{\mathrm{ee}}^{-}$ has
units of $\mathrm{erg} \ \mathrm{cm^{-3}} \ \mathrm{s^{-1}}$. With the
thermal bremsstrahlung cooling rates in hand, we may now write the total
emissivity in the fluid rest frame as
\begin{equation}
j_{\nu,\mathrm{br}} = \frac{1}{4\pi\nu}q_{\mathrm{br}}^{-} \ \! x \ \!
\mathrm{e}^{-x} \ \! \overline{g}(x) \,,
\end{equation}
where
\begin{equation}
x \equiv \frac{h_{\mathrm{P}}\nu}{k_{\mathrm{B}}T_{\mathrm{e}}} \,,
\end{equation}
and $h_{\mathrm{P}}$ is the Planck constant. The factor of $1/(4\pi)$
specifies isotropic emission in the fluid rest frame, and the mean Gaunt
factor, $\overline{g}(x)$, is given by
\begin{equation}
\overline{g}(x) \equiv 
\begin{cases}
  \dfrac{\sqrt{3}}{\pi}\ln\left(2.246 x^{-1}\right) \,, & \text{$x<1$}
  \,, \vspace*{2mm} \\
  \left( \dfrac{3}{\pi}x^{-1} \right)^{1/2} \,, & \text{$x>1$} \,.
\end{cases}
\end{equation}
In all calculations reported here, we assumed that the ionic contribution comes exclusively from protons. The simulation data provide only the
equilibrium temperature of electrons and protons, and not of individual
species, therefore we take $T_{\mathrm{e}}=T_{\mathrm{cgs}}$, assuming that the
local electron and proton temperatures do not differ significantly from
the local equilibrium temperature.

In 3D models where we considered the opacity of the emitting medium, we
assumed a modified Kramer opacity law as employed in
\cite{Schnittman2006} and \cite{Andersonetal10}, where
\begin{equation}
\alpha_{0,\nu} = 5\times 10^{24} \rho_{\mathrm{cgs}}^{2}T^{-7/2} \left(
\frac{1-\mathrm{e}^{-x}}{x^{3}} \right) \ \mathrm{cm}^{-1} \,.
\end{equation}
This opacity adds thermal radiation for optically thick regions, whilst
the optically thin regions radiate bremsstrahlung
\citep[see][]{Andersonetal10}.

%ssssssssssssssssssssssssssssssssssssssssssssssssssssssssssssssssss
\subsection{Recoiling black hole in 2D}\label{sec:ray-recoil2D}
%ssssssssssssssssssssssssssssssssssssssssssssssssssssssssssssssssss

The dynamics of the 2D recoiling black hole is ultimately that of a
planar flow in the equatorial ($\theta=\pi/2$) plane of the black
hole. When we ray-trace these simulations, we need only calculate the
intersection point of the ray with the equatorial plane, determining the
emitted spectrum at that particular pixel of the image. Since the ray
does not traverse the emitting medium, the emission is optically thick
and planar, thus the calculated results are scale-free and do not depend
on the mass of the black hole. To generate each image, we ray-traced a
grid of $2,\!500 \times 2,\!500$ photons, sampling 200 uniformly
logarithmically spaced frequency bins between $10^{5}\,{\rm Hz}$ and
$10^{25}\,{\rm Hz}$. Each pixel of the calculated images represents the
total frequency-integrated emission.

We also present calculations of light curves for different inclination
parameters, where the total integrated intensity over all frequencies and
over every pixel in an image (\ie flux) corresponds to that point in time
on the light curve. We recall that the intensity is the energy
  received per unit time, and we also refer to it as the ``flux'', which
  should not be confused with the ``fluxes'' introduced in
  Eq. \eqref{Eq:Flux}, however. 
  Since we stored the
entire spectrum for each pixel, we can also calculate the image and
light curve at specific observer frequencies, which is of practical
interest when comparing images from radiation transport calculations of
GRMHD simulation data with observations of Sagittarius A* (Sgr A*) at
$1.3\,{\rm mm,         }$ for instance.

%ffffffffffffffffffffffffffffffffffffffffffffffffffffffffffffffffff
\begin{figure}
\centering
\includegraphics[width=\hsize]{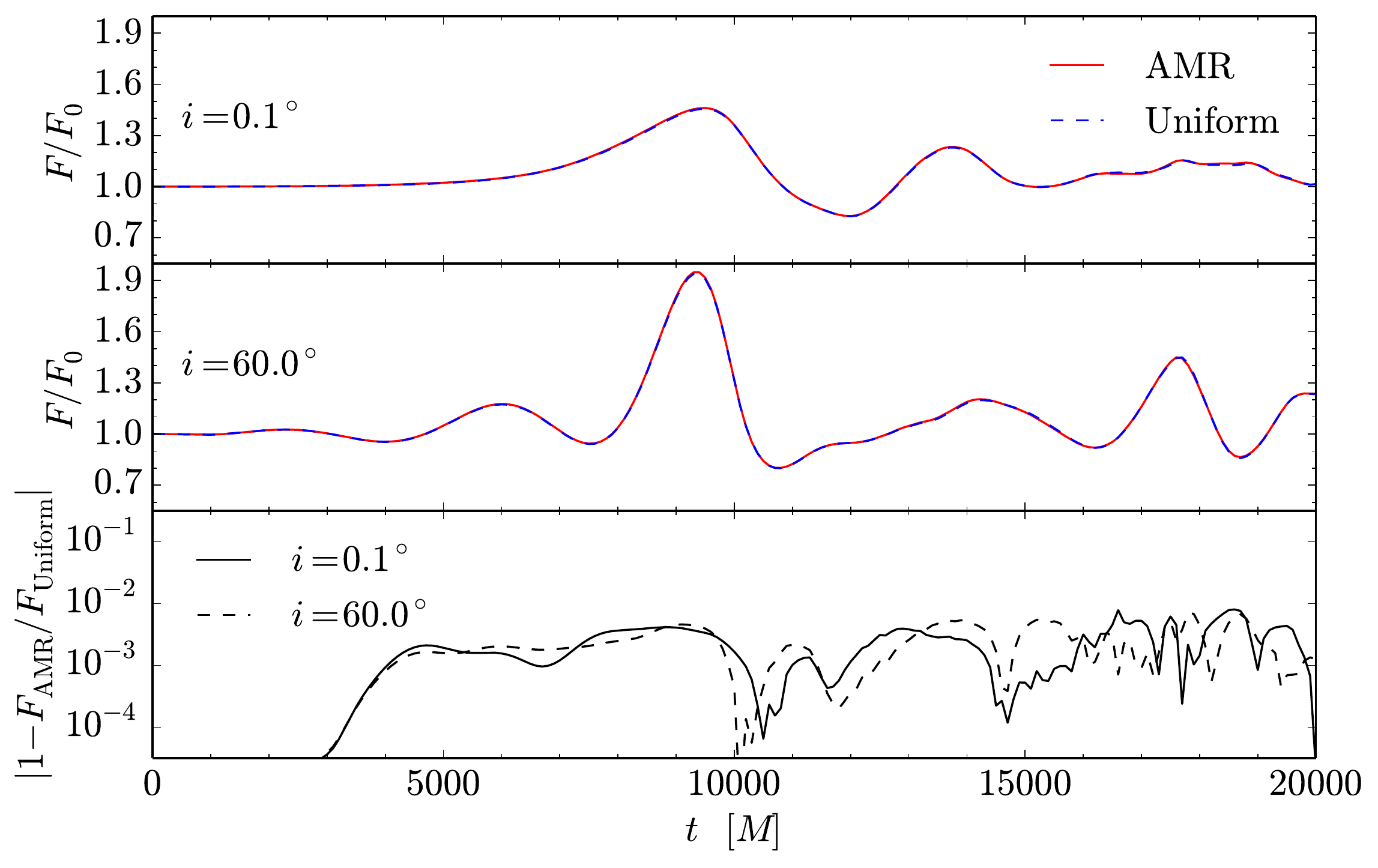}
\caption{Top and middle panels: normalised total flux light curves
  of the
  emission from the 2D recoiling black hole simulation for
  $\theta_{\mathrm{obs}} = 0.1^{\circ}$ and $\theta_{\mathrm{obs}} =
  60^{\circ}$ , respectively. Solid lines are for the simulation run with three
  AMR levels and a tolerance of $\varepsilon_t=0.005$. Dashed lines are for
  the uniform grid simulation run of equivalent resolution. Bottom: flux
  difference between the 2D AMR and uniform run light-curves for
  $\theta_{\mathrm{obs}}=0.1^{\circ}$ (solid) and
  $\theta_{\mathrm{obs}}=60^{\circ}$ (dashed).}
\label{fig:2D_lightcurve}
\end{figure}
%ffffffffffffffffffffffffffffffffffffffffffffffffffffffffffffffffff

Figure \ref{fig:2D_ray} presents radiation image calculations of the 2D
recoiling black hole simulation. For an inclination angle of
$\theta_{\mathrm{obs}}=0.1^{\circ}$, the structure of the flow is similar
to the renderings in Fig. \ref{fig:2Drecoil-rho2D}. One obvious
difference is the absence of emission from the innermost region in the
vicinity of the black hole event horizon. When instead
$\theta_{\mathrm{obs}}=60^{\circ}$, the image of the 2D recoiling black
hole is warped and has a smaller projected surface area, with the
approaching side of the flow being Doppler-boosted and projected along
the line of sight, and conversely for the receding side.

%ffffffffffffffffffffffffffffffffffffffffffffffffffffffffffffffffff
\begin{figure*}
\centering
\includegraphics[width=0.275\hsize]{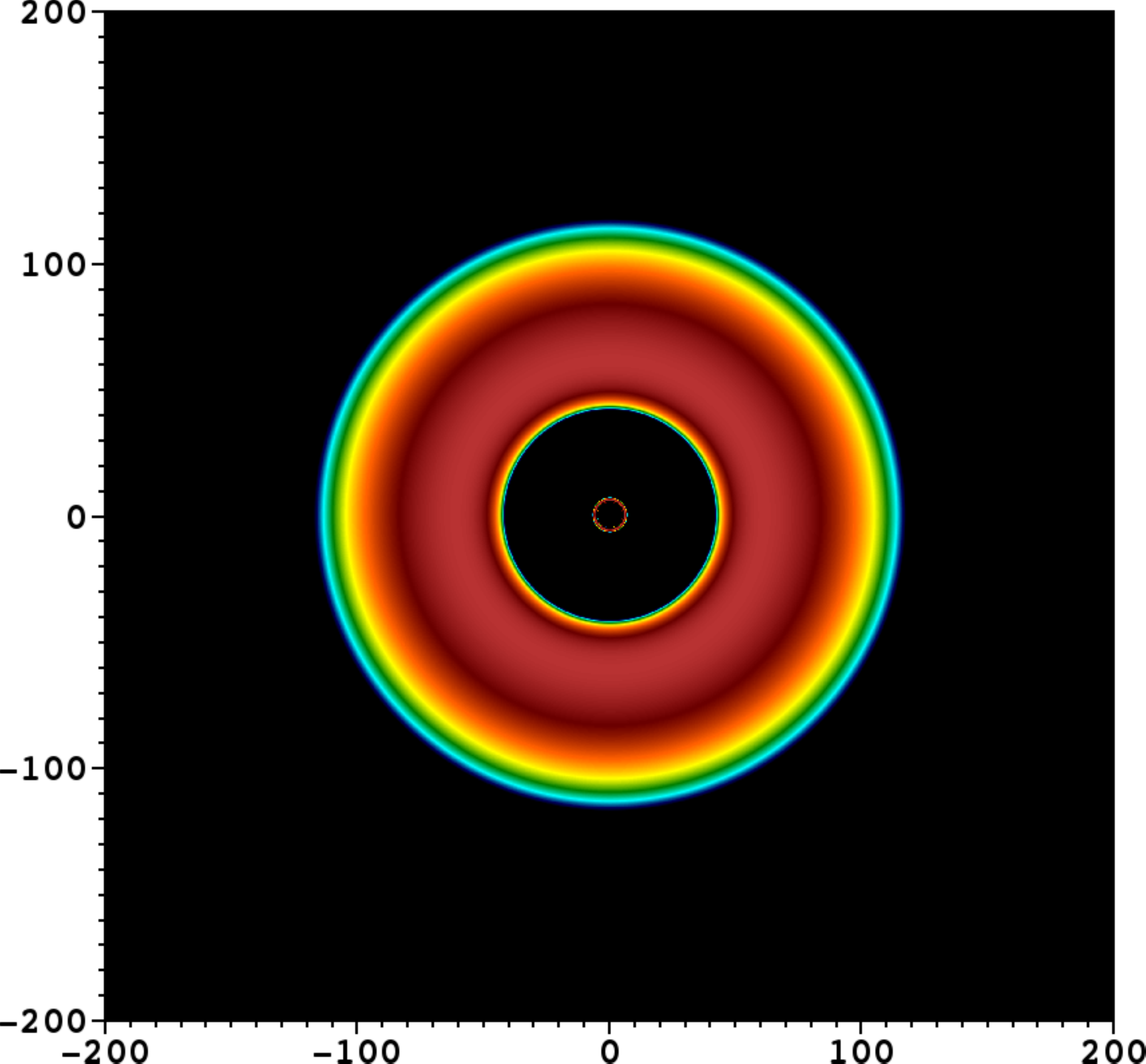}
\includegraphics[width=0.275\hsize]{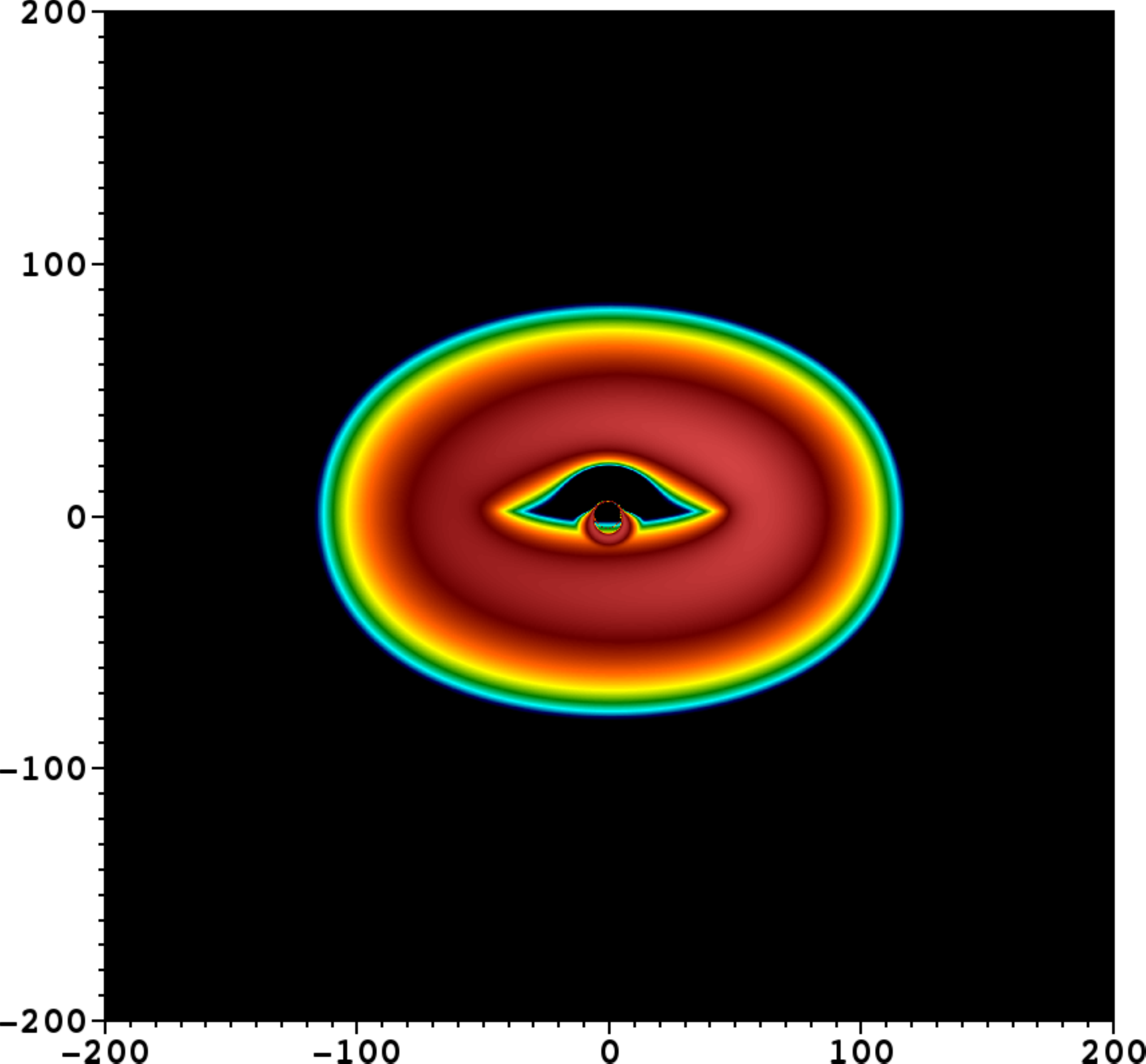} \vspace{4mm} \\
\includegraphics[width=0.275\hsize]{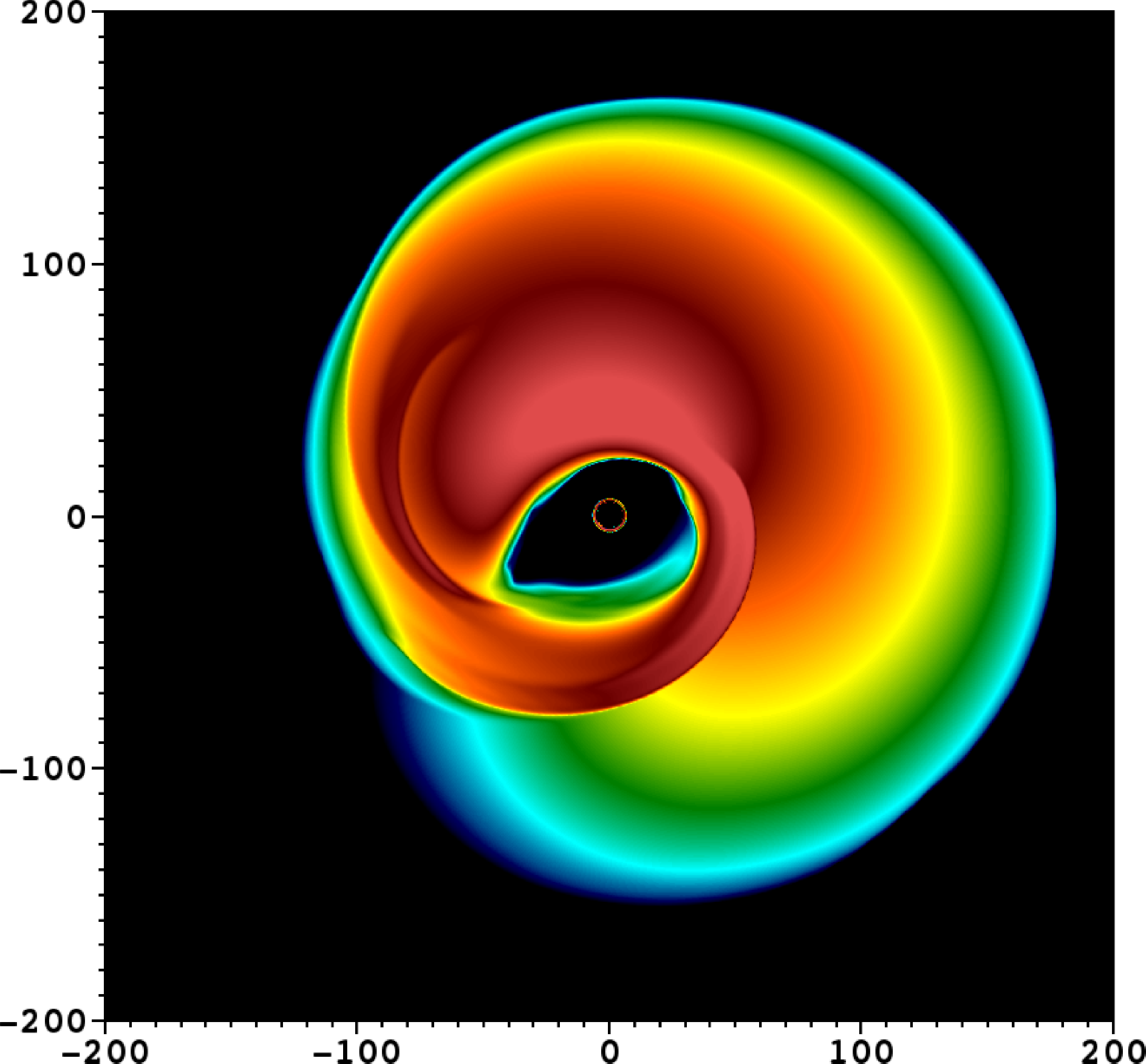} 
\includegraphics[width=0.275\hsize]{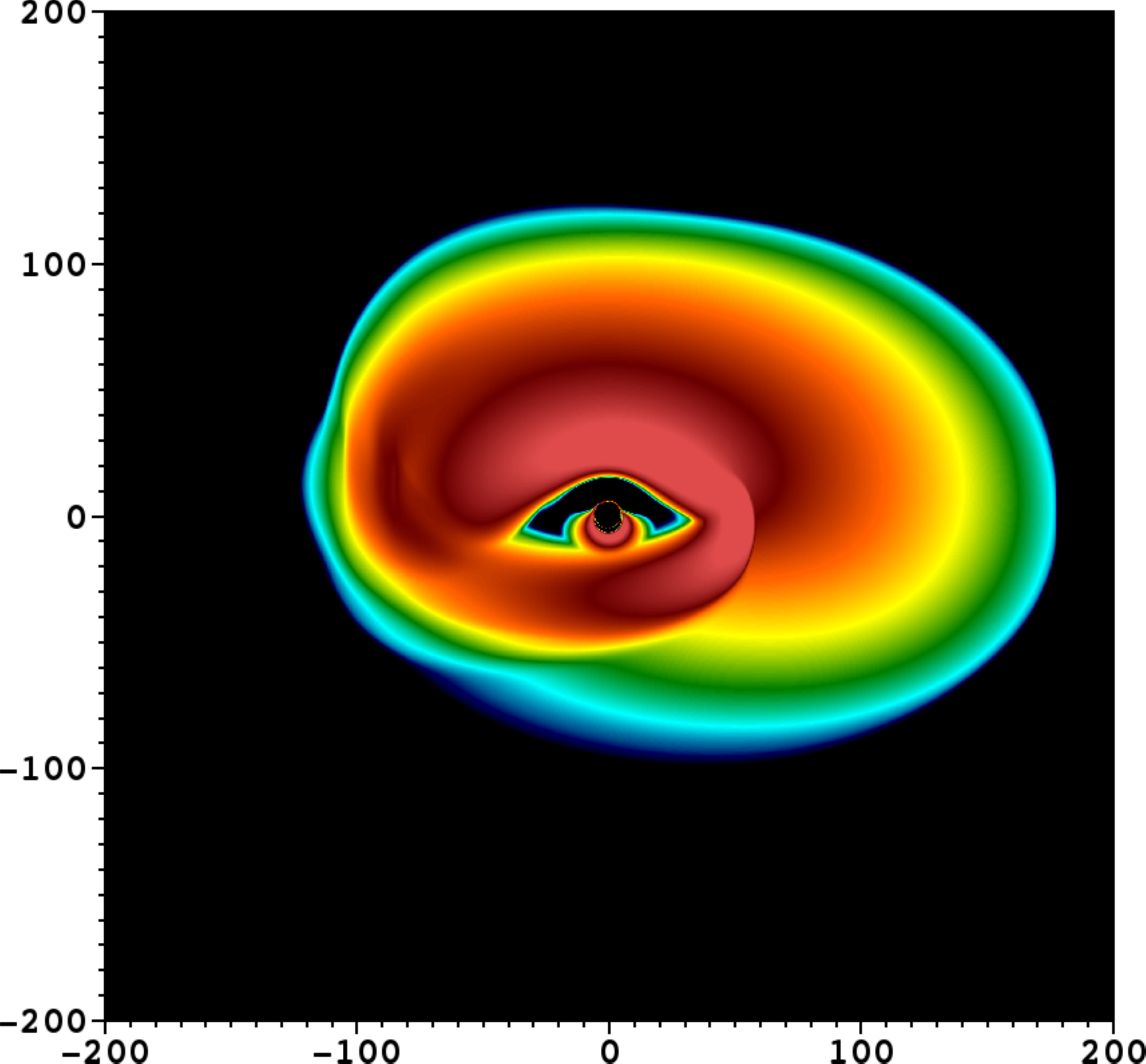} \vspace{4mm} \\
\includegraphics[width=0.275\hsize]{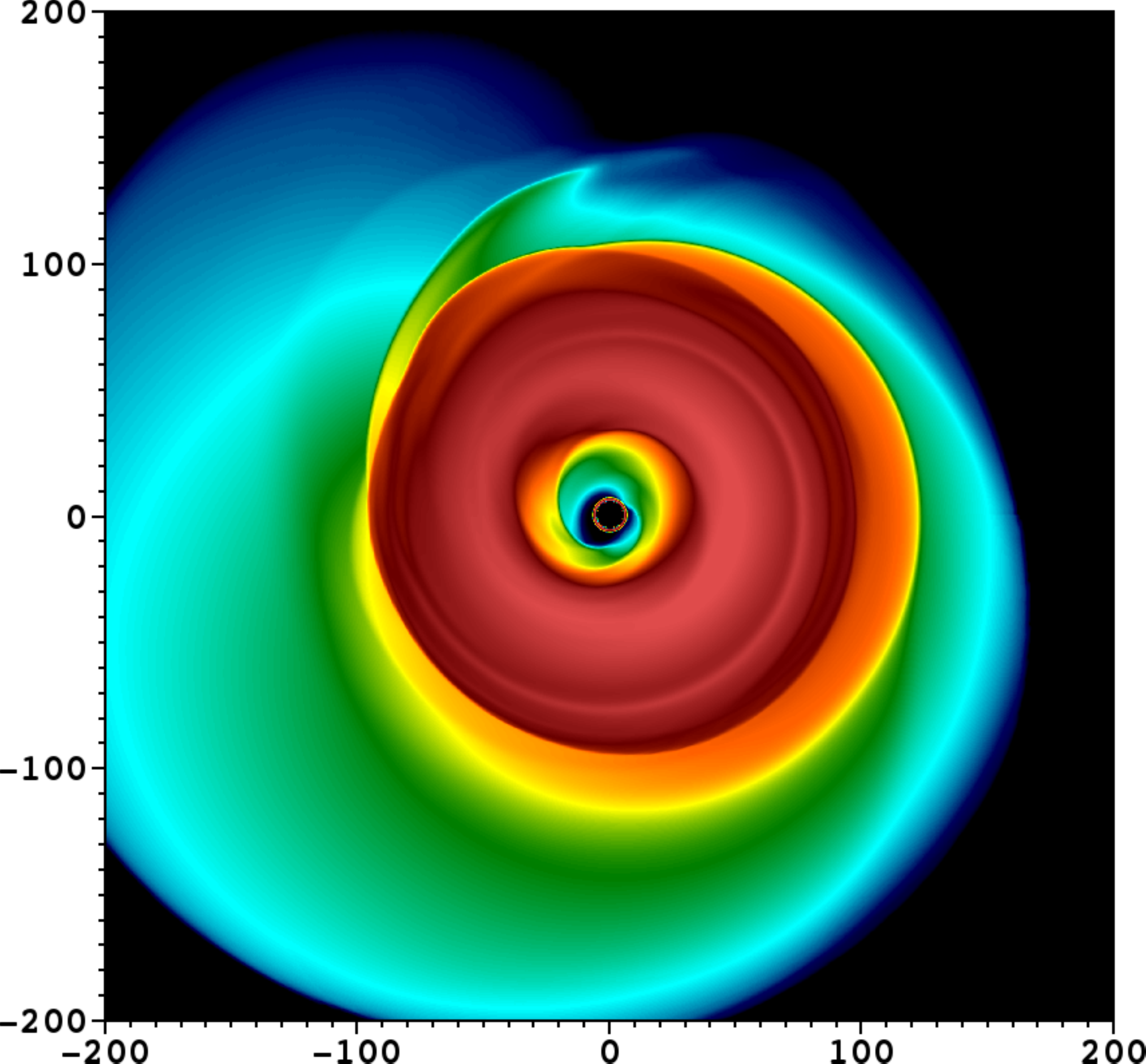}
\includegraphics[width=0.275\hsize]{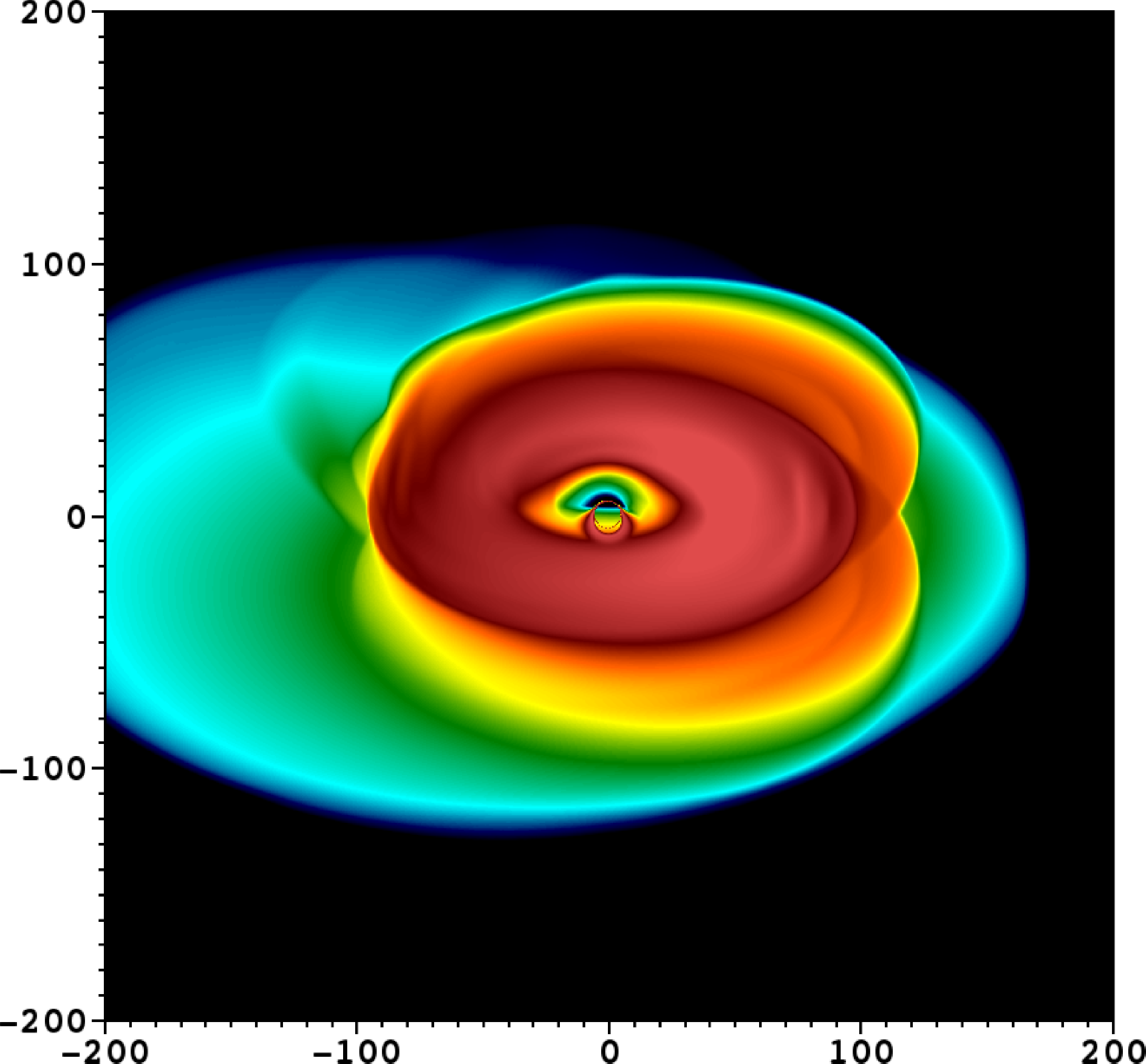} \vspace{4mm} \\
\includegraphics[width=0.275\hsize]{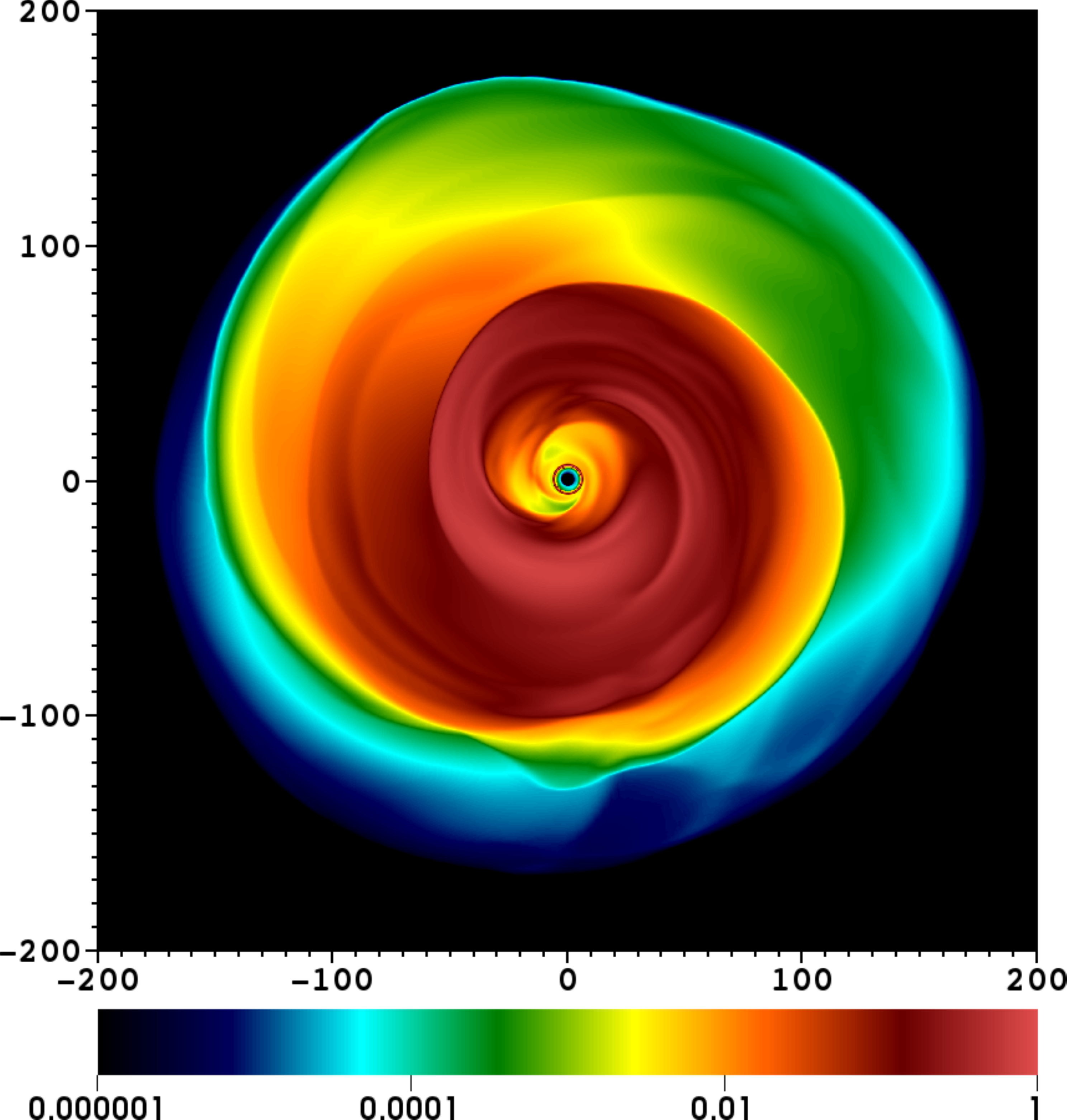}
\includegraphics[width=0.275\hsize]{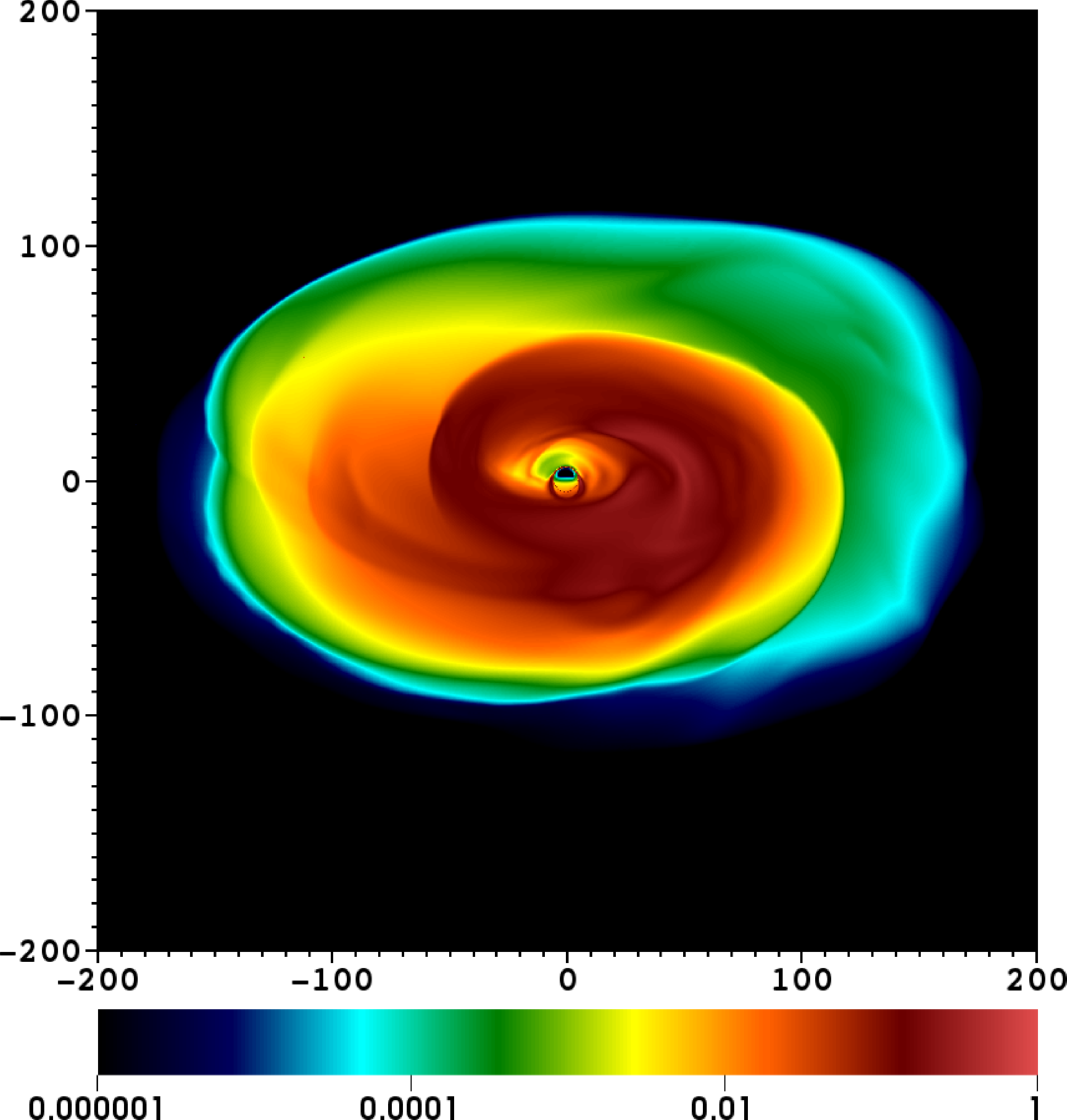}
\caption{Ray-tracing and radiative-transfer calculation of the
3D recoiling
  black hole simulation. Same panel descriptions as Fig.
  \ref{fig:2D_ray}, but now the colour scale is logarithmic in
  $I/I_{\mathrm{max}}$, \ie normalised to the peak intensity.}
\label{fig:3D_ray}
\end{figure*}
%ffffffffffffffffffffffffffffffffffffffffffffffffffffffffffffffffff

In Fig. \ref{fig:2D_lightcurve} we present light-curve calculations of
the AMR and uniform grid runs of the same 2D recoiling black hole
simulation. As in Fig. \ref{fig:2D_ray}, we considered two observer
inclination angles and calculated the light curves from the two sets of
simulation data. The AMR and uniform grid runs are in excellent
agreement, indicating that the AMR simulation captures both the
qualitative and quantitative aspects of the dynamics and thermodynamics
well, which is reflected in the light curves. The bottom panel reveals
that while the differences between the two runs vary, they always remain
below the 1\% level. For an observer at
$\theta_{\mathrm{obs}}=0.1^{\circ}$, the oscillations are almost
exclusively due to the growth and propagation of the large spiral shock,
since Doppler and aberrational effects are essentially uniform across the
entire image at such low inclinations. For an inclination
$\theta_{\mathrm{obs}} = 60^{\circ}$, the flux is lower (since the
projected surface area is smaller) and more oscillations of the
light-curve can be seen. These are due to the additional presence of
non-uniform Doppler boosting of the emission from the spiral-shocked
material as it approaches the observer (peaks) and as it moves away from
the observer (troughs).

%ffffffffffffffffffffffffffffffffffffffffffffffffffffffffffffffffff
\begin{figure}
\centering
\includegraphics[width=\hsize]{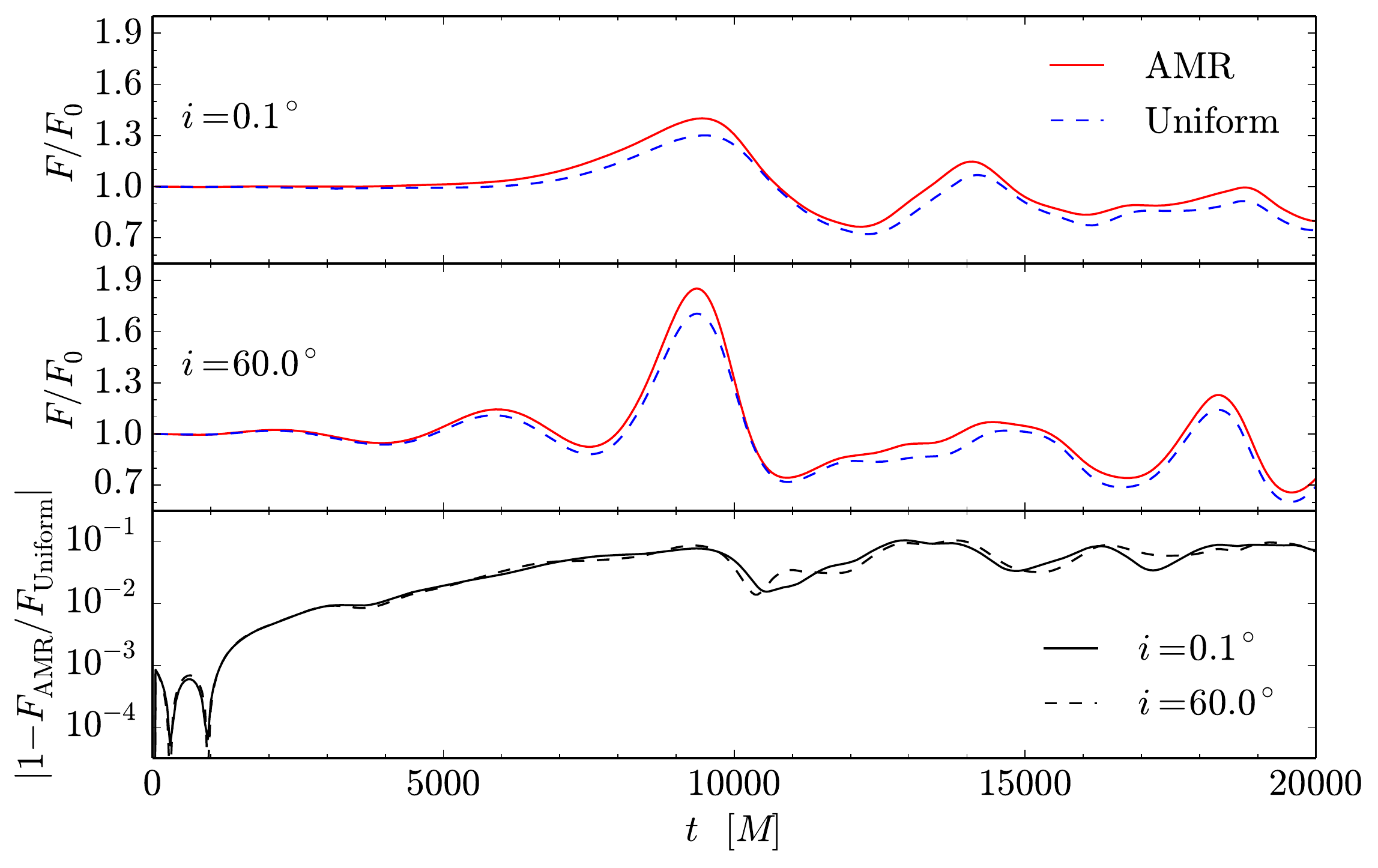}
\caption{As in Fig. \ref{fig:2D_lightcurve}, but now considering the 3D
  recoiling black hole simulation. The higher AMR level tolerance for the 3D
  run ($\varepsilon_t=0.1,0.3$) results in a less perfect agreement with the
  uniform-resolution run than for 2D ($\varepsilon_t=0.005$,
  Fig. \ref{fig:2D_lightcurve}), as is discussed in
  Sec. \ref{sec:recoilbhsetup3d}.}
\label{fig:3D_lightcurve}
\end{figure}
%ffffffffffffffffffffffffffffffffffffffffffffffffffffffffffffffffff

%ssssssssssssssssssssssssssssssssssssssssssssssssssssssssssssssssss
\subsection{Recoiling black hole in 3D}\label{sec:ray-recoil3D}
%ssssssssssssssssssssssssssssssssssssssssssssssssssssssssssssssssss

In analogy with \citet{Andersonetal10} and \citet{Zanotti2010}, we set
the initial rest-mass density at the centre of the torus to be
$\rho_{\mathrm{c}} = 1.38\times
10^{-10}~\mathrm{g}~\mathrm{cm}^{-3}$. The image parameters are identical
to the 2D case. Figure \ref{fig:3D_ray} presents radiation image
calculations of the 3D recoiling black hole simulation for the AMR
run. For $\theta_{\mathrm{obs}}=0.1^{\circ}$ the panels in
Figs. \ref{fig:2D_ray} and \ref{fig:3D_ray} appear similar, but there are
several differences.

The first difference is the near-absence of accretion at early times in
3D, occurring (much more slowly) at later times than in the 2D case.
The second difference is the lensed
emission from the torus, which manifests itself as the inner ring of
emission and is most distinct at $t=0$. This is due to rays that
traverse the entire 3D emitting medium multiple times before reaching the
observer. For an observer at $\theta_{\mathrm{obs}}=60^{\circ}$ , the front
limb of the torus obscures the central region of the black hole and
opacity effects dominate, giving rise to much stronger emission from this
region than in either the $\theta_{\mathrm{obs}}=0.1^{\circ}$ case in
3D or all viewing angles in the 2D case (where self-obscuration is
absent). A third major difference is that, particularly at late times,
shocked regions and the emission from near the vicinity of the event
horizon appear more optically thick because the flow is 3D and
absorptive as well as emissive.

In Fig. \ref{fig:3D_lightcurve} we present light-curve calculations of
the AMR and uniform grid runs of the same 3D recoiling black hole
simulation. Unlike the 2D case that we investigated previously, the tolerance
in this simulation was higher, \ie $\varepsilon_t=0.1$
(see discussion in Sect. \ref{sec:recoilbhsetup3d}). As expected,
the higher tolerance causes larger differences between the uniform and
AMR runs, the maximum difference being $\sim 9.8\%$. However, the
light curves still retain the same morphological profiles and relative
properties, with the AMR runs slightly overestimating the flux relative
to the uniform run. These differences are acceptable and do not change
the physical conclusions drawn from these calculations. Moreover, considering that
the difference in runtime between the uniform and AMR simulations was a
factor of $\sim 7$, this represents a significant speedup, allowing
us to
employ still higher resolutions.

%ssssssssssssssssssssssssssssssssssssssssssssssssssssssssssssssssss
\section{Conclusions}\label{sec:conclusions}
%ssssssssssssssssssssssssssssssssssssssssssssssssssssssssssssssssss

We have discussed results from a new 3D general-relativistic
hydrodynamics code with grid-based AMR capabilities, the motivation for
which arose mainly from our own continued efforts in augmenting the
wealth of community codes available for astrophysical research.

The code was tested in the general-relativistic regime by evolving a
number of stationary and non-stationary flows onto black hole space-times,
including the spherical (Michel) accretion onto a Schwarzschild black
hole and stationary tori with a constant angular momentum in a rotating
black hole space-time using Boyer-Lindquist and Kerr-Schild
coordinates. We further demonstrated that the code can be properly
employed in the consideration of other scientific applications.

A particularly critical test performed has involved the evolution
in 2D and in 3D of a black hole recoiling into a circumbinary accretion
disc, where both the nonlinearity of the dynamics and the development of
strong large-scale shocks have been tested, making use of the capabilities
of AMR. In particular, we have shown that AMR is essential for recoiling
black hole simulations because the dynamics of the kicked accretion disc
are very sensitive to the numerical resolution, and AMR has proven
effective in capturing and resolving the spiral shock structure that
develops in the accretion disc. AMR has also been shown to be very
economical, requiring only half of the computational grid and time
compared to the high-resolution case without AMR and still yielding
virtually unchanged results. 

Our relativistic hydrodynamics calculations have also been coupled to a
consistent treatment of the general-relativistic radiation-transport
equation to compute the electromagnetic emissions from the underlying
dynamics of the flow. The radiative-emission calculations were
performed in post-processing and combined with ray-tracing techniques to
obtain a somewhat realistic representation of the
electromagnetic emission from this process for the first time. 

In summary, the work presented here lays the ground for the development
of a generic computational infrastructure to accurately and
self-consistently calculate accretion flows onto compact objects, either black
holes or neutron stars, and to compute with an increased degree of
precision the associated electromagnetic emission from these scenarios. This could have a direct effect on collaborative efforts such as the
Event Horizon Telescope
Collaboration\footnote{\url{http://www.eventhorizontelescope.org/.}}
\citep{Doeleman2009a} or the Black Hole Camera
project\footnote{\url{http://www.blackholecam.org/.}} \citep{Goddi2016}.
Work is already ongoing to include the effects of magnetic fields in
the ideal-magnetohydrodynamics limit and will be presented in a
forthcoming publication.

\begin{acknowledgements}
     It is a pleasure to thank M. De Laurentis and C. Fromm for
     discussion and comments. This research is supported by the ERC
     Synergy Grant ``BlackHoleCam -- Imaging the Event Horizon of Black
     Holes'' (Grant 610058).  ZY is supported by an Alexander von
     Humboldt Fellowship. HO gratefully acknowledges the support from a 
     CONACYT-DAAD scholarship.
     The simulations were performed on LOEWE at the CSC-Frankfurt.

\end{acknowledgements}

\bibliographystyle{aa}
\bibliography{aeireferences}

%aaaaaaaaaaaaaaaaaaaaaaaaaaaaaaaaaaaaaaaaaaaaaaaaaaaaaaaaaaaaaaaaaa
\begin{appendix}
%ssssssssssssssssssssssssssssssssssssssssssssssssssssssssssssssssss
\section{Convergence tests}\label{sec:convergence}
%ssssssssssssssssssssssssssssssssssssssssssssssssssssssssssssssssss

We have measured the order convergence of the code by studying the norms
of the ``errors''. Since we employed a finite-volume scheme, the values of
the conserved variables $\boldsymbol{u}_h$ at each cell are spatial
volume averages of the numerical solution computed at a resolution with
cell width $h$. When an exact solution $\boldsymbol{{u}}$ exists to the
problem at hand, a natural way to quantify the errors is by comparing the
computed value of a quantity at each cell with the volume average of the
exact solution in the same cell $\boldsymbol{\bar{u}}$. Formally, the
$L_1$ norm of the error $\epsilon_h$ for a resolution $h$ is computed as
\begin{eqnarray}
\label{L1_norm}
||\epsilon_h||_{1} &=& ||\bar{\boldsymbol{u}}_h - \bar{\boldsymbol{u}}||_{1}
\\
&=&\sum_{i,j,k}\left|\bar{\boldsymbol{u}}_h -
\frac{1}{V_{i,j,k}}\int_{V_{i,j,k}}\boldsymbol{u}
\sqrt{\gamma}dx^1dx^2dx^3\right|\\
&=& \sum_{i,j,k}\left|\bar{\boldsymbol{u}}_h - \bar{\boldsymbol{u}}\right|,\end{eqnarray}
where $V_{i,j,k}$ is the proper volume of the cell $i,j,k$, and the sum
is performed on all the cells, except those containing the atmosphere, as they are not genuine solutions of the system of partial differential
equations. Equivalent relations can be given for the $L_{\infty}$ norm
that keeps track of the maximum error in the domain. The numerical order
of convergence of the simulation is then calculated as \citep[see,
  \eg][]{Rezzolla_book:2013}
\begin{equation}
\label{p_num}
\tilde{p} = \log\left(
\frac{||\epsilon_h||}{||\epsilon_k||}\right)/\log(h/k) \,.
\end{equation}
If the code is convergent, $\tilde{p}$ must be close to the nominal
order of accuracy $p$ of the numerical method, which is here $2$ for all
cases.  To simplify the calculations even more, we adopted the refinement factor
$h/k=2$ in this work. In the case of the Michel accretion accretion
problem or of a stationary torus, the exact solution is known, so that
the relations above at two resolutions can be employed to calculate the
convergence order at any time during the evolution.

On the other hand, in the far more common case in which an exact solution
is not known, as is the case for the simulations of recoiling black
holes, a self-convergence needs to be performed. This requires three
different estimates of the errors and the cancellation of the
higher-order terms, so that Eq. \eqref{p_num} becomes \citep[see,
  \eg][]{Rezzolla_book:2013}

\begin{equation}
\label{self_conv}
\tilde{p} = \log{\frac{||\bar{\boldsymbol{u}}_1 -
    \bar{\boldsymbol{u}}_2||}{||\bar{\boldsymbol{u}}_2 - 
\bar{\boldsymbol{u}}_3||}}/\log{2}\,
,\end{equation}
where
\begin{equation}
\boldsymbol{\bar{u}}_{2,i,j,k} =
\frac{1}{V_{i,j,k}}\int_{i,j,k}\boldsymbol{u}_2
\sqrt{\gamma}dx^1dx^2dx^3\,, 
\end{equation}
and
\begin{equation}
\boldsymbol{\bar{u}}_{3,i,j,k} =
\frac{1}{V_{i,j,k}}\int_{i,j,k}\boldsymbol{u}_3
\sqrt{\gamma}dx^1dx^2dx^3 \,.
\end{equation}
We note that in the expression above, the indices $i,j,k$ refer to the cells
of the simulation with the lowest resolution. It is also important to remark
that the time-dependent exponent $\tilde{p}=\tilde{p}(t)$ is a genuine
measure of the convergence order of our code and provides a far more
severe assessment of the convergence properties than the instantaneous
measurement shown in Fig. \ref{fig:error-stationary}. While in this work
we have presented both approaches to assess the convergence order, measurements of $\tilde{p}=\tilde{p}(t)$ should accompany
any work where the convergence properties of a numerical code are
presented \citep[see also][]{Radice2013b,Tsokaros2016}.

The infrastructure for refining or coarsening that is present in the code
greatly simplifies the task of performing the convergence tests. Since at
each refinement level the cell widths are halved with respect to those of
the previous level, simulations with higher resolution can be obtained by
enforcing a higher level. In practice, the volume averages for
Eq. \eqref{self_conv} are computed through coarsening each snapshot of
the data to a lower level. Moreover, when comparing the convergence of the
simulations using AMR with that of the uniform cases, a simulation with
three AMR levels was taken as equivalent to a uniform simulation with
the same resolution of the highest AMR level, and the same averages of
Eq. \eqref{self_conv} were then employed for the convergence test.

%ffffffffffffffffffffffffffffffffffffffffffffffffffffffffffffffffff
\begin{figure}
\centering
\includegraphics[width=\hsize]{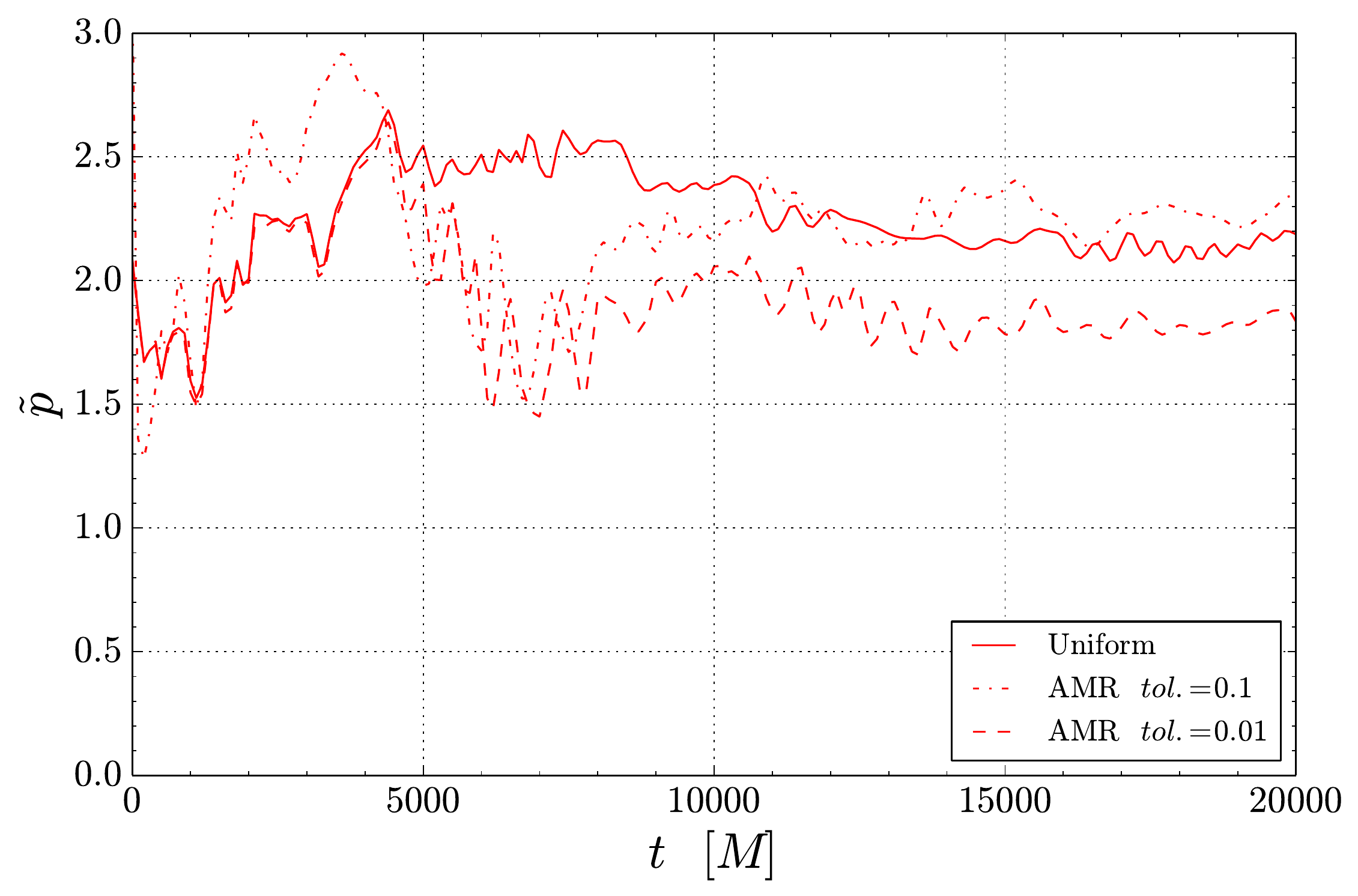}
\caption{Convergence order for the $L_1$ norm of the 2D recoiling black
  hole setup with the kick velocity set to zero. We show the uniform grid
  (solid) and AMR (L\"ohner scheme) with tolerances of $\varepsilon_t=0.1$
  (dashed) and $\varepsilon_t = 0.01$ (dash-dotted). In the uniform grid
  case we used three different resolutions: low ($128 \times 128$), medium
  ($256 \times 256$), and high ($512 \times 512$). In the AMR case, the
  simulation data used were uniform ($128 \times 128$, level 1), and two
  and three AMR levels.}
\label{fig:2Dstationary_conv_AMRL1}
\end{figure}
%ffffffffffffffffffffffffffffffffffffffffffffffffffffffffffffffffff

Figure \ref{fig:2Dstationary_conv_AMRL1} shows the evolution of the
convergence order $\tilde{p}$ of the stationary torus, and where
we compare simulations with uniform grid and those with two AMR
realisations employing a L\"ohner scheme with tolerances $\varepsilon_t=0.1$
and $\varepsilon_t=0.01$.
In
  practice we considered the same setup used when considering a recoiling
  black hole, but then imposed the kick velocity to be zero.
In the uniform-grid case, the resolutions used
are $(N_{r}\times N_{\phi})$: low ($128 \times 128$), medium ($256 \times
256$), and high ($512 \times 512$). In the AMR case, the same
low-resolution case ($128 \times 128$) was employed, so that the medium
and high resolutions are achieved by allowing two and three mesh
refinements, respectively.

In this test case, where the torus is stationary and the solution is
smooth everywhere except for the torus surface, the convergence order
settles to $\sim 2.2$ in the long-time evolution for the uniform-grid
case. This is in good agreement with our expected convergence order,
since we have here employed Koren's slope limiter \citep{Koren1993},
which has third-order spatial accuracy in the absence of extrema. For the
high-tolerance case with $\varepsilon_t=0.1$, the AMR run displays a
convergence index of only $\tilde{p}\approx1.8$ in the long-time
evolution. However, lowering the tolerance to $\varepsilon_t=0.01$, we
recover the higher-convergence order measured in the uniform-grid case.

%ffffffffffffffffffffffffffffffffffffffffffffffffffffffffffffffffff
\begin{figure}
\centering
\includegraphics[width=\hsize]{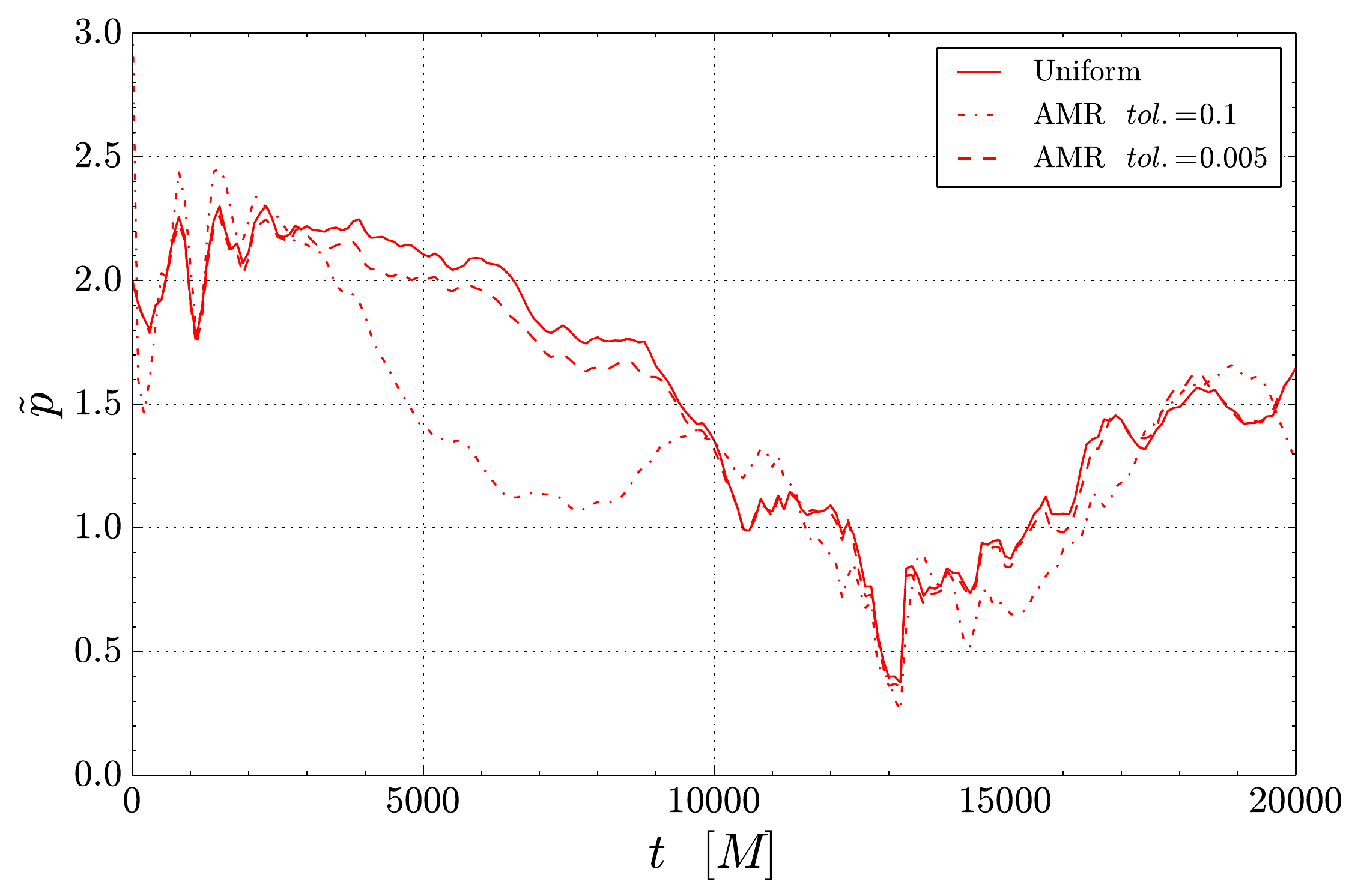}
\caption{Convergence ($L_1$ norm) of the 2D recoiling black hole setup
  with kick velocity set to $v_{\rm recoil}=0.001\rm c$. We show the
  uniform grid (solid) and AMR (L\"ohner scheme) with tolerances
  $\varepsilon_t=0.1$ (dashed) and $\varepsilon_t=0.005$ (dash-dotted). In
  the uniform grid case we used three different resolutions: low ($256
  \times 128$), medium ($512 \times 256$), and high ($1024 \times
  512$). In the AMR case, the simulation data used were uniform ($256
  \times 128$, level one), AMR with two levels, and AMR with three
  levels.}
\label{fig:2Drecoil_conv_AMRL1}
\end{figure}
%ffffffffffffffffffffffffffffffffffffffffffffffffffffffffffffffffff

Figure \ref{fig:2Drecoil_conv_AMRL1} shows the corresponding convergence
order when a kick velocity of $v_{\rm R}=10^{-3}c$ is used. Again, we
show a uniform case and two AMR cases with tolerances of
$\varepsilon_t=0.1$ and $\varepsilon_t=0.005$. In the uniform-grid case,
three different resolutions are employed, with $N_{r}\times N_{\phi}=256
\times 128$ (low), $512 \times 256$ (medium), and $1024 \times 512$
(high); equivalent AMR realisations are generated allowing one, two, and
three mesh refinements. 

As clearly shown in Fig. \ref{fig:2Drecoil_conv_AMRL1}, the convergence
order remains higher than 2 in the early stages of the simulation, when
the black hole has not yet interacted with the torus matter and the
spiral shocks have not yet developed. Most of the simulation region is
smooth, hence yielding a high convergence order. In the ensuing stage,
the strong shock has developed in the accreting disc and leads to a
deterioration of the convergence order, which decreases to being $\sim
1$, as expected from Godunov's theorem \citep[see
  \eg][]{Rezzolla_book:2013}. After this stage, the spiral shock expands
and weakens, and the convergence order increases as the simulation
progresses. At the end of the simulation, the convergence order has
recovered its stationary value of $\sim 2$. The AMR simulations show a
similar trend and, in particular, the low-tolerance case is very close to
the uniform-grid case at all times.

\begin{figure}
\centering
\includegraphics[width=\hsize]{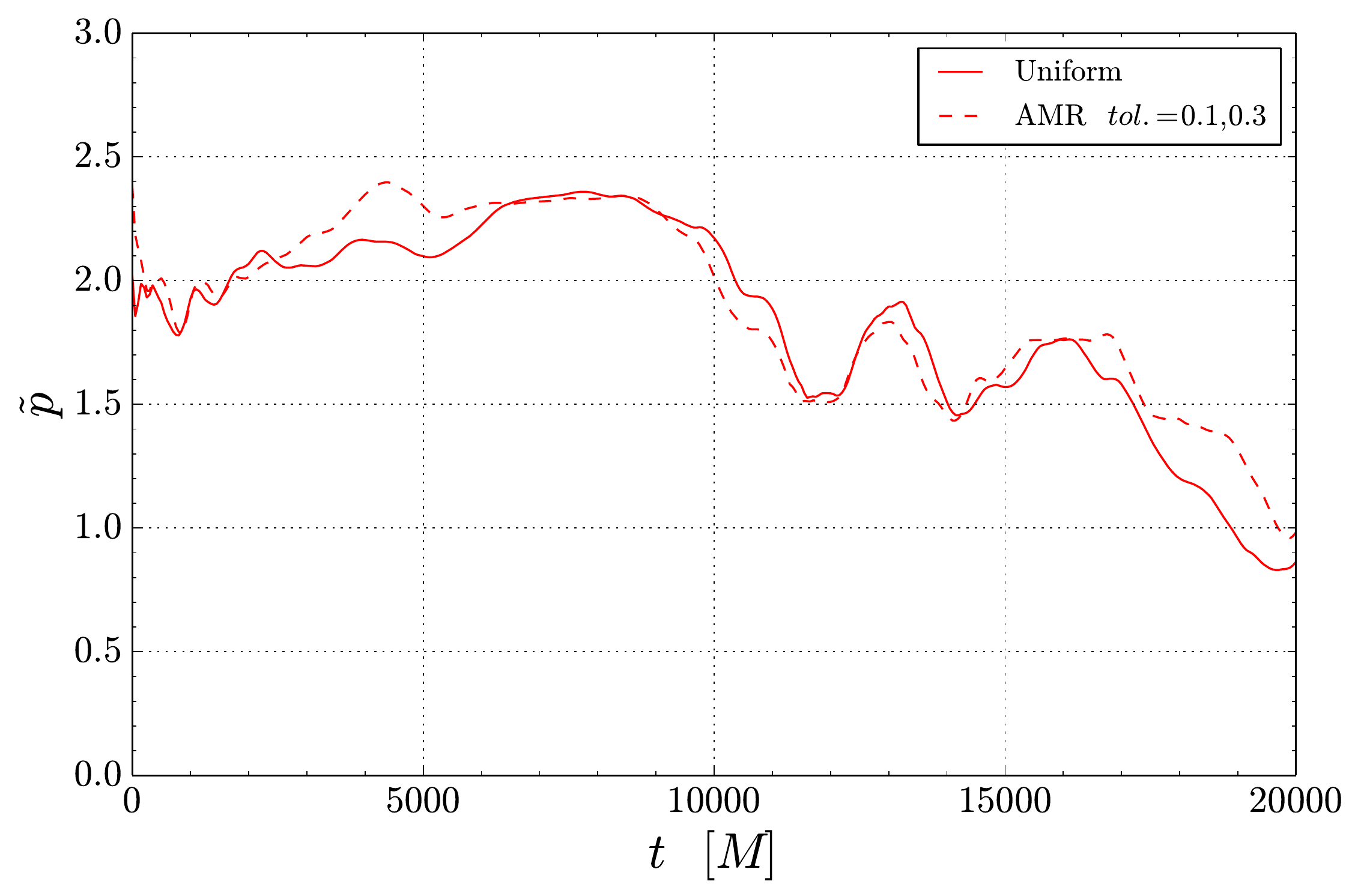}
\caption{Numerical order of accuracy of the 3D recoiling black hole
  simulation calculated from the $L_1$ norm using grid (uniform) and AMR
  (dashed) cases. In the uniform grid case, we used three different
  resolutions, low ($128 \times 64 \times 16$), medium ($256 \times 128
  \times 32$), and high ($512 \times 256 \times 64$). In the AMR runs,
  tolerances of $\varepsilon_t=0.1$ and $\varepsilon_t=0.3$ are set to
  trigger the second and third refinement levels, respectively. The
  simulation data used were uniform ($128 \times 64 \times 16$, level 1),
  AMR with two levels, and AMR with three levels.}
\label{fig:3Drecoil_conv_AMRL1}
\end{figure}

Finally, the convergence results for the 3D recoiling black hole
simulations are shown in Fig. \ref{fig:3Drecoil_conv_AMRL1}. As seen in
the 2D case, the convergence order remains higher than $\sim 2$ in the
early stages of the simulation as most of the solution is smooth. After
$t=10,\!000\,M$, the convergence order gradually decreases, but remains
higher than for the 2D case. This is a consequence of the slightly
different dynamics of the 3D case. Even though a large spiral
shock develops in 3D at around $t=1,\!000\,M$, the actual accretion onto
the black hole, which is responsible for the formation of much of the
shock structure seen in 2D, starts only much later at $t\approx
19,\!500\,M$ reported here. As a result, at the end of the simulation,
$\tilde{p}$ approaches $\sim 1$. The evolution of the convergence order
$\tilde{p}$ was also computed for the uniform and AMR simulations, in
the same way as described for the 2D case. Clearly, the evolution
$\tilde{p}$ for the AMR simulations closely follows that of the
uniform cases.

%ffffffffffffffffffffffffffffffffffffffffffffffffffffffffffffffffff
%ffffffffffffffffffffffffffffffffffffffffffffffffffffffffffffffffff

\end{appendix}
%aaaaaaaaaaaaaaaaaaaaaaaaaaaaaaaaaaaaaaaaaaaaaaaaaaaaaaaaaaaaaaaaaa

\end{document}